\documentclass[useAMS,usenatbib]{mn2e}

\usepackage{times}
\usepackage{graphicx}
\usepackage{subfigure}
\usepackage[fleqn]{amsmath}

\newcommand{\gsim}{\mathrel{\hbox{\rlap{\lower.55ex \hbox {$\sim$}}
                   \kern-.3em \raise.4ex \hbox{$>$}}}}
\newcommand{\lsim}{\mathrel{\hbox{\rlap{\lower.55ex \hbox {$\sim$}}
                   \kern-.3em \raise.4ex \hbox{$<$}}}}

\title[Radiation magnetohydrodynamic outflows]{Collapse of a molecular cloud core to stellar densities: stellar core and outflow formation in radiation magnetohydrodynamics simulations}
\author[M.R. Bate, T.S. Tricco \& D.J. Price ]{Matthew R. Bate$^{1}$\thanks{E-mail:
mbate@astro.ex.ac.uk}, Terrence S. Tricco$^2$ and Daniel J. Price$^2$ \\ $^1$School of Physics and Astronomy, University of Exeter, Stocker
Road, Exeter EX4 4QL \\ $^2$Monash Centre for Astrophysics, School of Mathematical Sciences, Monash University, Clayton, Vic 3800, Australia}

\date{\today}
\begin{document}
\maketitle
\begin{abstract}
We have performed smoothed particle radiation magnetohydrodynamics (SPRMHD) simulations of the collapse of rotating, magnetised molecular cloud cores to form protostars.  The calculations follow the formation and evolution of the first hydrostatic core, the collapse to form a stellar core, the launching of outflows from both the first hydrostatic core and stellar cores, and the breakout of the stellar outflow from the remnant of the first core.  We investigate the roles of magnetic fields and thermal feedback on the outflow launching process, finding that both magnetic and thermal forces contribute to the launching of the stellar outflow.  We also follow the stellar cores until they grow to masses of up to 20 Jupiter-masses, and determine their properties.  We find that at this early stage, before fusion begins, the stellar cores have radii of $\approx 3$~R$_\odot$ with radial entropy profiles that increase outward (i.e. are convectively stable) and minimum entropies per baryon of $s/k_{\rm B} \approx 14$ in their interiors.  The structure of the stellar cores is found to be insensitive to variations in the initial magnetic field strength.  With reasonably strong initial magnetic fields, accretion on to the stellar cores occurs through inspiralling magnetised pseudo-discs with negligible radiative losses, as opposed to first cores which effectively radiate away the energy liberated in the accretion shocks at their surfaces. We find that magnetic field strengths of $>$ 10~kG can be implanted in stellar cores at birth.
\end{abstract}

\section{Introduction}
\label{introduction}

More than four decades ago, \cite{Larson1969} performed the first numerical calculations of the collapse of a molecular cloud core to stellar core formation and beyond. These spherically-symmetric, radiation hydrodynamical calculations revealed the main stages of protostar formation: an almost isothermal collapse until the inner regions become optically thick; the almost adiabatic formation of the first hydrostatic core (typical radius $\approx 5$~au and initial mass $\approx 5$ Jupiter-masses [$M_{\rm J}$]); growth of this core as it accreted from the infalling envelope; the second collapse within this core triggered by the dissociation of molecular hydrogen; the formation of the stellar core (initial radius $\approx 2$~R$_\odot$ and mass $\approx 1.5~M_{\rm J}$) and, lastly, the long accretion phase of the stellar core to its final mass. More recent one-dimensional \citep{MasInu2000,Commerconetal2011b,Vaytetetal2012,Vaytetetal2013} and even multi-dimensional studies have not altered this general picture.

Multi-dimensional calculations do, however, allow for the effects of rotation and magnetic fields to be investigated.  Rotation alters the structure of the first hydrostatic core and, potentially, allows fragmentation.  As the degree of rotation is increased, two-dimensional calculations show that the first core becomes more disc-like in structure \citep{Larson1972, Tscharnuter1987, Tscharnuteretal2009}.  Three-dimensional calculations show that if a first core rotates rapidly enough it becomes dynamically unstable to a bar-mode, leading to the formation of trailing spiral arms \citep*{Bate1998, SaiTom2006, SaiTomMat2008, MacInuMat2010, Bate2010, Bate2011}.   Gravitational torques remove angular momentum and rotational support from the inner regions of the first core, quickening the onset of the second collapse and preventing fragmentation during the second collapse to form close binaries \citep{Bate1998}.  Since rapidly-rotating first cores are disc-like in structure, when the stellar core forms due to the second collapse it is already embedded within a disc --- the disc forms before the star \citep{Bate1998, Bate2011, MacInuMat2010}.

Magnetic fields provide another mechanism for angular momentum transport and also drive outflows.  Outflows with typical speeds of $v \sim 2$~km/s are found to be launched from the first core \citep{Tomisaka2002, Machidaetal2005, BanPud2006, MacInuMat2006, MacInuMat2008, HenFro2008, Commerconetal2010, Burzleetal2011b, PriTriBat2012}, while the formation of the stellar core allows faster outflows to be driven with typical speeds of $v=10-30$~km/s \citep{BanPud2006, MacInuMat2006, MacInuMat2008}.

Although \citeauthor{Larson1969}'s original paper included radiative transfer, three-dimensional calculations of the collapse of molecular cloud cores have only included radiative transfer relatively recently.  \cite{WhiBat2006} and \cite{Stamatellosetal2007} followed collapse to the point of stellar core formation, but not beyond.   \cite{Bate2010, Bate2011} found that heating of the inner regions of the first core due to the high accretion rates during stellar core formation was sufficient to launch short-lived bipolar outflows even without magnetic fields \cite[see also][]{SchTsc2011}.  \citeauthor{Tomidaetal2010a} (\citeyear{Tomidaetal2010a}a, \citeyear{Tomidaetal2010b}b) and \citeauthor{Commerconetal2010} (\citeyear{Commerconetal2010}, \citeyear{Commerconetal2012}) performed radiation magnetohydrodynamical (RMHD) calculations of first core formation, again demonstrating the launching of outflows from the first core, but treating the thermodynamics more realistically.  Most recently, \cite{Tomidaetal2013} have performed radiation magnetohydrodynamical calculations that follow the collapse to stellar core formation and the launching of the faster outflow from the vicinity of the stellar core.  However, they were only able to follow the fast outflow for a fraction of an AU.

In this paper, we follow up our previous paper which reported the first long-lived magnetised jet calculations performed using the smoothed particle magnetohydrodynamics (SPMHD) method \citep{PriTriBat2012}.  
These earlier calculations used a simple barotropic equation of state and only resolved the collapse down to scales of $\sim 1$~AU.  
In this paper, we combine the radiation hydrodynamical method of \cite{WhiBatMon2005} and the magnetohydrodynamical method of \cite{TriPri2012}, to perform the first smoothed particle radiation magnetohydrodynamical calculations (SPRMHD) of protostellar collapse to stellar core formation and beyond.  For the first time, we also follow RMHD calculations long enough for the fast outflow to exit the first core ($\approx 4$~AU).  We also resolve the stellar cores, follow their growth in mass up to 0.02~M$_\odot$, and measure their entropy which is of crucial importance for the initial condition of pre-main-sequence stellar evolution models.

\begin{table*}
\begin{tabular}{lccccc}\hline
Calculation & Mass-to-flux ratio & Initial magnetic & First core  & Stellar core & Length of calculation   \\
&  & field strength & outflow speed & outflow speed & beyond stellar core formation  \\
 & $\mu$ & $\mu$G  & km/s & km/s & yr \\ \hline
Hydro & $\infty$ & 0  & None & --- & 22~~~~~~\\ 
MF100 & 100 & 8.1 & None & --- & 2.5 \\ 
MF20 & 20 & 41 & 0.8-1.8 & 9 & 1.6  \\ 
MF10 & 10 & 81 & 1.2-1.8 & 9-10 & 1.9 \\  
MF05 & 5 & 163 & 1-2.5 & 10-11 & 2.0 \\ \hline 
\end{tabular}
\caption{\label{table1} The parameters and a summary of the main results for the radiation magnetohydrodynamical calculations carried out for this paper.  The initial conditions for all calculations were identical except for the magnetic field strength.  The calculations were run for different numbers of years beyond stellar core formation due to computational limitations, but all of the magnetised calculations were followed until the stellar jet had broken out of the remains of the first hydrostatic core. We also give the speeds of the outflows found from the first hydrostatic core and the stellar core.}
\end{table*}

\section{Computational method}
\label{method}

The calculations presented here were performed 
using a three-dimensional smoothed particle
hydrodynamics (SPH) code based on the original 
version of \citeauthor{Benz1990} 
(\citeyear{Benz1990}; \citealt{Benzetal1990}), but substantially
modified as described in \citet{BatBonPri1995},
\citet*{WhiBatMon2005}, \citet{WhiBat2006}, \cite{PriMon2007},
\cite{PriBat2007}, \cite{TriPri2012}, and 
parallelised using both OpenMP and MPI.

\subsection{Equations of radiation magnetohydrodynamics}

We solve the equations of self-gravitating, ideal MHD with flux-limited radiation hydrodynamics
given by
\begin{eqnarray}
\frac{D\rho}{Dt} & = & -\rho \nabla\cdot {\bf \mbox{\boldmath $v$} }, \label{eq:cty} \\
\frac{D{\bf \mbox{\boldmath $v$} }}{Dt} & = & -\frac{1}{\rho}\nabla\left(P + \frac12 \frac{B^{2}}{\mu_{0}} - \frac{\mbox{\boldmath $B$}\mbox{\boldmath $B$}}{\mu_{0}}\right) - \nabla\phi + \frac{\kappa \mbox{\boldmath$F$}}{c}, \label{eq:mom} \\
\frac{D}{Dt} \left(\frac{\bf \mbox{ \boldmath $B$} }{\rho} \right) & = & \left( \frac{{\bf \mbox{ \boldmath $B$} }}{\rho}\cdot \nabla \right) \mbox{\boldmath $v$} \label{eq:ind}, \\
\rho \frac{D}{Dt}\left( \frac{E}{\rho}\right) & = & -\mbox{\boldmath $\nabla\cdot F$} - \mbox{\boldmath $\nabla v${\bf :P}} + 4\pi \kappa \rho B - c \kappa \rho E~, \label{eq:radiation} \\
\rho \frac{Du}{Dt} & = & -p \mbox{\boldmath $\nabla\cdot v$} - 4\pi \kappa \rho B + c \kappa \rho E~, \label{eq:matter} \\
\nabla^{2}\phi & = & 4\pi G\rho, \label{eq:grav}
\end{eqnarray}
where $D/Dt \equiv \partial/\partial t + \mbox{\boldmath $v \cdot \nabla$}$ is the convective derivative, $\rho$ is the density, $\mbox{\boldmath $v$}$ is the velocity, $u$ is the specific energy of the gas, $P$ is the hydrodynamic pressure, $\mbox{\boldmath $B$}$ is the magnetic field, $\phi$ is the gravitational potential, $E$ is the radiation energy density, $\mbox{\boldmath $F$}$ is the radiative flux, {\bf P} is the radiation pressure tensor, $\mu_{0}$ is the permeability of free space, $c$ is the speed of light, and $G$ is the gravitational constant. The assumption of local thermodynamic equilibrium (LTE) allows us to use the 
Planck function $B=(\sigma_{\rm \scriptscriptstyle B}/\pi)T_{\rm
g}^4$ to describe the emission of radiation from the matter,  where $T_{\rm g}$ is the gas temperature and $\sigma_{\rm \scriptscriptstyle B}$ is the Stefan-Boltzmann
constant. The radiation energy density also has an associated temperature
$T_{\rm r}$  from the equation $E=4 \sigma_{\rm \scriptscriptstyle B} T_{\rm
r}^4/c$.  Equations \ref{eq:mom}, \ref{eq:radiation}, and \ref{eq:matter} have been integrated over frequency and the opacity $\kappa$ is assumed to be independent of frequency and there is no distinction between absorption and total opacities (i.e. absorption plus scattering).

The density of each SPH particle is computed by summation over nearest neighbouring particles.  The smoothing length of each particle is variable in time and space, iteratively solving $h = 1.2 (m/\rho)^{1/3}$ where $m$ and $\rho$ are the SPH particles mass and density, respectively. 

Gravitational forces are calculated using a binary tree.  The gravitational potential is softened using the SPH kernel such that the softening varies with the smoothing length  \cite[see][for further details]{PriMon2007}.

We solve the MHD equations using a standard smoothed particle magnetohydrodynamics (SPMHD) scheme, evolving $\mbox{\boldmath $B$}/\rho$ as the magnetic field variable (Eq.~\ref{eq:ind}), using the \citet*{BorOmaTru2001} source-term approach for stability, and with artificial viscosity and resistivity terms added to capture shocks and magnetic discontinuities, respectively \citep{PriMon2005,Price2012}.  The artificial viscosity and resistivity parameters are spatially varying and time dependent as described in \citet{Price2012}, using the \citet{MorMon1997} method for viscosity and a new method for resistivity whereby the resistivity parameter is set as $\alpha_{\rm B}=h \vert \nabla \mbox{\boldmath $B$} \vert / \vert \mbox{\boldmath $B$} \vert$ \citep{TriPri2013}. This approach is well suited for tracking magnetic shocks as the magnetic field grows in strength.  We use values of $\alpha_{\rm AV} \in [0.1,1]$ and $\alpha_{\rm B} \in [0,1]$. 

The solenoidal constraint on the magnetic field is maintained by using the constrained hyperbolic divergence cleaning method of \citet{TriPri2012}, an adaptation of a similar method developed for grid-based codes \citep{Dedneretal2002}.  The method removes divergence of the magnetic field by coupling a scalar field, $\psi$, to the magnetic field according to
\begin{align}
 \frac{{\rm d}\mbox{\boldmath $B$}}{{\rm d}t} =& - \nabla \psi , \label{eq:clean1} \\
\frac{{\rm d}\psi}{{\rm d}t} =& - c_h^2 \nabla \cdot \mbox{\boldmath $B$} - \frac{\psi}{\tau} - \tfrac{1}{2} \psi \left(\nabla \cdot \mbox{\boldmath $v$}\right) , \label{eq:clean2}
\end{align}
which propagates divergence through the magnetic field as a series of damped waves.  \emph{Constrained} divergence cleaning means that the discrete operators used to compute $\nabla\psi$ (Eq.~\ref{eq:clean1}) and $\nabla\cdot{\bf B}$ (Eq.~\ref{eq:clean2}) are constructed so that the non-dissipative hyperbolic part of the cleaning scheme is conservative \citep{TriPri2012}.
 The cleaning wave speed, $c_h$, is typically set to the fast MHD wave speed, but for these simulations we used 30 times this value, which significantly enhances the effectiveness of the method, but with a corresponding reduction in timestep. The damping timescale is $\tau = h / \sigma c_h$, where we use $\sigma=0.8$ which is in the range of critical damping. 


The radiation hydrodynamic terms in the above equations describe two-temperature (matter and radiation) radiative transfer in the flux-limited diffusion approximation in which
\begin{equation}
\label{eddington}
\mbox{\boldmath $F$} = -\frac{c \lambda}{\kappa \rho} \nabla E,
\end{equation}
and we use the flux limiter of \cite{LevPom1981}
\begin{equation}
\lambda(R) = \frac{2+R}{6 + 3R + R^2},
\end{equation}
where $R$ is the dimensionless quantity $R = |\nabla E|/(\kappa \rho E)$.
The radiation pressure tensor may then be written in terms of the radiation
energy density as
\begin{equation}
\label{fld3}
\mbox{\rm \bf P} = \mbox{ \rm \bf f} E,
\end{equation}
where the components of the Eddington tensor, {\bf f}, are given by
\begin{equation}
\label{fld4}
\mbox{\rm \bf f} = \frac{1}{2}(1-f)\mbox{\bf I} + \frac{1}{2}(3f-1)\mbox{\boldmath $\hat{n}\hat{n}$},
\end{equation}
where $\mbox{\boldmath $\hat{n}$}=\nabla E/|\nabla E|$ is the unit vector
in the direction of the radiation energy density gradient and the dimensionless
scalar function $f(E)$ is called the Eddington factor.  The flux limiter
and the Eddington factor are related by
\begin{equation}
\label{fld5}
f = \lambda + \lambda^2 R^2.
\end{equation}
The matter and radiation energy equations (\ref{eq:radiation},\ref{eq:matter}) are solved using the method of \citet{WhiBatMon2005} and \citet{WhiBat2006}, except that the standard explicit SPH contributions to the gas energy equation due to the work and artificial viscosity are used when solving the (semi-)implicit energy equations to provide better energy conservation. 

The SPH equations are integrated using a second-order Runge-Kutta-Fehlberg integrator with individual time steps for each particle
\citep{BatBonPri1995}.

\subsection{Equation of state and opacities}

The calculations presented in this paper were performed using an ideal gas equation of state for the gas pressure
$p= u \rho$.  At low gas temperatures ($T_{\rm g} \lsim 60$~K) $u = T_{\rm g} \cal{R}/\mu_{\rm g}$, where
$\mu_{\rm g}$ is the mean molecular weight of the gas,
and $\cal{R}$ is the gas constant.  
However, the thermal evolution takes into account the translational,
rotational, and vibrational degrees of freedom of molecular hydrogen 
(assuming a 3:1 mix of ortho- and para-hydrogen; see
\citealt{Boleyetal2007}) so, for example, as the temperature rises above 60~K the internal energy increases more quickly with temperature than a linear dependence because rotational modes are excited.  The thermal evolution also includes molecular
hydrogen dissociation, and the ionisations of hydrogen and helium.  
The hydrogen and helium mass fractions are $X=0.70$ and 
$Y=0.28$, respectively.
The contribution of metals to the equation of state is neglected.
For this composition, the mean molecular weight of the gas is initially
$\mu_{\rm g} = 2.38$.

The radiative transfer scheme assumes that the gas and dust temperatures are the same.  Taking solar metallicity gas, the opacity is set to be the maximum of the interstellar grain opacity tables of \citet{PolMcKChr1985} and, at higher temperatures when the dust has been destroyed, the gas opacity tables of \citet{Alexander1975} (the IVa King model)  \citep[see][for further details]{WhiBat2006}.

\subsection{Initial conditions}
\label{initialconditions}

The initial conditions for the calculations are identical to those employed by \citet{PriTriBat2012}.
A 1~M$_{\odot}$ dense, cold, spherical, uniform density and slowly rotating core is placed in pressure equilibrium with a warm, low-density ambient medium. The core has radius $R_{c} = 4 \times 10^{16} $cm ($2.7 \times 10^{3}$ AU), giving an initial density of $\rho_{0} = 7.4 \times 10^{-18}$ g~cm$^{-3}$ and a gravitational free-fall time of $t_{\rm ff} = 2.4 \times 10^{4}$ yrs. We use an (isothermal) sound speed $c_{\rm s} = \sqrt{p/\rho}= 2.2\times 10^{4}$~cm~s$^{-1}$, which corresponds to $u=4.8 \times 10^8$ erg~g$^{-1}$. The core is placed inside a larger, cubic domain with a side length of $8 \times 10^{16}$ cm and a $30:1$ density ratio between the core and the ambient medium, in pressure equilibrium. For simplicity we use periodic but non-self-gravitating boundary conditions on the global domain. The core is set in solid body rotation about the $z$-axis with $\Omega = 1.77 \times 10^{-13} $rad s$^{-1}$,  corresponding to a ratio of rotational to gravitational energy $\beta_{\rm r} \simeq 0.005$ and $\Omega t_{\rm ff} = 0.14$. 
  
The magnetic field is initially uniform in the $z$-direction, with strength $B_{0}$ characterised by the parameter $\mu$, specifying the mass-to-magnetic flux ratio ($M/\Phi$) in units of the critical value for a uniform spherical cloud \citep[e.g.][]{Mestel1999,MacKle2004},
\begin{equation}
\mu \equiv \left(\frac{M}{\Phi}\right) / \left(\frac{M}{\Phi}\right)_{\rm crit},
\end{equation}
where
\begin{equation}
\left(\frac{M}{\Phi}\right) \equiv \frac{M}{\pi R_{\rm c}^{2} B_{0}}; \hspace{5mm} \left(\frac{M}{\Phi}\right)_{\rm crit} = \frac{2 c_{1}}{3} \sqrt{\frac{5}{\pi G \mu_{0} }},
\label{eq:mphicrit}
\end{equation}
where $c_{1}$ is a parameter determined numerically by \citet{MouSpi1976} to be $c_{1} \simeq 0.53$. We have performed simulations with five magnetic field strengths ($\mu = \infty, 100, 20, 10, 5$), with the values given in Table \ref{table1}.

The initial structure described above results in gas temperatures of $T_{\rm g}=14$~K in the dense core and $T_{\rm g}=323$~K in the low-density ambient medium.  These temperatures differ to those used by \cite{PriTriBat2012} because of our change from a simple barotropic equation of state to a realistic equation of state with a different value of $\mu_{\rm g}$.  Consequently, although the dense core and ambient medium are in pressure equilibrium with one another (i.e. $u\rho$ is a constant), the ratio of the temperatures is less than the density ratio of 30:1 because the ambient medium is hot enough for rotational and vibrational modes of hydrogen to be excited.  The initial radiation energy density in the dense core is set such that it is in equilibrium with the gas (i.e. $T_{\rm r}=T_{\rm g} = 14$~K). The radiative transfer should allow loss of the radiation from the material in dense core as it collapses.  To achieve this, we set the radiation temperature of the ambient medium equal to the initial value in the dense core (i.e. $T_{\rm r}=14$~K).  Thus, the gas and radiation are initially in equilibrium with each other in the dense core, and the dense core is initially in radiative equilibrium with the ambient medium in which it is embedded.  Finally, neither the gas or radiation temperatures of the particles modelling the ambient medium evolve --- their internal energies and radiation temperatures are fixed.

\begin{figure}
\centering \vspace{-0cm}\hspace{-0.3cm}
    \includegraphics[width=8.5cm]{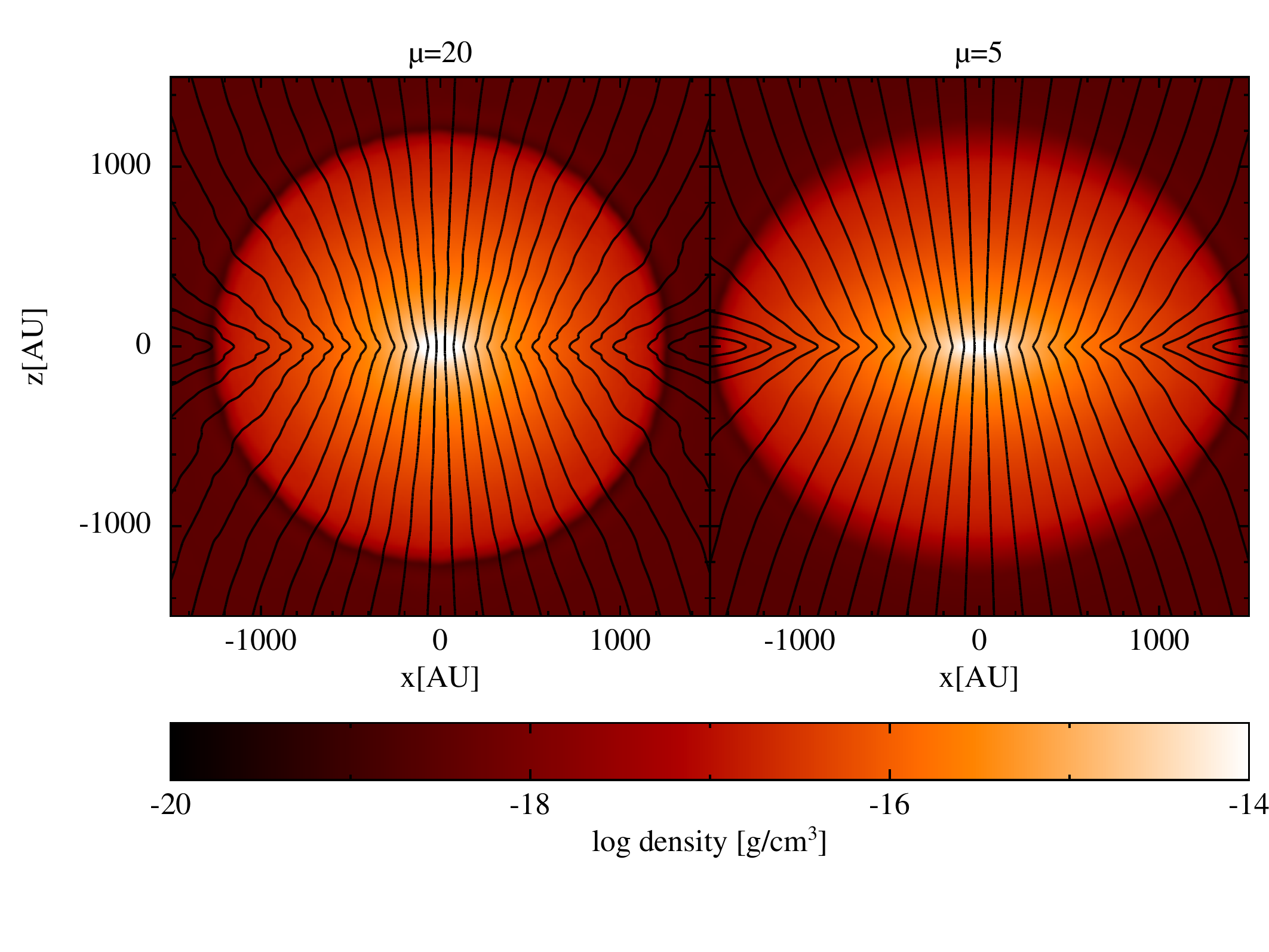}\vspace{-0.5cm}
\caption{The density structure and magnetic field line pinching in the calculations with initial mass-to-flux ratios of 20 (left) and 5 (right) times critical at one initial free-fall time of the molecular cloud core.  The molecular cloud core with a weak field collapses almost spherically symmetrically, whereas with the strongest magnetic field collapse in the $x$-$y$-plane is inhibited and the cloud core becomes oblate.  The solid lines give the direction of the field in the $x$-$z$-plane, clearly showing the expected hour-glass morphology and pinching of the field lines in the $x$-$y$-plane.  Note that the plotted lines are not magnetic field lines, since the field lines have a helical component (which is not shown) due to the rotation of the molecular cloud core. }
\label{fig:global}
\end{figure}

\begin{figure}
\centering \hspace{0.0cm}
    \includegraphics[width=15.0cm]{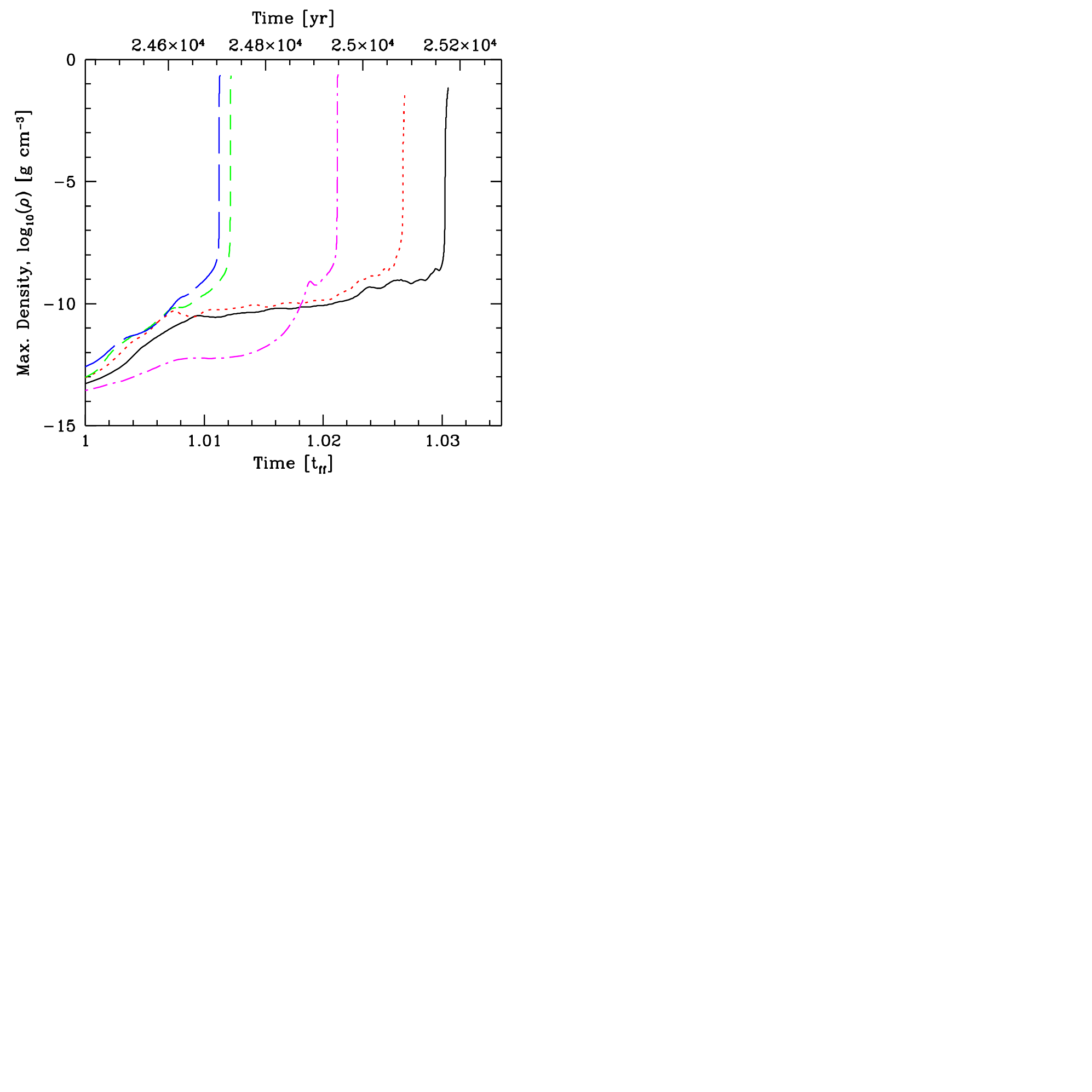}\vspace{-8.5cm}
\caption{The time evolution of the maximum density during the radiation magnetohydrodynamical calculations of the collapse of molecular cloud cores.  The different lines are for cloud cores with different initial mass-to-flux ratios: $\mu=\infty$ (i.e.\@ no magnetic field; black solid), $\mu=100$ (red dotted), $\mu=20$ (green short-dashed), $\mu=10$ (blue long-dashed), and $\mu=5$ (magenta dot-dashed). The free-fall time of the initial cloud core, $t_{\rm ff}=7.71\times 10^{11}$~s (24,430 yrs).  All magnetised calculations form the star more quickly than the unmagnetised calculation due to the angular momentum transport driven by the magnetic field.  Stronger magnetic fields lead to more rapid formation of the stellar core (due to the increased angular momentum transport) with the exception of the most strongly magnetised calculation which takes the second longest of the magnetised calculations due to the magnetic pressure which provides substantial support during the initial collapse of the molecular cloud core.
}
\label{fig:timeevolution}
\end{figure}

\begin{figure*}
\centering \vspace{-0.5cm}
    \includegraphics[width=17.0cm]{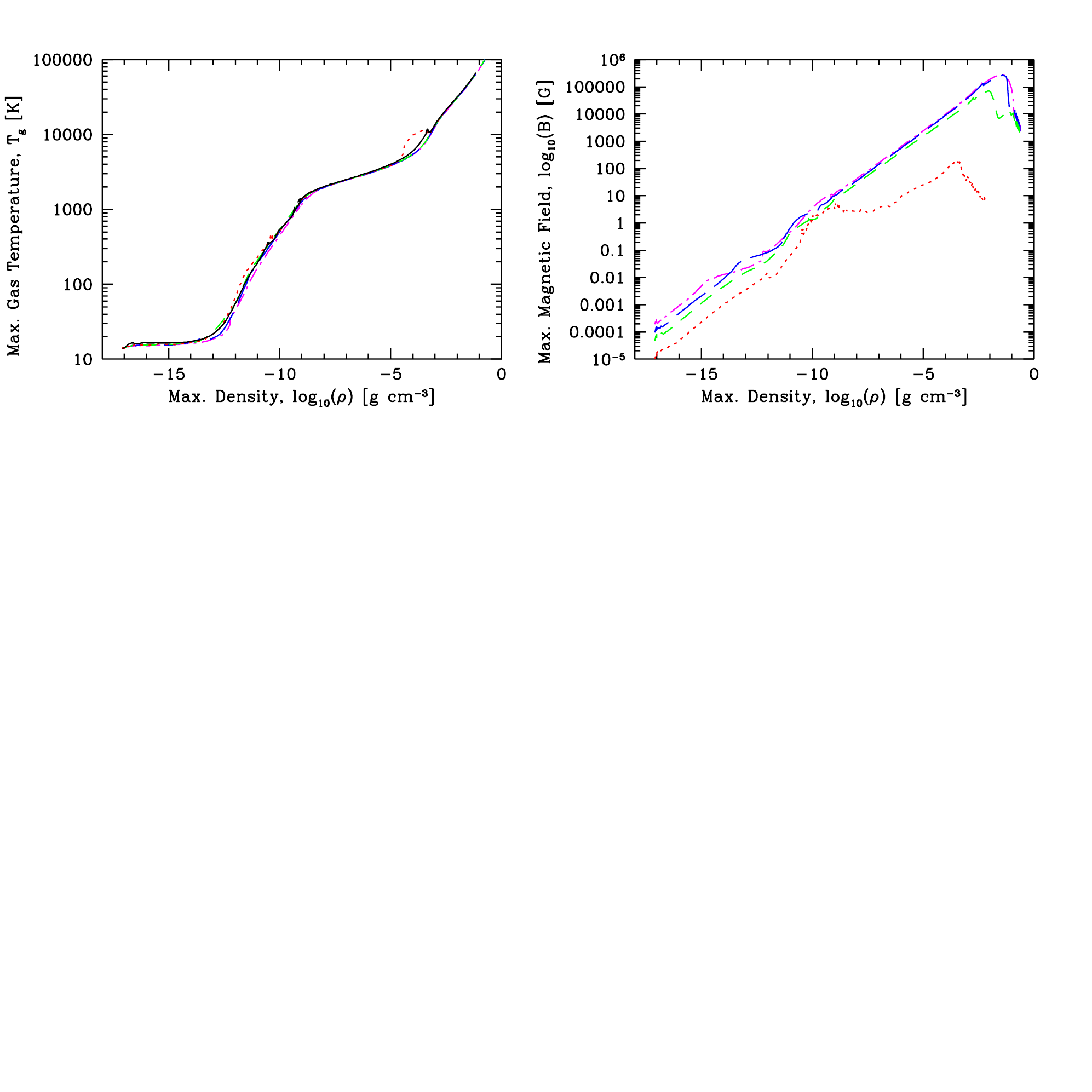}\vspace{-10.5cm}
\caption{The evolution of the maximum gas temperature (left) and maximum magnetic field strength (right) versus maximum density for the RMHD calculations of the collapse of rotating molecular cloud cores with different initial magnetic field strengths.  The initial mass-to-flux ratios, $\mu$, are infinity (i.e. no magnetic field) (black solid lines), 100 (red dotted lines), 20 (green short-dashed lines), 10 (blue long-dashed lines), and 5 (magenta dot-dashed lines) times the critical value.  There is very little dependence of the temperature-density relation on field strength, though the central temperatures of the first hydrostatic cores are slightly cooler at a given density when the magnetic field is stronger. The maximum temperature-density curves after stellar core formation ($\rho \gsim 10^{-3}$~g~cm$^{-3}$) are indistinguishable.  The small departures from a smooth curve for the unmagnetised and $\mu=100$ cases are due to oscillations of the first core and shock formation during stellar core formation.  The maximum magnetic field strength scales with the maximum density approximately as $B_{\rm max} \propto  \rho_{\rm max}^{0.6}$ until after the stellar core forms.  Beyond this point, the field in the star decays, presumably due to the artificial resistivity in the calculations.}
\label{fig:maxdensitytemp}
\end{figure*}

\subsection{Resolution}
\label{resolution}

The results presented in this paper were obtained from calculations performed using equal-mass SPH particles with $3 \times 10^{6}$ particles in the core, and $1.44 \times 10^{6}$ particles in the surrounding medium.  Resolving the Jeans length throughout the collapse, according to the \citet{BatBur1997} criterion, requires $\gsim 3 \times 10^{4}$ particles per solar mass.  Thus, the Jeans mass is well resolved at all times.  However, we note that numerical studies of MHD turbulence have shown that magnetised calculations may require substantially higher resolution than that required to resolve the Jeans mass \citep{Federrathetal2011}.  Therefore, in addition to our $3 \times 10^{6}$ particle calculations, lower resolution calculations (using $1 \times 10^{6}$ particles in the core) were performed with $\mu=5$ and $\mu=10$, and a higher resolution calculation (using $1 \times 10^{7}$ particles in the core) was performed with $\mu=5$ to investigate the dependence of the results on numerical resolution. These gave qualitatively similar results to the results presented in the main text of this paper (see the Appendix for further details), including the structure and speeds of the outflows that were produced, but the higher resolution calculation was not able to be followed far after stellar core formation due the increased computational expense.

The calculations were performed on the University of Exeter Supercomputer, an SGI Altix ICE 8200, and the STFC DiRAC Complexity machine.  Using 64--96 compute cores, the 3 million particle magnetised calculations each required $\approx 150,000$ core-hours (i.e. around 3 months of wall time).  In each case, most of this time was spent after stellar core formation.

\begin{figure*}
\centering \vspace{-0.0cm}
    \includegraphics[width=7.6cm]{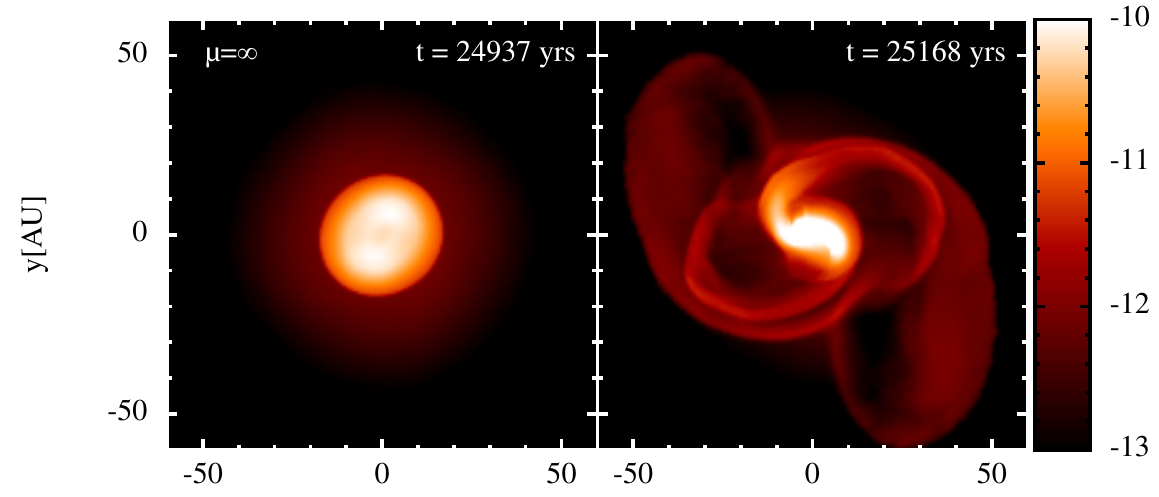}\vspace{-0.0cm}\hspace{0.5cm}
    \includegraphics[width=8.4cm]{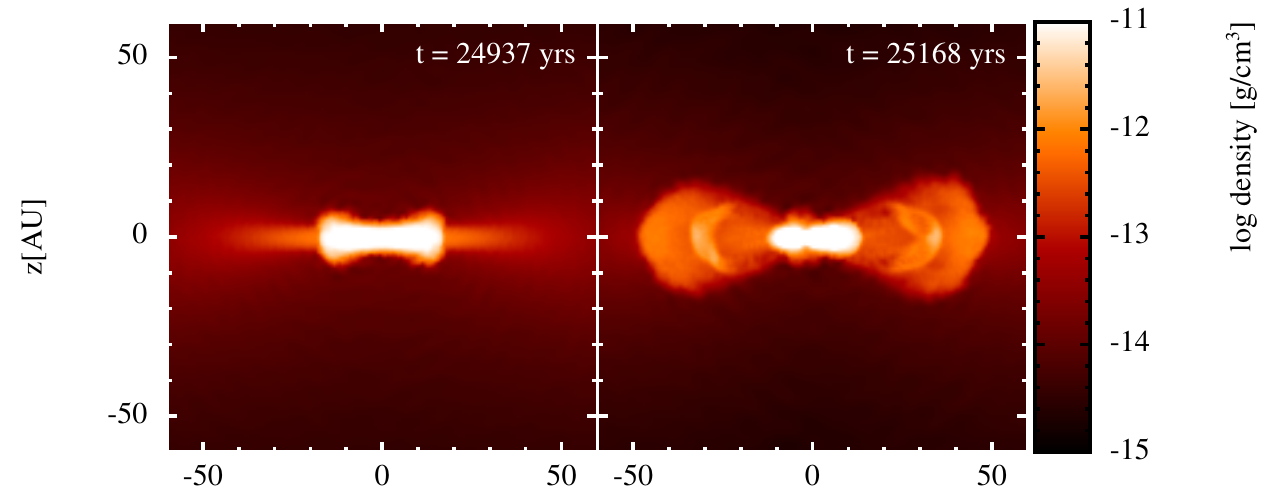}\vspace{-0.1cm}
    \includegraphics[width=7.6cm]{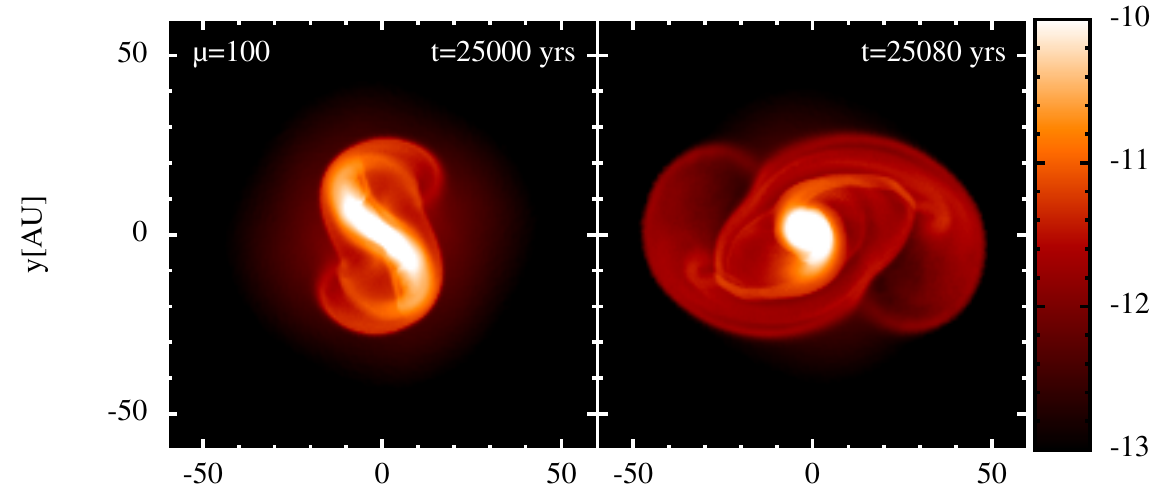}\vspace{-0.0cm}\hspace{0.5cm}
    \includegraphics[width=8.4cm]{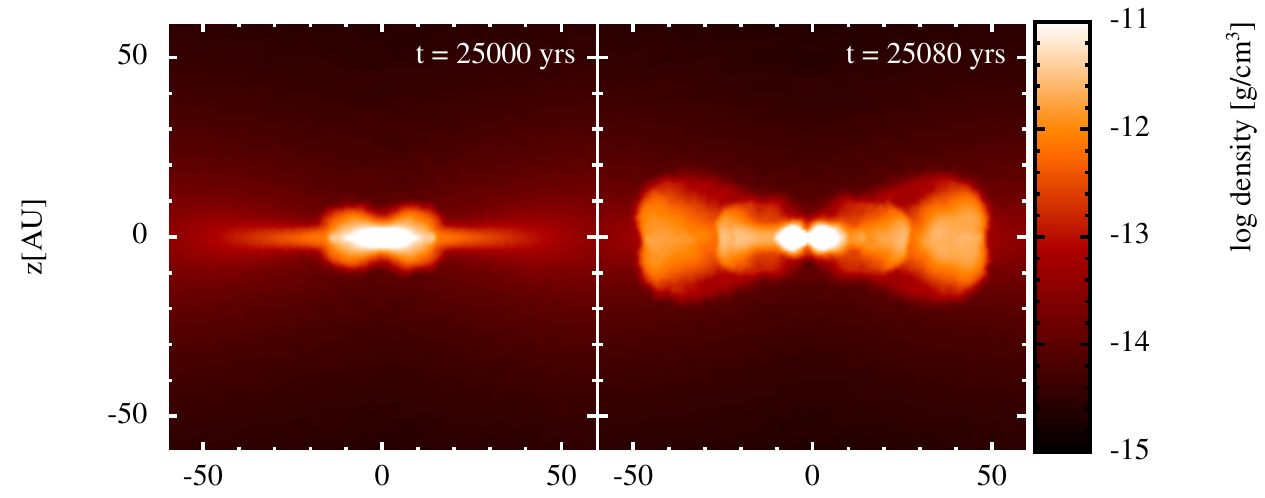}\vspace{-0.1cm}
    \includegraphics[width=7.6cm]{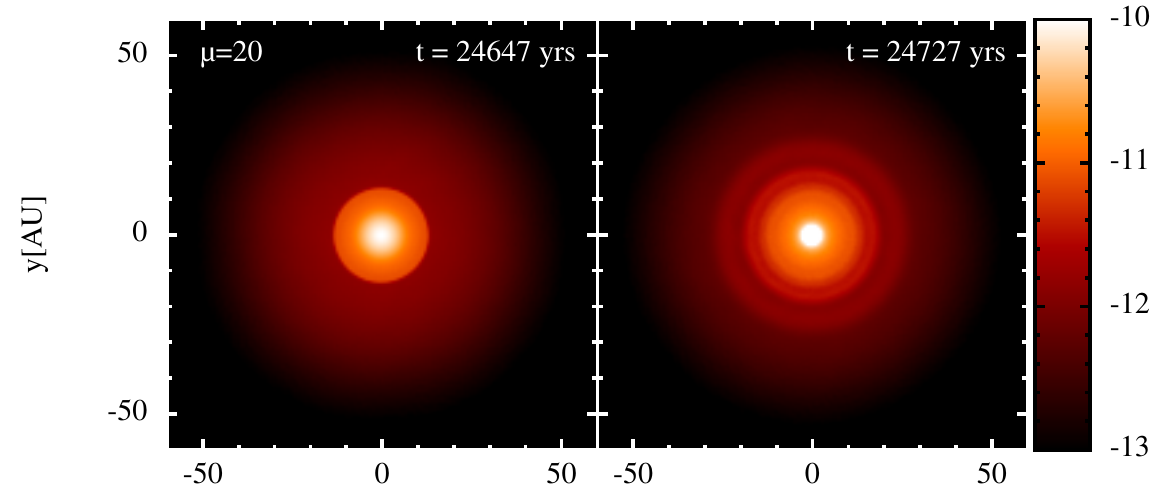}\vspace{-0.0cm}\hspace{0.5cm}
    \includegraphics[width=8.4cm]{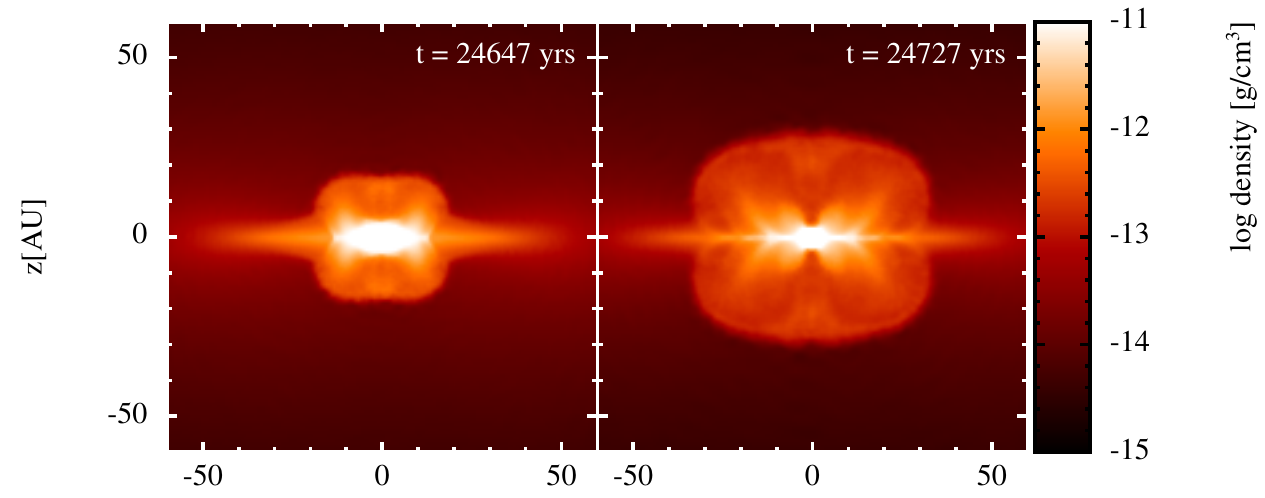}\vspace{-0.1cm}
    \includegraphics[width=7.6cm]{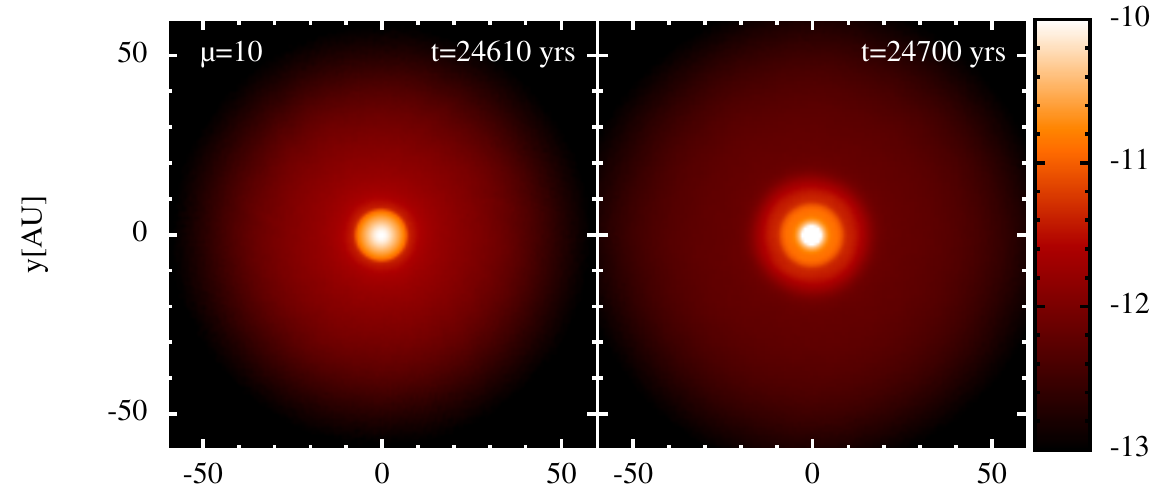}\vspace{-0.0cm}\hspace{0.5cm}
    \includegraphics[width=8.4cm]{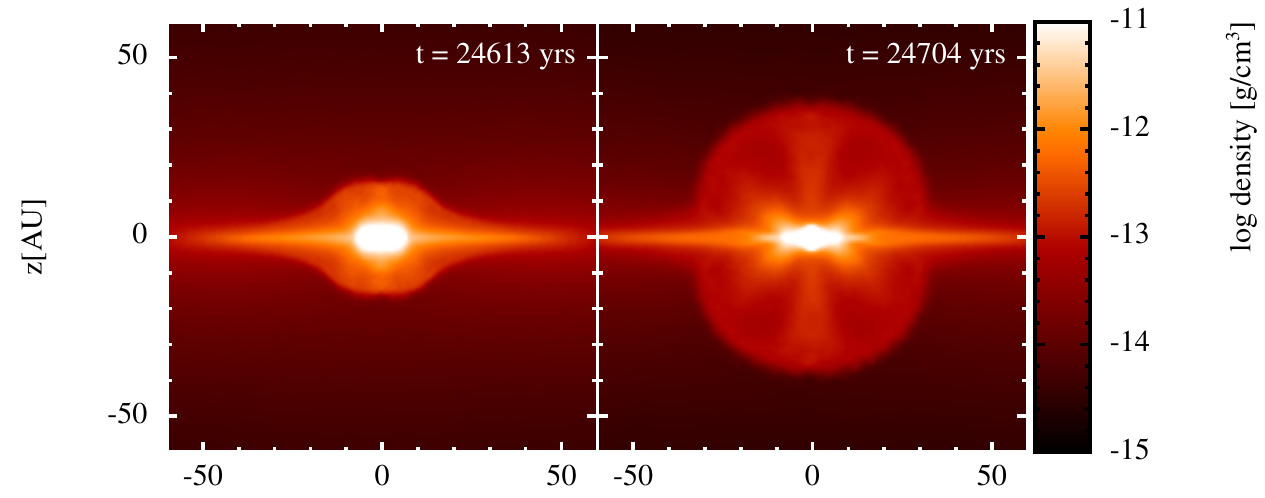}\vspace{-0.1cm}
    \includegraphics[width=7.6cm]{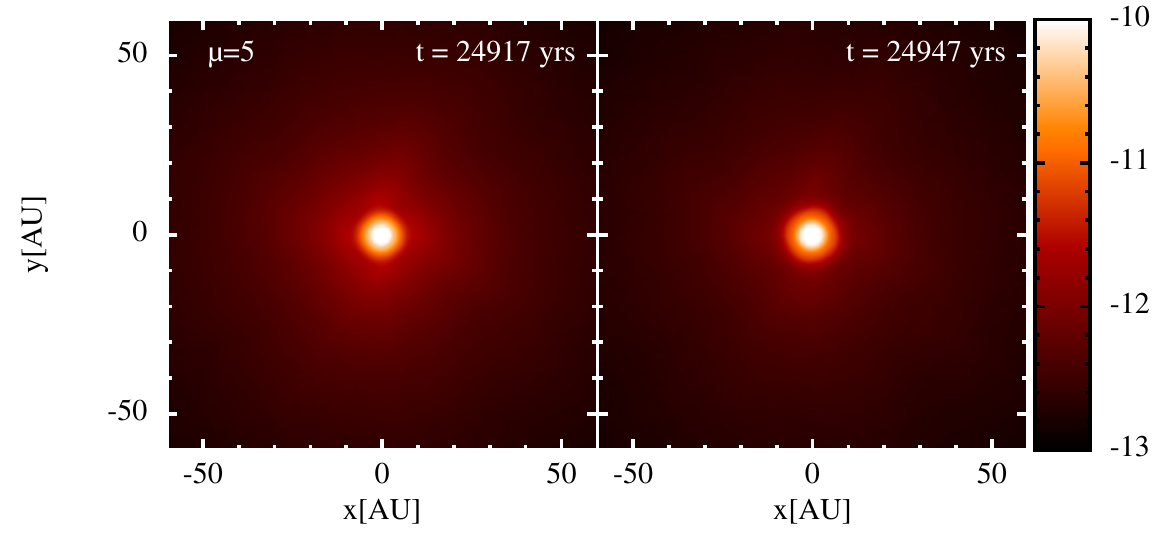}\vspace{-0.0cm}\hspace{0.5cm}
    \includegraphics[width=8.4cm]{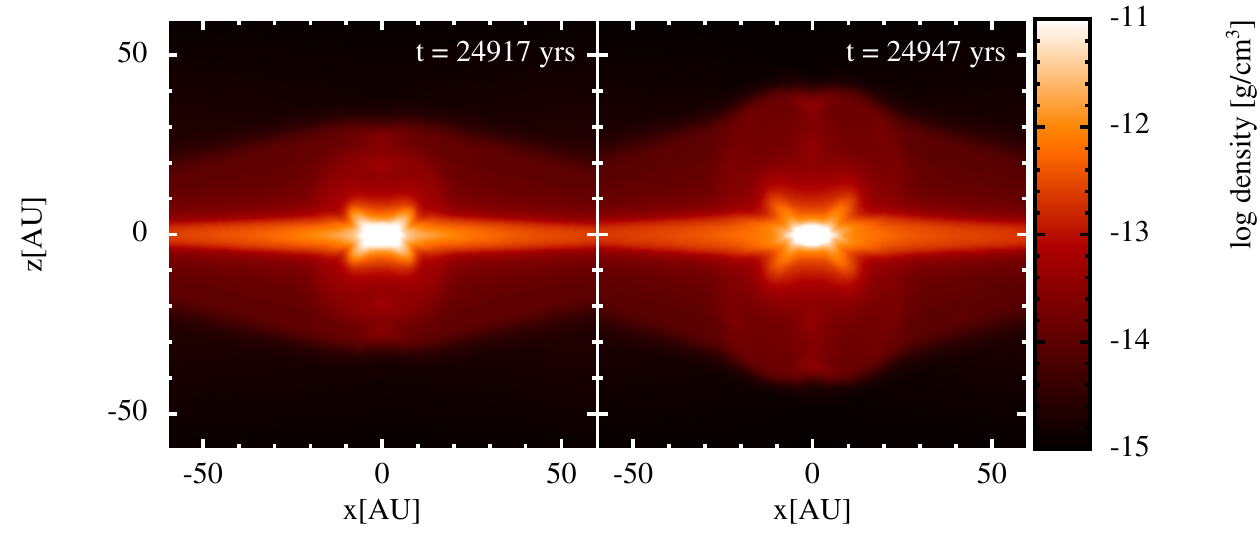}\vspace{-0.0cm}
\caption{Snapshots of the density on slices perpendicular to the rotation axis (in the $x$-$y$-plane; left panels) and down the rotation axis (in the $x$-$z$-plane; right panels) during the evolution of the radiation hydrodynamical calculations of the collapse of molecular cloud cores with different initial magnetic field strengths.  From top to bottom, the different rows are for cloud cores with initial mass-to-flux ratios of $\mu= \infty$, 100, 20, 10, and 5 times critical.  From left to right, the snapshots are taken when the maximum density reaches 10$^{-10}$, and $10^{-7}$~g~cm$^{-3}$ except for the first panel of the $\mu=100$ case for which the maximum density is 10$^{-9}$~g~cm$^{-3}$  (to show the bar-mode instability) and the first panel of the $\mu=5$ case for which the maximum density is 10$^{-9}$~g~cm$^{-3}$ (to better show the development of the pseudo-disc and the outflow from the first core). }
\label{fig:images_xy}
\end{figure*}

\begin{figure}
\centering \vspace{-0.0cm}
    \includegraphics[width=8.4cm]{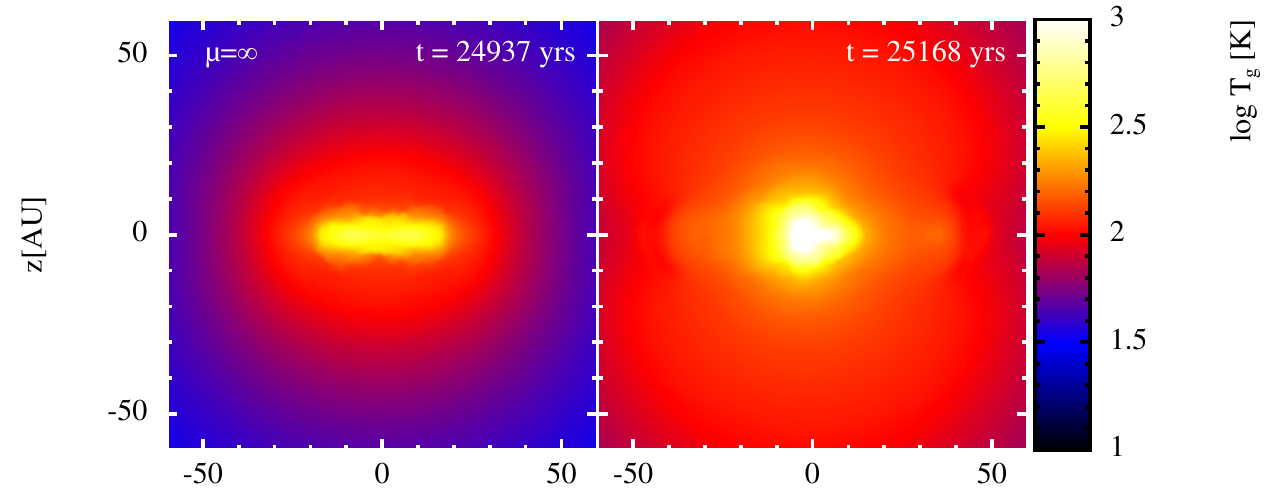}\vspace{-0.1cm}
    \includegraphics[width=8.4cm]{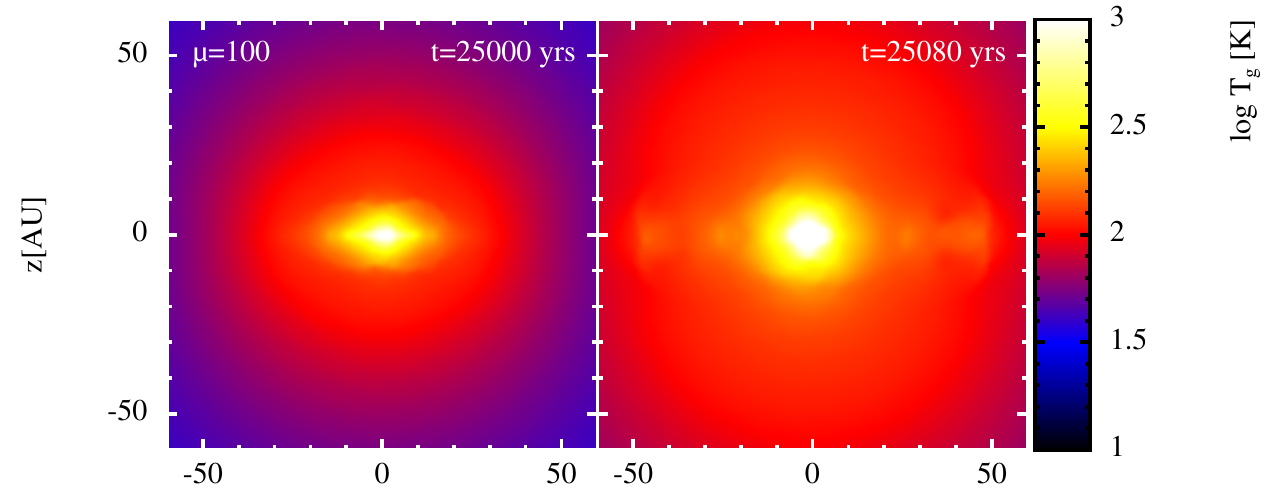}\vspace{-0.1cm}
    \includegraphics[width=8.4cm]{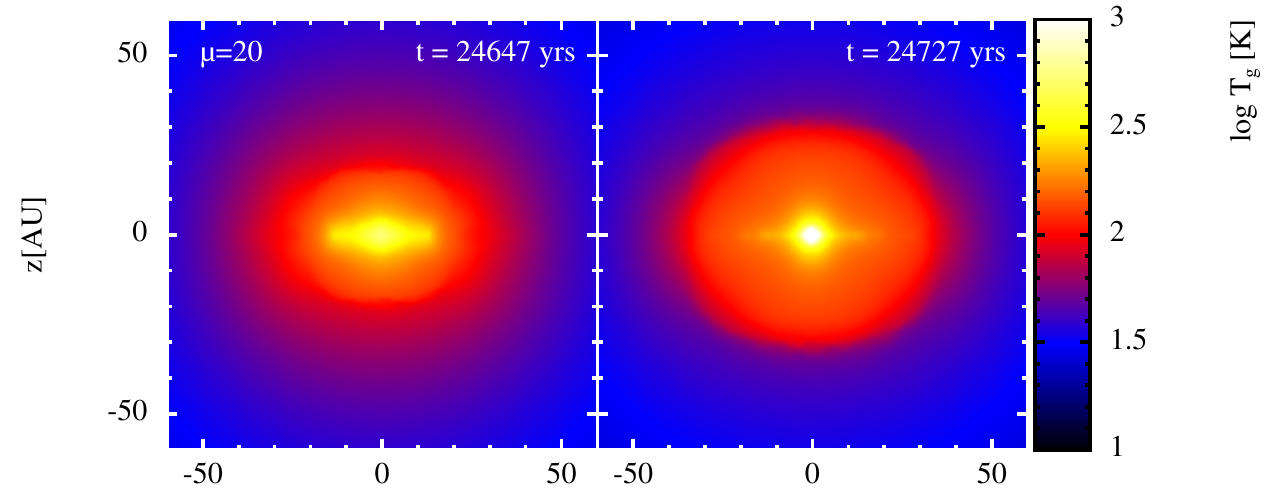}\vspace{-0.1cm}
    \includegraphics[width=8.4cm]{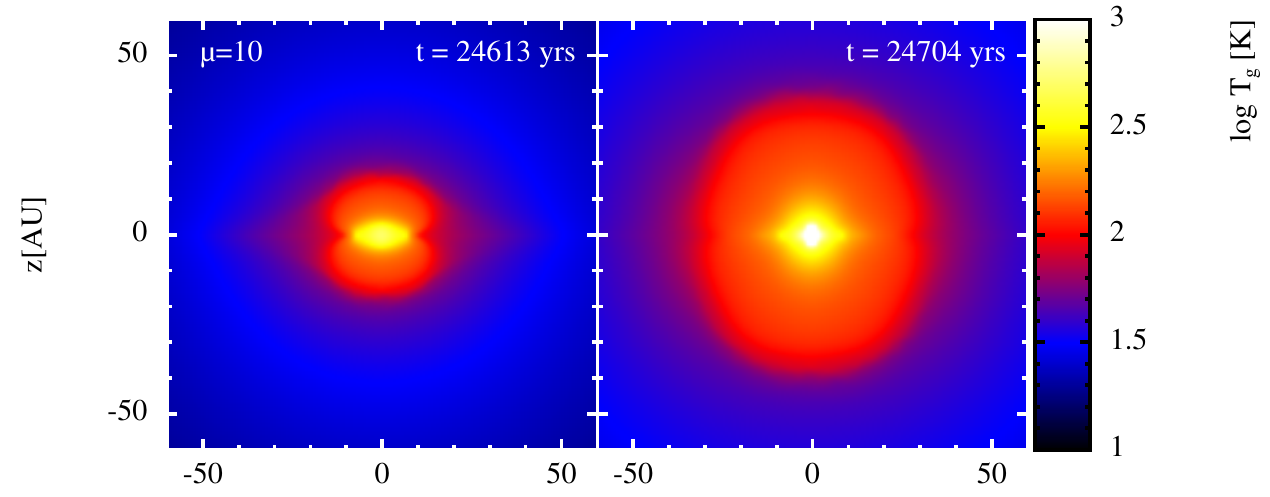}\vspace{-0.1cm}
    \includegraphics[width=8.4cm]{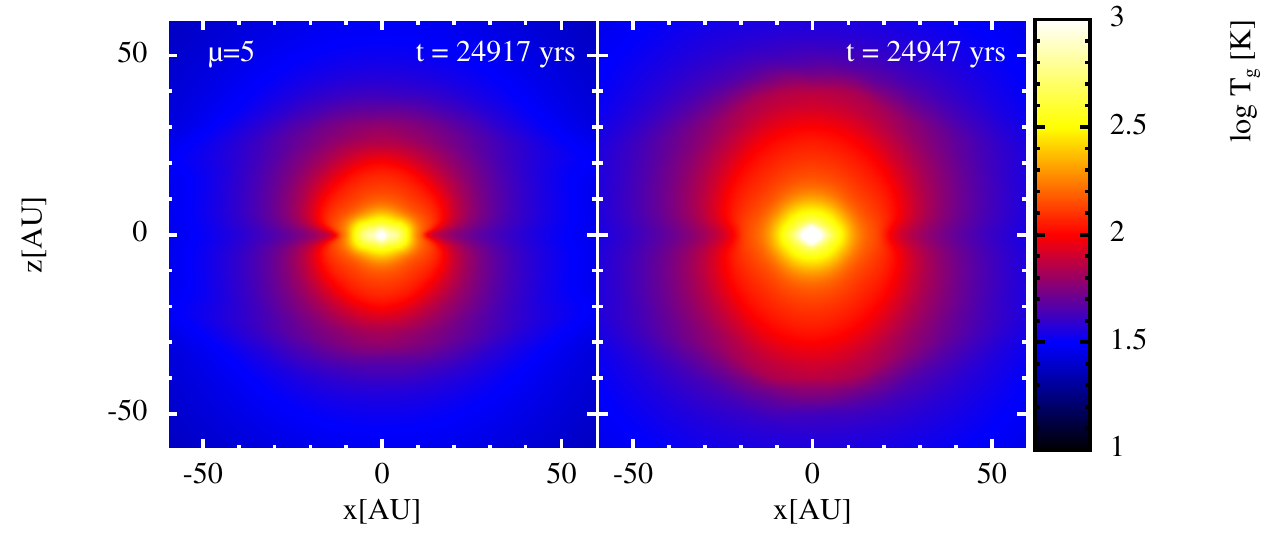}\vspace{0.0cm}
\caption{Snapshots of the temperature on a slice down the rotation axis (in the $x$-$z$-plane) during the evolution of the radiation hydrodynamical calculations of the collapse of molecular cloud cores with different initial magnetic field strengths.  From top to bottom, the different rows are for cloud cores with initial mass-to-flux ratios of $\mu= \infty$, 100, 20, 10, and 5 times critical.  The times at which the snaphots are taken are the same as in Fig.~4.  Note that in the cases with the strongest two magnetic fields the temperature at a given distance is lowest in the plane of the pseudo-disc due to the higher densities.  Also note that in the calculations with outflows the hot gas ($T_{\rm g}>100$~K)  is contained within the outflows whereas without outflows hot gas is distributed further from the protostars.
}
\label{fig:temp_xz}
\end{figure}

\section{Results}
\label{results}

In past papers, we have presented radiation hydrodynamical (RHD) calculations of the collapse of molecular cloud cores to stellar core formation.  Spherically-symmetric (i.e. non-rotating) models were presented by \cite{WhiBat2006} and compared with earlier one-dimensional and three-dimensional models.  \cite{Bate2010, Bate2011} presented results from the first rotating RHD models to follow the collapse beyond stellar core formation, along with a non-rotating case.  The new ingredient in these models is the addition of magnetic fields.

The initial collapse is similar to the unmagnetised cases in that the collapse proceeds almost isothermally.  The initial rotation rate of the cores is relatively low so without the magnetic field the collapse is almost spherically symmetric until the first hydrostatic core begins to form on length-scales of $\sim 10$~AU.  The main difference on large scales with magnetic fields is that during the initial collapse the core becomes distinctly oblate with strong magnetic fields ($\mu \lsim 10$) because collapse is inhibited across field lines, i.e. in the $x$-$y$-plane (Fig.~\ref{fig:global}).  For the same reason, a pseudo-disc is created in the magnetised calculations which is perpendicular to the magnetic field and rotation axis (see below).

In Fig.~\ref{fig:timeevolution} we show the evolution of the maximum density during the calculations, from one initial cloud free-fall time onwards.  It is apparent that the dynamical collapse stalls at densities $\sim 10^{-12}$~g~cm$^{-3}$, with a slower contraction phase to densities $\sim 10^{-8}$~g~cm$^{-3}$, followed by a rapid collapse toward stellar densities $\gsim 0.01$~g~cm$^{-3}$.  The slow contraction phase is associated with the formation and evolution of the first hydrostatic core as it grows in mass (temperatures $\approx 100-2000$~K) and the rapid second collapse begins when the central temperature of the first core exceeds $\approx 2000$~K and molecular hydrogen begins to dissociate \citep{Larson1969}.

\subsection{The first hydrostatic core and its outflow}

As in calculations performed without magnetic fields, the formation of the first hydrostatic core begins at densities $\sim 10^{-13}$~g~cm$^{-3}$ when the isothermal collapse at the centre of the cloud ends and the gas begins to heat up because the compressional heating rate exceeds the rate at which the gas can cool \citep{Larson1969, MasInu1999}.  This transition is clearly shown in the left panel of Fig.~\ref{fig:maxdensitytemp} which shows the evolution of the maximum temperature versus maximum density for all five of the calculations.  In the right panel of Fig.~\ref{fig:maxdensitytemp} we plot the maximum magnetic field strength versus maximum density for each of the calculations.  The maximum field strength scales with maximum density approximately as $B_{\rm max} \propto \rho_{\rm max}^{0.6}$ until after the stellar core forms.  The decay of the field after this point is dominated by the artificial resistivity used in the calculations  (see the Appendix).  During the first core phase the field strengths range from $\approx 0.1-10$~G.

The thermal evolution of the gas during the collapse is largely independent of the magnetic field strength (left panel of Fig.~\ref{fig:maxdensitytemp}).  In Fig.~\ref{fig:images_xy} we show the density evolution of the first hydrostatic cores in each of the five calculations.  The left panels of Fig~\ref{fig:images_xy} display the density in slices perpendicular to the rotation axis, while the right panels of Fig.~\ref{fig:images_xy} show the vertical structure of the first cores in slices that run along the rotation axis.  In the strongly magnetised cases a pseudo-disc is present even before the first hydrostatic core forms.  In Fig.~\ref{fig:temp_xz} we also show the temperature evolution of the first hydrostatic cores in vertical slices.  In the two most magnetised calculations, the temperature at a given radius is lowest in the plane of the pseudo-disc due to the higher optical depths which shield the gas and dust from the first core's radiation.  In the strongest magnetic field case there is a shock above and below the pseudo-disc.  This shock is visible in the density (Fig.~\ref{fig:images_xy}, bottom right).  Before the development of the outflow, the temperature in this shock is hotter ($\approx 30$~K) than the temperature of the cold infalling gas.  Similar shock heating was seen in the radiative magnetohydrodynamical calculations of \citet{Commerconetal2010}.  

The radii of the first hydrostatic cores are smaller with stronger magnetic fields (Fig.~\ref{fig:images_xy}, second-to-last panel in each row) due to the more efficient angular momentum transport.  In fact, without the magnetic field the first core is rotating rapidly enough to become bar-unstable and form a gravitationally unstable pre-stellar disc with a radius of $\approx 50$~AU that is dominated by spiral arms (Fig.~\ref{fig:images_xy}, top-right panel).  This dynamical rotational instability to form non-axisymmetric structure is a common feature of the collapse of unmagnetised molecular cloud cores that are rotating sufficiently rapidly initially, both without \citep{Bate1998, SaiTom2006, SaiTomMat2008, MacMat2011} and with radiative transfer \citep{Bate2010, Bate2011}.  In the magnetised cases, the weakest magnetic field ($\mu=100$) does not provide sufficient angular momentum transport for the first core to avoid the rotational instability.   For $\mu \lsim 20$, magnetic torques do remove enough angular momentum so that the cores remain axisymmetric (Fig.~\ref{fig:images_xy}) with radii ranging from $\approx 5$~AU with the strongest magnetic field (and almost spherical) to $\approx 20$~AU (and highly oblate) in the $\mu=20$ case.

Perhaps the greatest difference between the magnetised and unmagnetised calculations at the first core stage, however, is the fact that magnetically-driven outflows are launched vertically from the first core in all but the $\mu=100$ case.  These are clearly visible in the lower-right panels of Fig.~\ref{fig:images_xy}.  The outflows are slow, with speeds ranging from $\approx 1-2.5$~km/s and the magnetic field strengths range from $\approx 0.1-3$~G.  We find only a weak correlation of the speed of the outflow with the initial magnetic field strength (Table \ref{table1}) such that stronger initial fields produce slightly faster outflows.  A stronger effect is that the outflows are broader with weaker initial magnetic fields in the same way that the first hydrostatic cores have larger radii.  When the outflows reach a distance of 30~AU, their radii are $\approx 20$~AU, 25~AU, and 30~AU for $\mu = 5$, 10, and 20, respectively.

\begin{figure*}
\centering \vspace{-0.0cm}
    \includegraphics[height=5.7cm]{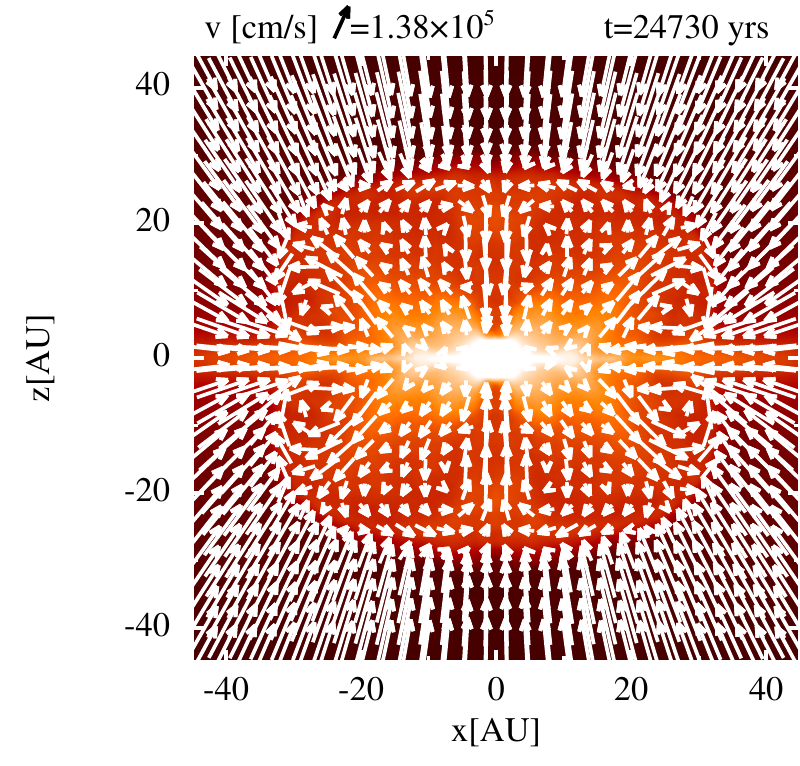}\vspace{0cm}
    \includegraphics[height=5.7cm]{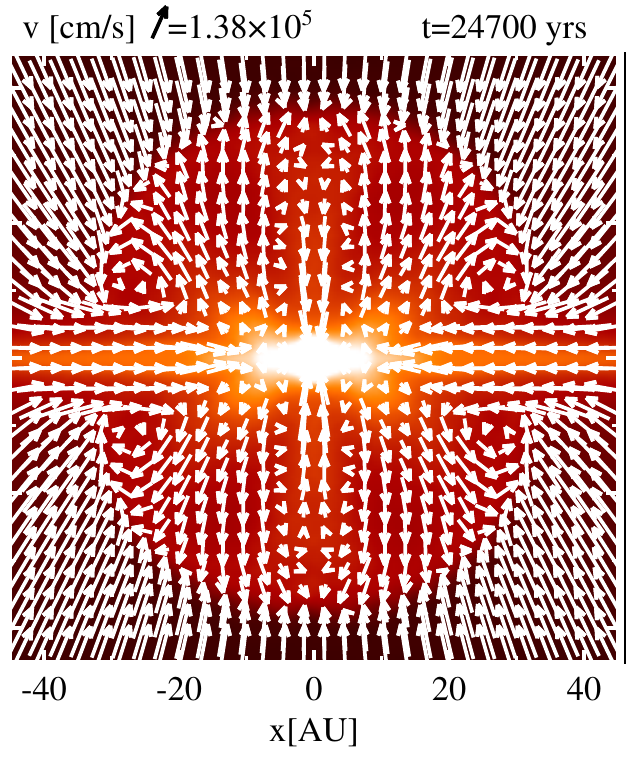}\vspace{0cm}
    \includegraphics[height=5.7cm]{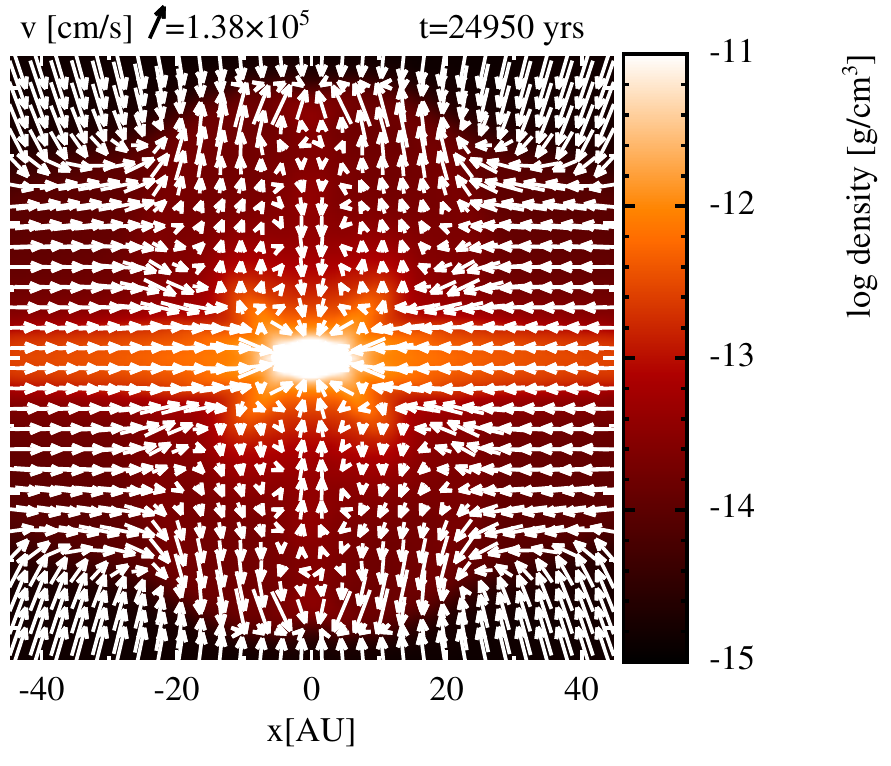}\vspace{0cm}
\caption{Snapshots of the density and velocity vectors on a slice parallel to the rotation axis showing the development of the outflows that are launched from the first hydrostatic cores for each of the three mass-to-flux ratios: $\mu=20$ (left), $\mu=10$ (centre), $\mu=5$ (right). In each case the outflowing material surrounds the vertical (rotation) axis, but along the rotation axis itself the gas is still infalling. Both the first hydrostatic cores and the outflows have smaller diameters for clouds that are initially more magnetised, presumably due to the greater angular momentum transport that occurs as the magnetic field strength is increased.  However, the maximum outflow speed is $\approx 2$~km/s in all cases.}
\label{fig:1stOutflow}
\end{figure*}

The structure of the outflows is complicated (Fig.~\ref{fig:1stOutflow}).  The outflowing gas is concentrated in a conical region and moves at an angle of $\approx 45$ degrees to the vertical axis from the outer parts of each of the first cores, which explains the wide opening angle of the outflows.  Along the rotation axis itself, the gas is primarily falling inwards onto the surfaces of the first cores.  This is also seen in many of the first core outflows that were produced in the calculations of \cite{Tomisaka2002} and \cite{MacTomMat2004, Machidaetal2005}.

\begin{figure}
\centering \vspace{-0cm}\hspace{0cm}
    \includegraphics[height=13cm]{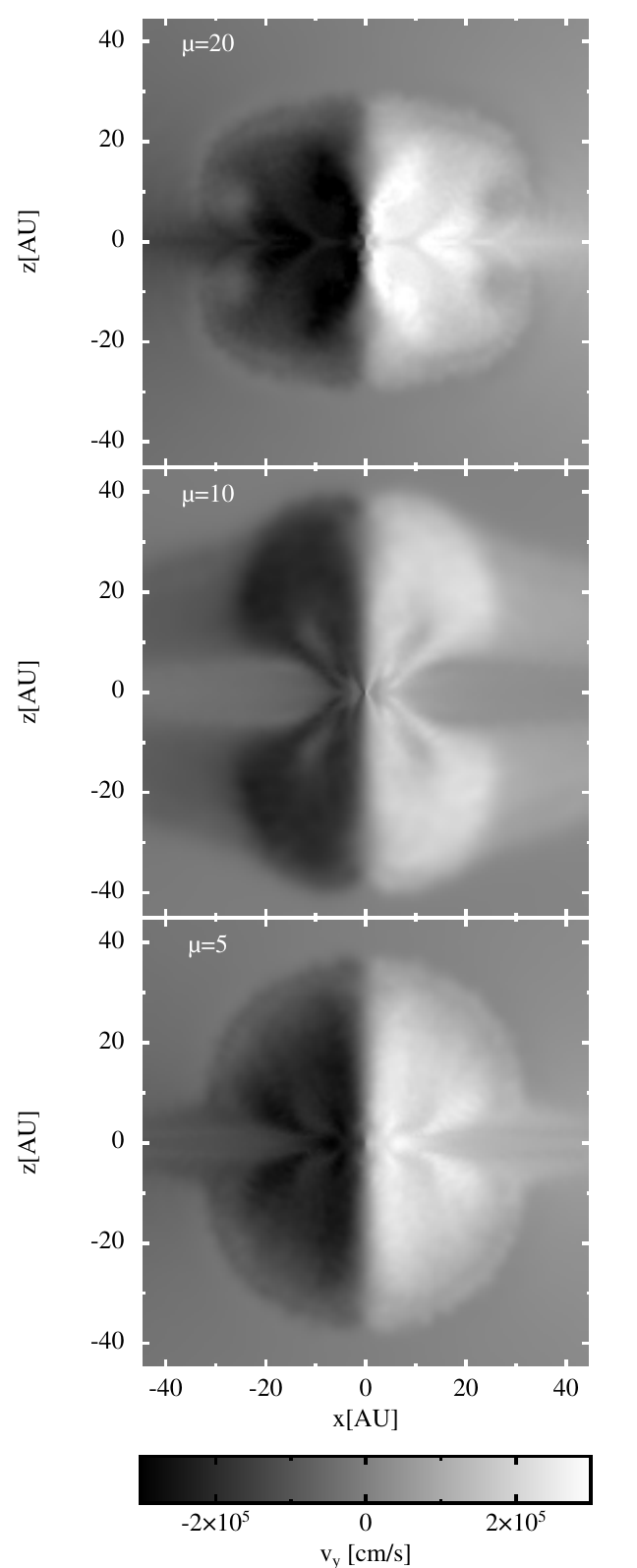}\vspace{0.0cm}
\caption{The velocity perpendicular to a slice through the outflow from the first hydrostatic core (i.e. $v_{\rm y}$) for each of the magnetised calculations that produces an outflow, just before each calculation forms a stellar core.  The outflows are clearly rotating in the same sense as the progenitor dense cores, with rotation speeds that are similar to the outflow speeds.}
\label{fig:1stOutflow_rotation}
\end{figure}

In Fig.~\ref{fig:1stOutflow_rotation} we depict the rotational velocity of each of the first-core outflows.  The outflows are clearly rotating in the same sense as the initial molecular cloud cores (and the first cores) and the tangential speeds are similar to the vertical outflow speeds (i.e.~$1-3$~km/s) and to the rotation speeds of the outer parts of the first cores themselves.  It is interesting to note that the rotation speeds are greater for calculations with weaker initial magnetic fields.  This is likely to be due to the weaker magnetic braking of the first cores.

\begin{figure*}
\centering \vspace{-0.0cm}
    \includegraphics[width=16.5cm]{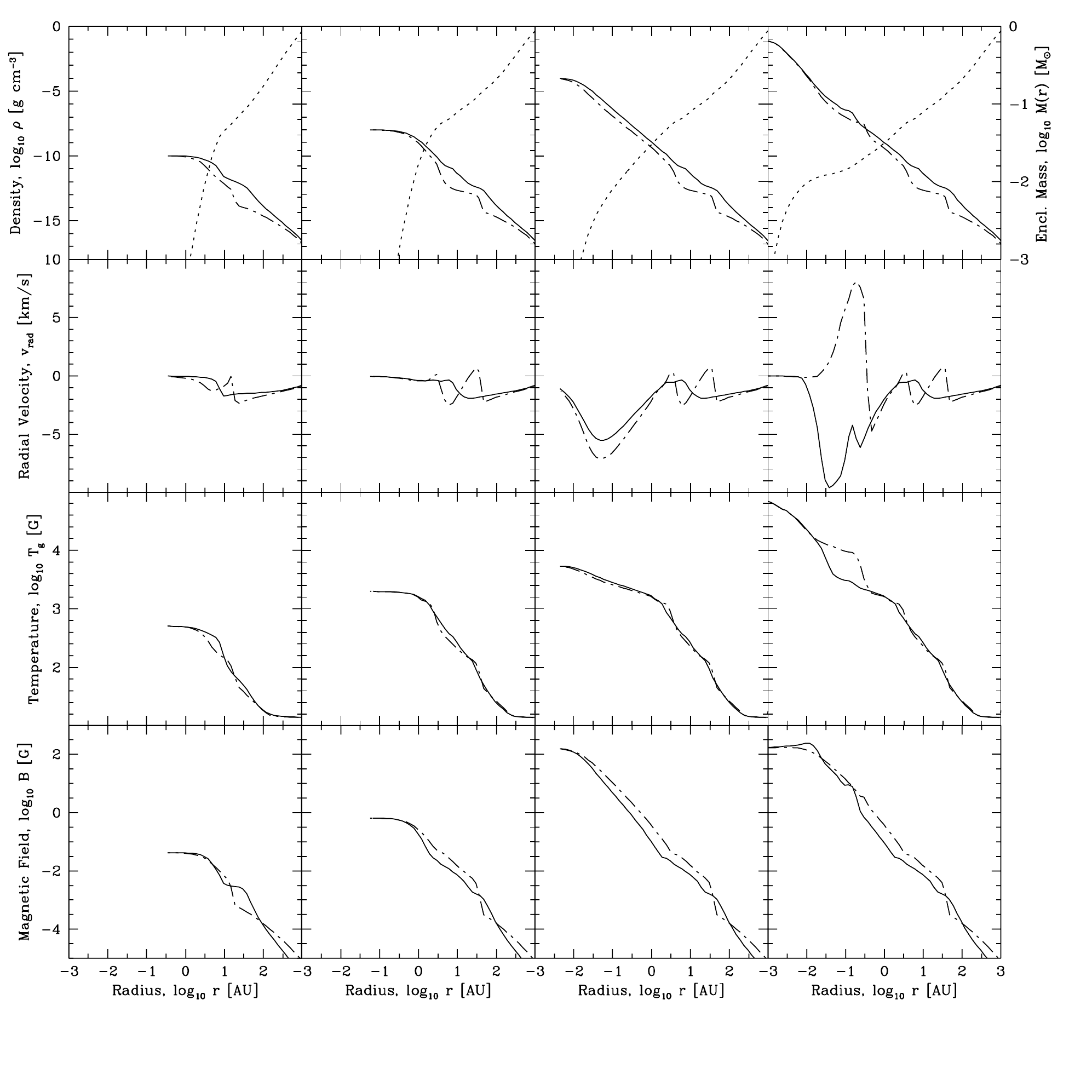}\vspace{-1cm}
\caption{The evolution of the radiation magnetohydrodynamical calculation with initial mass-to-flux ratio ten times critical ($\mu=10$).  Each of the four columns shows the state of the protostar at a different time.  The top row of panels provide the radial density profile perpendicular to the rotation axis (solid line, averaged in azimuth), the density profile along the rotation axis (dot-dashed line), and the cumulative mass profile (dotted line).  The second row of panels give the radial velocity profiles perpendicular to (solid line) or along (dot-dashed line) the rotation axis.  The third row of panels show the radial temperature profiles perpendicular to (solid line) or along (dot-dashed line) the rotation axis.  The bottom row of panels provide the radial magnetic field profiles perpendicular to (solid line) or along (dot-dashed line) the rotation axis.  The first and second columns are during the first core phase, the third column is during the second collapse phase, and the fourth column is after the formation of the stellar core and generation of the stellar outflow.}
\label{fig:profileMF10}
\end{figure*}

In Fig.~\ref{fig:profileMF10} we give plots of the radial profiles of enclosed mass, density, radial velocity, temperature, and magnetic field along the (vertical) rotation axis and in the (horizontal) plane perpendicular to the rotation axis at four times during the evolution of the calculation with $\mu=10$.  The profiles are averages of the quantities within angles of 20 and 10 degrees from the vertical and horizontal directions, respectively.  The wider angle along the vertical axis is taken because of the complex structure of the outflow from the first core (i.e. the fact that gas along the axis itself is actually infalling).  The outflow from the first core has a maximum radial velocity in Fig.~\ref{fig:profileMF10} of only $\approx 0.8$~km/s because the plotted value is an average over both outflowing and infalling gas whereas, in fact, the maximum outflow speed is $\approx 2$~km/s .  The mass of the first core increases from $\approx 0.015$~M$_\odot$ soon after it begins to form to $\approx 0.05$~M$_\odot$ by the time the second collapse begins (top panels, dotted lines).  The onset of the second collapse which occurs when the central temperature exceeds $\approx 2000$~K (third row of panels) can be seen in the third radial velocity panel (inward velocities peaking at a radius of $\approx 0.03$~AU).  The magnetic field strength generally increases with decreasing radius (and increasing density) to in excess of 1~G before the onset of the second collapse.  We only provide profiles for the $\mu =10$ for the sake of brevity because the $\mu=5$ and $\mu =20$ quantities look very similar on such plots.

\subsection{The stellar core and its outflow}

\begin{figure*}
\centering \vspace{0.0cm}
    \includegraphics[width=17.0cm]{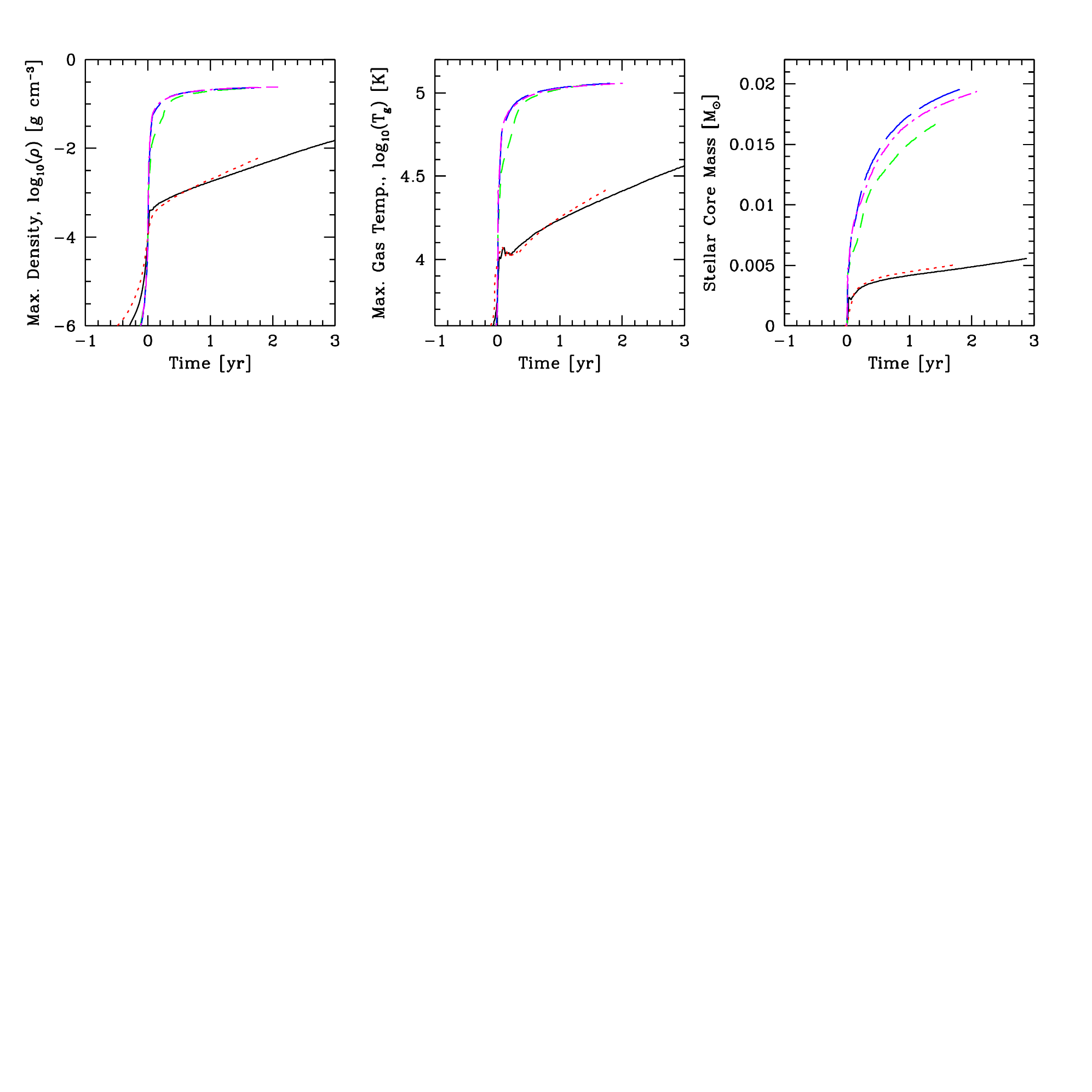}\vspace{-11cm}
\caption{The time evolution of the maximum density (left panel), maximum gas temperature (centre panel), and stellar core mass (right panel) during the radiation magnetohydrodynamical calculations of the collapse of molecular cloud cores.  The different lines are for cloud cores with different initial mass-to-flux ratios: $\mu=\infty$ (i.e.\@ no magnetic field; black solid), $\mu=100$ (red dotted), $\mu=20$ (green short-dashed), $\mu=10$ (blue long-dashed), and $\mu=5$ (magenta dot-dashed). The time is measured in years from the formation of the stellar core, which is defined as being the moment when the maximum density reaches $10^{-4}$~g~cm$^{-3}$.  The stellar core grows much more rapidly in the strongly magnetised calculations than the unmagnetised and weakly magnetised calculations due to the angular momentum transport driven by the magnetic field.  However, there is little difference between the two most highly magnetised cases.
}
\label{fig:densitytempzoom}
\end{figure*}

The onset of molecular hydrogen dissociation at $\approx 2000$~K triggers a rapid second phase of collapse within the first hydrostatic core at densities of $\sim 10^{-8}$~g~cm$^{-3}$ (Figures \ref{fig:timeevolution} and \ref{fig:maxdensitytemp}).  This collapse continues until the dissociation is complete at densities of $\sim 10^{-4}$~g~cm$^{-3}$ and temperatures of $\approx 5000$~K.  The second hydrostatic, or stellar, core is subsequently formed \citep{Larson1969}.  In Fig.~\ref{fig:densitytempzoom} we plot the evolution of maximum density, temperature, and total mass of the stellar core versus time for each of the simulations, with the time set to zero when the density exceeds $10^{-4}$~g~cm$^{-3}$ (i.e. at stellar core formation).  The evolutionary curves from the two most strongly magnetised simulations lie almost on top of each other, while the density and temperature increase more slowly in the weak field and unmagnetised simulations.  The rapid growth rate of the stellar core in the strongly magnetised simulations is no doubt due to the efficient angular momentum transport provided by the magnetic field and, hence, rapid accretion. The accretion rates are $\sim 10^{-2}$~M$_\odot$/yr in the first year of stellar core formation for the strongly magnetised cases.  Although we only follow the calculations for $\approx 2$ years after stellar core formation, the masses of the stellar cores grow to $\approx 20$ Jupiter-masses in the strongly magnetised cases.

Returning to Fig.~\ref{fig:profileMF10} which gives radial profiles of various quantities during the $\mu=10$ calculation, we can examine the properties of the stellar core.  The stellar core begins with a radius of $\approx 2$~R$_\odot$.  It can be seen that an outflow is quickly launched from the vicinity of the stellar core with a speed of $\approx 10$~km/s.  It is also clear that the gas in the outflow is hot, with temperatures up to $\approx 10^4$~K.  The accretion onto the stellar core continues in the equatorial plane through a pseudo-disc that is not in Keplerian rotation, but is in fact falling inwards with speeds of $\approx 5-10$~km/s.  The stellar cores produced by the other calculations have similar radii and in all  three of the most highly magnetised calculations magnetically driven outflows are launched from the vicinities of the stellar cores and the stellar cores are surrounded by infalling pseudo-discs.  With $\mu=100$ and in the unmagnetised case these outflows are absent and the discs surrounding the stellar cores are close to Keplerian.

\begin{figure*}
\centering \vspace{-0.0cm}
    \includegraphics[width=8.8cm]{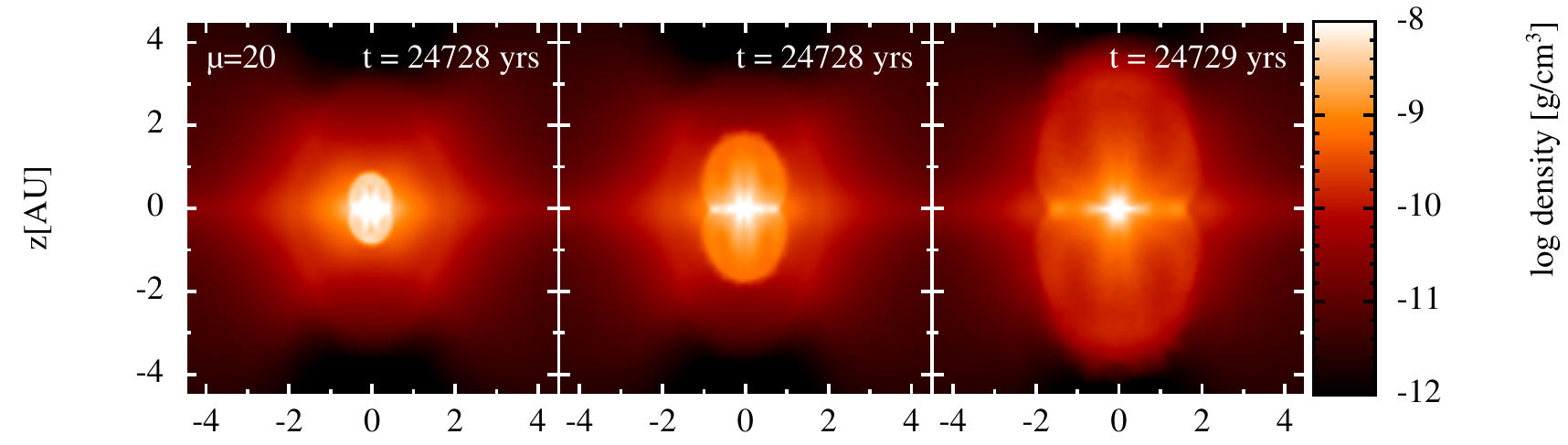}\hspace{0.3cm}
    \includegraphics[width=8.5cm]{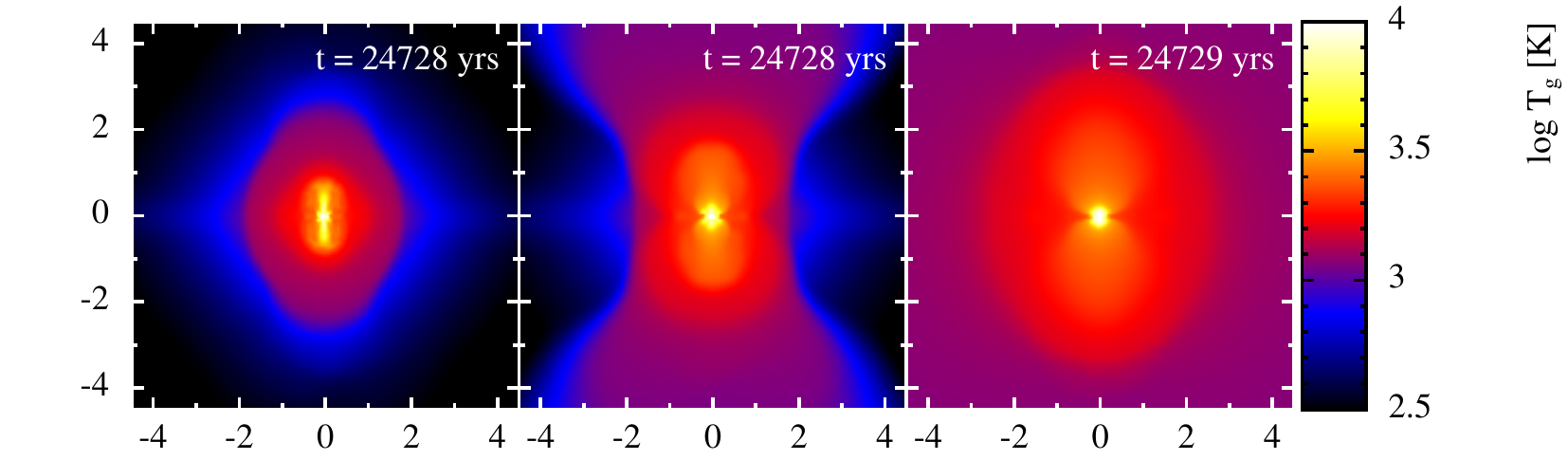}\vspace{-0.0cm}
    \includegraphics[width=8.8cm]{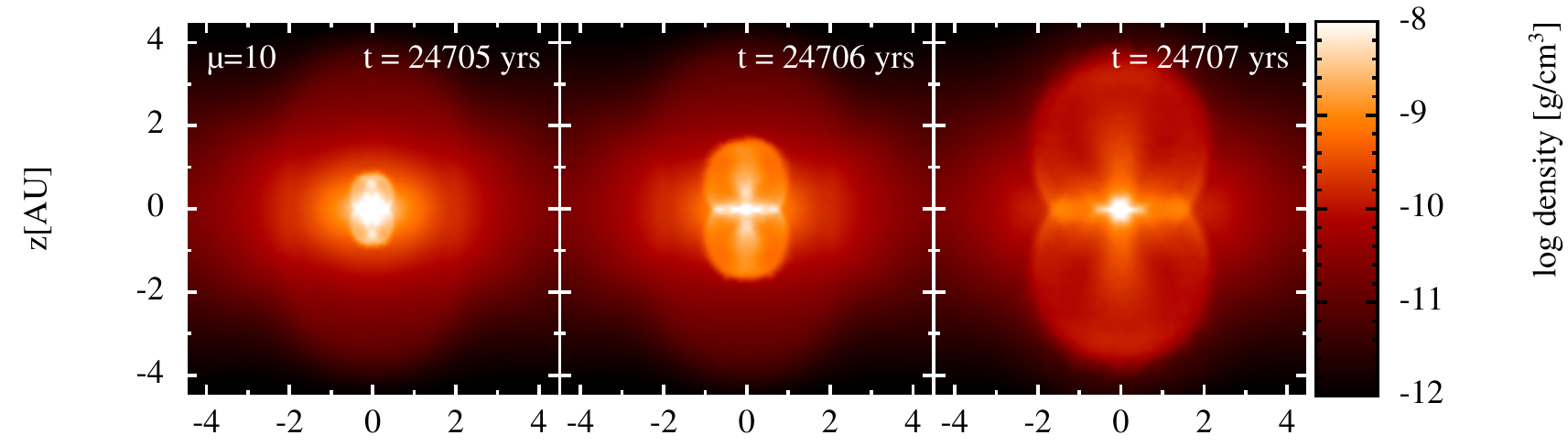}\hspace{0.3cm}
    \includegraphics[width=8.5cm]{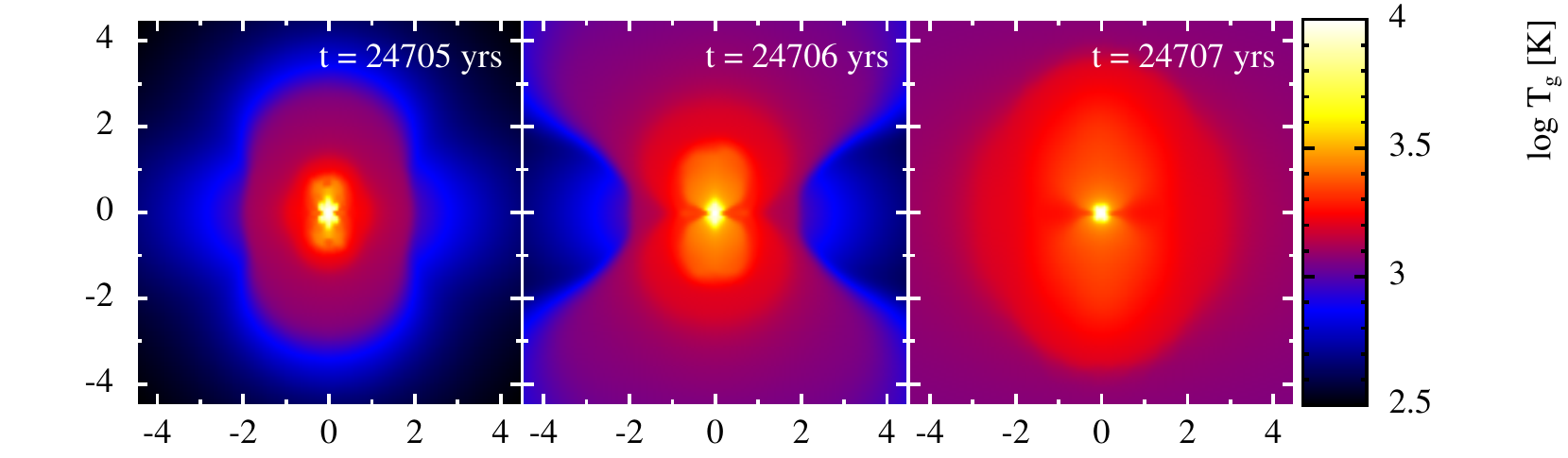}\vspace{-0.0cm}
    \includegraphics[width=8.8cm]{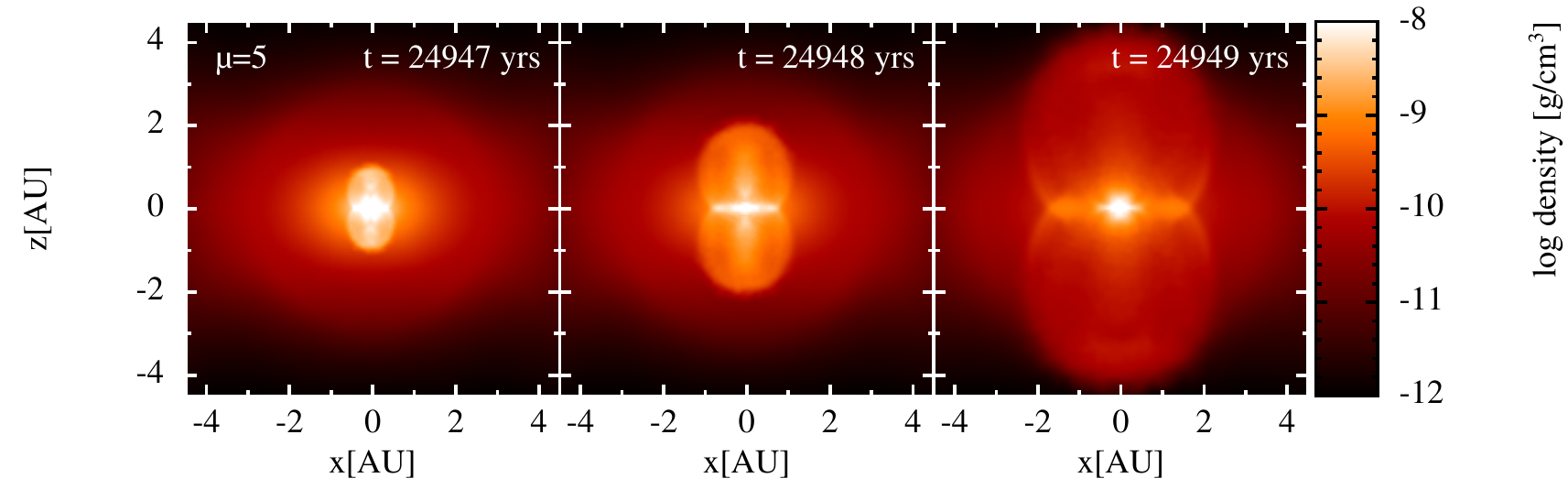}\hspace{0.3cm}
    \includegraphics[width=8.5cm]{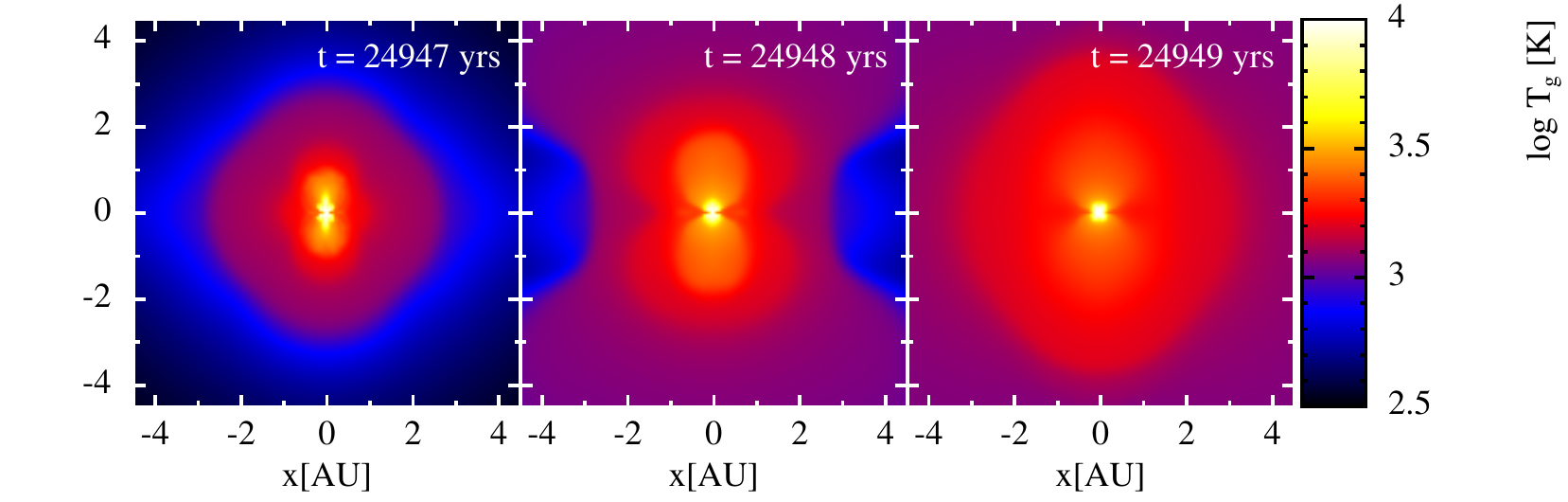}\vspace{-0.0cm}
\caption{Snapshots of the density (left panels) and temperature (right panels) on slices parallel to the rotation axis showing the development of the outflows that are launched from the stellar cores in the three most magnetised calculations.  From top to bottom, each row is for cloud cores with initial mass-to-flux ratios of $\mu=20$, 10, and 5 times critical, respectively.  For each case, we show snapshots 0.5, 1.0, and 2.0 yrs after stellar core formation.  The structure of the stellar outflows is similar for each initial magnetic field strength, but it appears that weaker initial fields produce slightly better collimation. The remnants of the first hydrostatic cores are clearly visible in the density slices (thickness $|z| \approx 3$~AU), and the stellar outflows are followed until just after they have broken out of the first core in each case.  }
\label{fig:images_xz2}
\end{figure*}

\begin{figure*}
\centering \vspace{-0.0cm}
    \includegraphics[height=5.5cm]{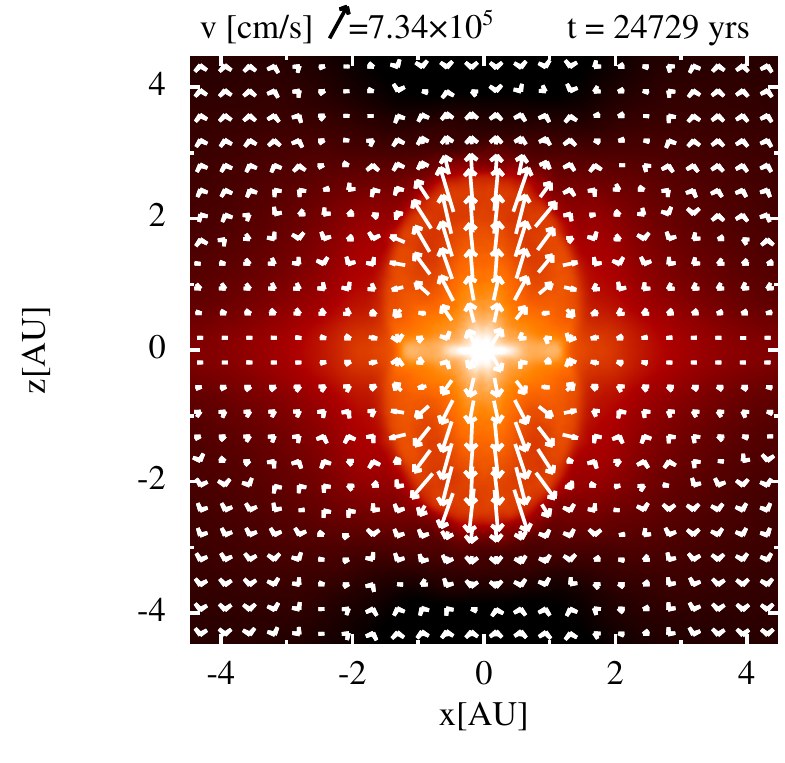}\vspace{0cm}
    \includegraphics[height=5.5cm]{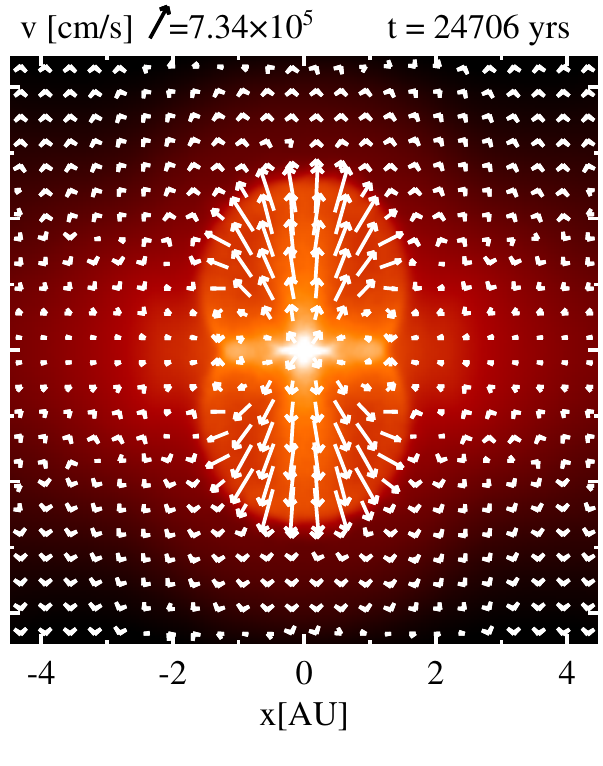}\vspace{0cm}
    \includegraphics[height=5.5cm]{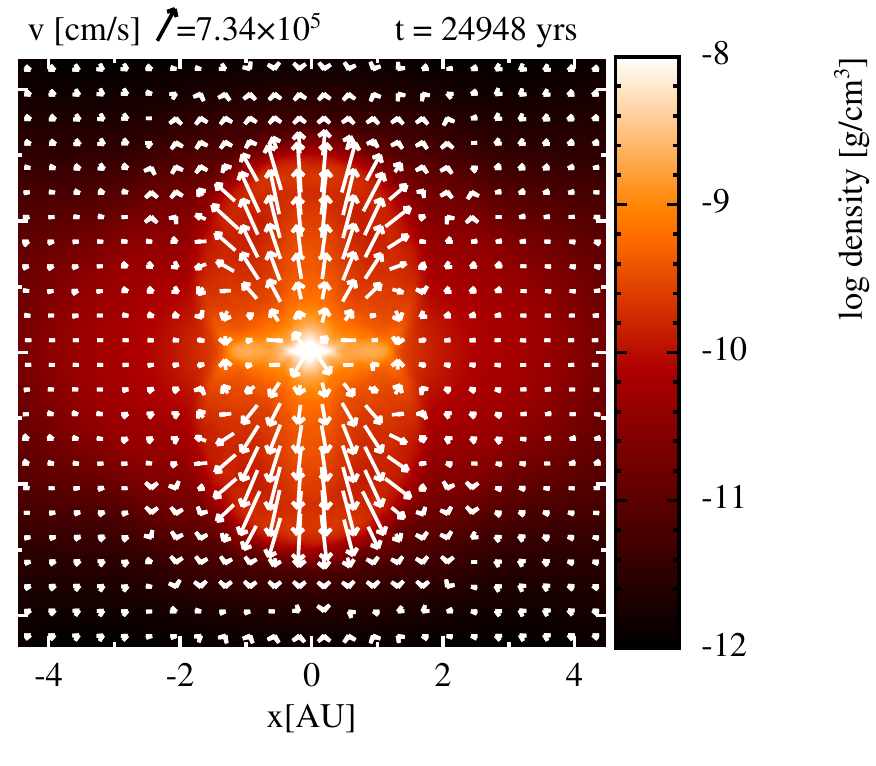}\vspace{-0.5cm}
\caption{Snapshots of the density and velocity vectors on a slice parallel to the rotation axis showing the development of the outflows that are launched from the stellar cores 1.5 yrs after stellar core formation for each of the mass-to-flux ratios: 
$\mu=20$ (left), $\mu=10$ (centre), $\mu=5$ (right). Unlike the outflows launched from the first hydrostatic cores, outflowing material is concentrated on the vertical (rotation) axis.  }
\label{fig:2ndOutflow}
\end{figure*}

In Fig.~\ref{fig:images_xz2} we show the density and temperature evolution of the second outflow launched from the vicinity of the stellar core, while in Fig.~\ref{fig:2ndOutflow} we plot the velocity structure in slices through the outflows for each magnetised calculation. The pseudo-discs that feed the stellar cores are clearly visible in these figures.  They grow in radius with time, extending to radii of $\approx 1$~AU one year after stellar core formation.  We follow each of the stellar outflows until they have broken out of the first core remnants (which each have vertical extents of $\approx 3$~AU).  The breaking out of the outflows is accompanied by strong heating of the gas surrounding the first cores as it becomes exposed to the hotter interiors (Fig.~\ref{fig:images_xz2}).

In addition to being much faster, the structure of the stellar outflow is quite different to that of the outflow from the first core.  The most rapidly moving material in the stellar outflow is found along the rotation (vertical) axis (Fig.~\ref{fig:2ndOutflow}) as opposed to the conical flows found from the first cores.  In this respect it is more jet-like and less like a disc wind.  The stellar outflows are almost identical for all of the magnetised calculations, whereas the first cores and their outflows were clearly broader with weaker magnetic fields.  We attribute this to the fact that the radii of the first cores were larger with weaker magnetic fields due to the reduced angular momentum transport, while the stellar cores are all essentially spherical and are not rotating particularly rapidly.  However, the stellar outflows are still not particularly well collimated, although it appears that the stellar outflow in the $\mu=20$ calculation is slightly more collimated than in the more magnetised cases.

\begin{figure}
\centering \vspace{-0cm}\hspace{0cm}
    \includegraphics[height=11cm]{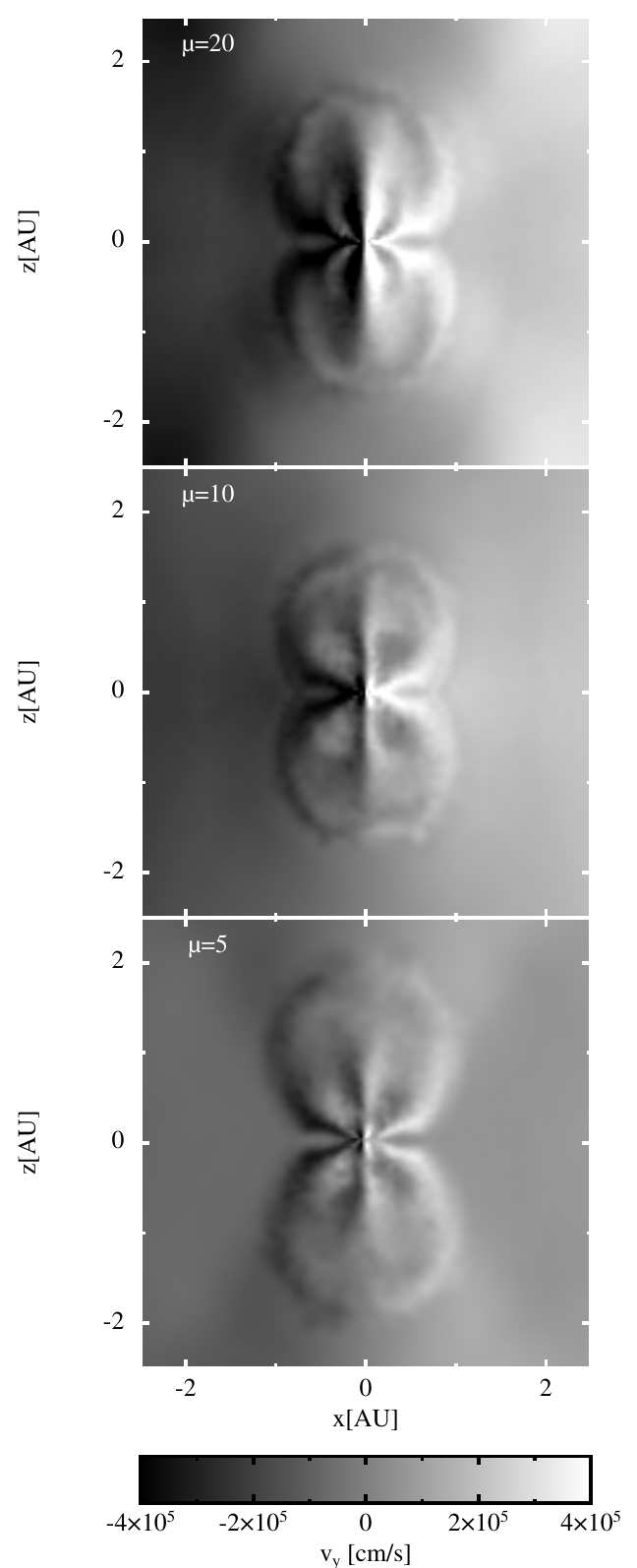}\vspace{0.0cm}
\caption{The velocity perpendicular to a slice through the outflow from the stellar core (i.e. $v_{\rm y}$) for each of the magnetised calculations that produces a stellar outflow, one year after stellar core formation in each case.  As with the outflows from the first cores, the stellar outflows are clearly rotating in the same sense as the progenitor dense cores and the rotation rates are greater for calculations with lower initial magnetic field strengths.  However, the rotation speeds are significantly lower than the outflow speeds.}
\label{fig:2ndOutflow_rotation}
\end{figure}

In Fig.~\ref{fig:2ndOutflow_rotation} we provide cross-sections of the rotational velocity of each of the stellar-core outflows.  As was found for the first-core outflows, they are rotating in the same sense as the initial molecular cloud cores and the rotation speeds are greater in calculations with weaker initial magnetic fields.  Unlike the outflows from the first cores, however, the tangential speeds ($2-4$~km/s) are much lower than the vertical outflow speeds (i.e.~$\approx 10$~km/s).

\section{Discussion}

\subsection{The structure of the stellar cores}

Recently, there has been a lot of interest in the initial conditions for pre-main-sequence stellar evolution codes and the effects of accretion on pre-main-sequence evolution because the properties of young pre-main-sequence stars can vary greatly depending on the assumptions that are made  \citep{HarCasKen1997, TouLivBon1999, BarChaGal2009, HosOffKru2011}.  Along with those of \cite{Tomidaetal2013}, the calculations presented here are the only self-consistent three-dimensional radiation magnetohydrodynamic calculations of stellar core formation performed to date.  Therefore, we can directly determine the initial structure of the stellar cores and examine the entropy of the material being accreted onto them.

\begin{figure*}
\centering \vspace{0.0cm}
    \includegraphics[width=14.5cm]{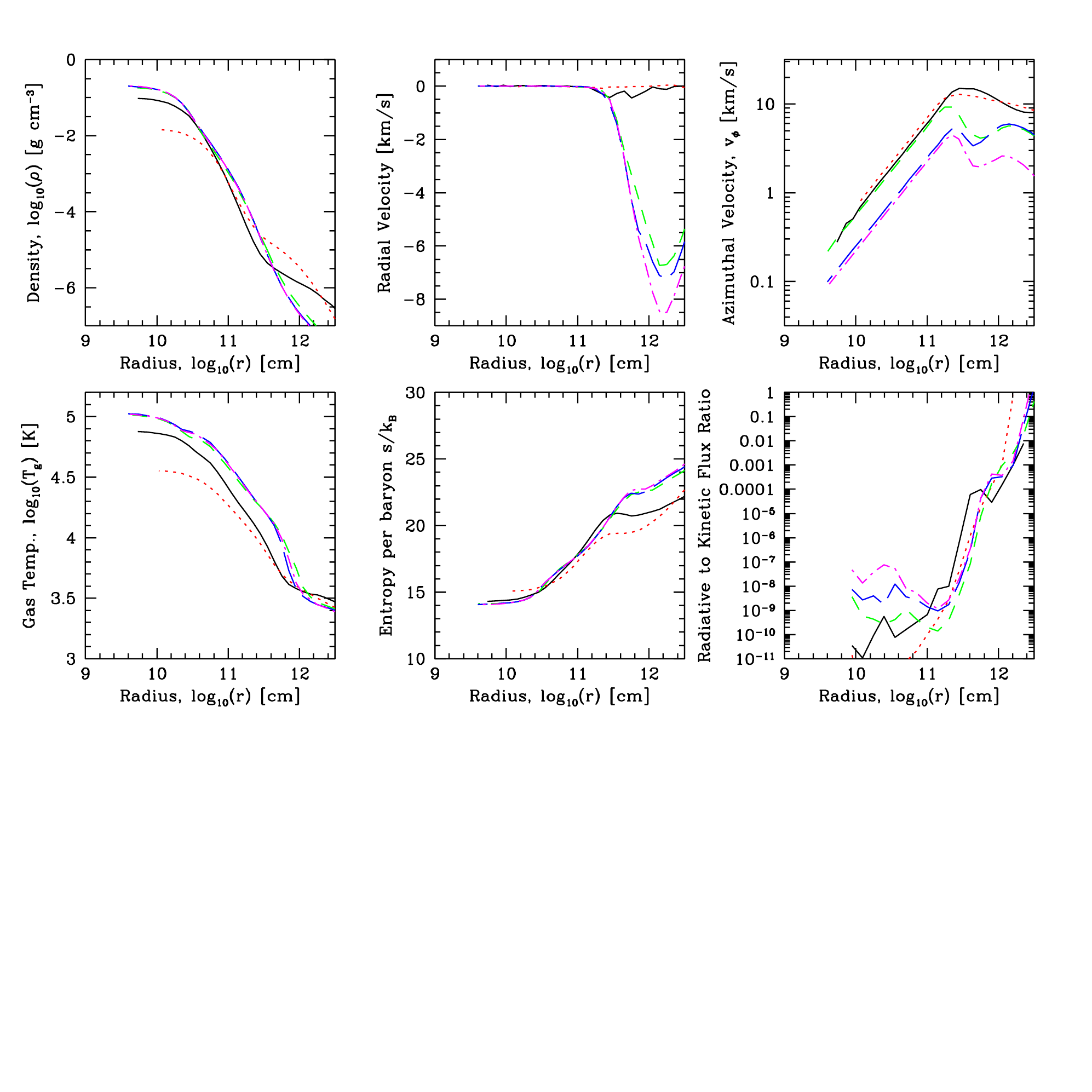}\vspace{-5.2cm}
\caption{The structures of the stellar cores formed in the radiation magnetohydrodynamical calculations with initial mass-to-flux ratios: $\mu=\infty$ (i.e.\@ no magnetic field; black solid lines), $\mu=100$ (red dotted lines), $\mu=20$ (green short-dashed lines), $\mu=10$ (blue long-dashed lines), and $\mu=5$ (magenta dot-dashed lines).  The structure is shown 1 year after stellar core formation for $\mu=5-20$, 2.5 years after stellar core formation for $\mu=100$, and 10 years after stellar core formation for $\mu=\infty$ because the growth rates of the stellar cores in the weakly magnetised and unmagnetised calculations are much lower.  We plot the azimuthally-averaged midplane density, radial velocity, azimuthal velocity, gas temperature, and entropy per baryon.  In the bottom right figure, we plot the ratio of the radiative to the kinetic flux calculated on spherical shells as functions of radius (see the main text for further details).
}
\label{fig:stellarcoreprop}
\end{figure*}

Defining the radius of a stellar core is non-trivial in that, as discussed above, it is being fed by an equatorial pseudo-disc and has no distinct surface.  In Fig.~\ref{fig:stellarcoreprop} we plot radial profiles of many quantities in the equatorial plane for each of the stellar cores, one year after they each began to form (i.e.\@ one year after the maximum density exceeded $10^{-4}$~g~cm$^{-3}$).  Although we only plot quantities within the equatorial plane, it should be noted that within the stellar cores the density, temperature, and entropy profiles are spherically symmetric.  It can be seen that there is little variation of the stellar core properties between the calculations.  The radii of the stellar cores is $\approx 2\times 10^{11}$~cm or $\approx 3$~R$_\odot$.  This radius is the location of  the peak in the angular velocity with radius (i.e. marking the location of the boundary layer with the pseudo-disc), and where the radial velocity drops to below 0.1~km/s (about 1/2000 of the local sound speed).  Note that the stellar cores are in solid-body rotation --- this could be physical due to the magnetic field, but it may also be due to the effects of the artificial viscosity.  They are rotating much slower than the breakup velocity (the escape velocity at the surface is approximately 50 km/s).

In the bottom centre panel of Fig.~\ref{fig:stellarcoreprop} we plot the entropy per baryon for the protons as a function of radius which, from the statistical physics of a perfect gas of non-relativistic and non-degenerate fermions, can be expressed as
\begin{equation}
s/k_{\rm B} = 2.5 - \ln \left[ \frac{n}{g} \left( \frac{ 2 \pi \hbar^2}{m_{\rm p} k_{\rm B} T_{\rm g}} \right)^{3/2}  \right]
\end{equation}
where $g=2$ is the spin degeneracy of a proton, $\hbar$ is the reduced Planck's constant, and $k_{\rm B}$ is Boltzmann's constant, and $m_{\rm p}$ is the mass of a proton.  For simplicity, we neglect the effect on the number density of helium and heavier elements (i.e. we simply set the number density $n=\rho/m_{\rm p}$).  Note that for atomic hydrogen ($\mu=1$), this can also be expressed in the familiar form of
\begin{equation}
s/k_{\rm B} = K + \frac{1}{\gamma - 1} \ln \left( \frac{P}{\rho^{\gamma}} \right)
\end{equation}
where $\gamma=5/3$, $P=n k_{\rm B} T$, and $K$ is a constant \citep[e.g.][]{Commerconetal2011b}.

Note that the entropy increases outward, meaning that the stellar cores are convectively stable at this point.  Nuclear fusion would not yet have begun as the central temperature is only $\approx 10^5$~K.

Recently, \cite{Commerconetal2011b} and \cite{Vaytetetal2012} examined the accretion of material onto the first hydrostatic core in one-dimensional calculations.  They found that the accretion shock was supercritical and that essentially all of the energy from the accretion shock was radiated away.  By contrast \cite{Vaytetetal2013}, using one-dimensional multi-group calculations, found that the accretion shock onto the stellar core is strongly sub-critical, with  all the accretion energy being transferred to the core.

\begin{figure*}
\centering \vspace{0.0cm}
    \includegraphics[width=14.5cm]{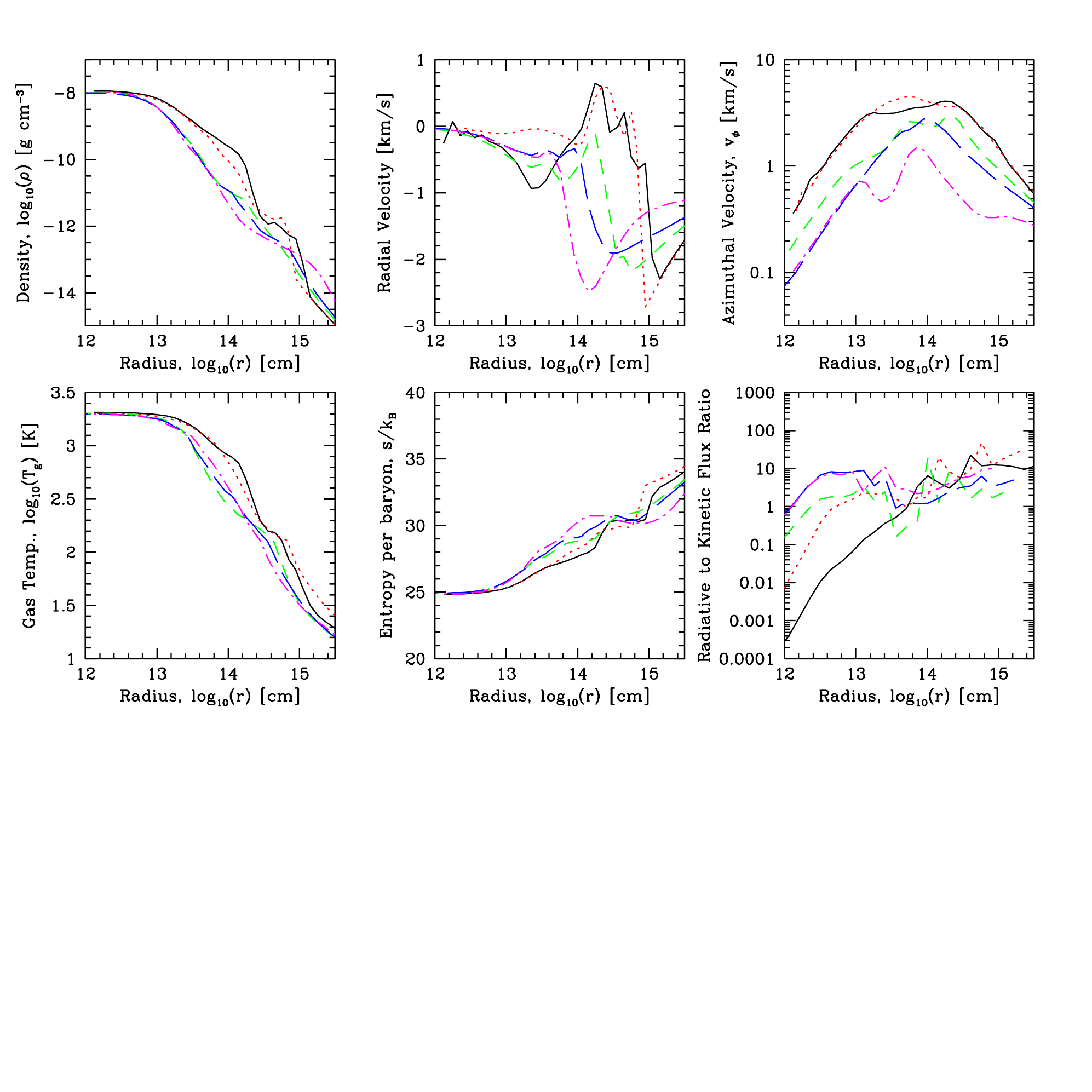}\vspace{-5.2cm}
\caption{The structures of the first cores formed in the radiation magnetohydrodynamical calculations with initial mass-to-flux ratios: $\mu=\infty$ (i.e.\@ no magnetic field; black solid lines), $\mu=100$ (red dotted lines), $\mu=20$ (green short-dashed lines), $\mu=10$ (blue long-dashed lines), and $\mu=5$ (magenta dot-dashed lines).  In each case the plots are made when the maximum density reaches $10^{-8}$~g~cm$^{-3}$ (before the onset of the second collapse). We plot the azimuthally-averaged midplane density, radial velocity, azimuthal velocity, gas temperature, and entropy per baryon.  In the bottom right figure, we plot the ratio of the radiative to the kinetic flux calculated on spherical shells as functions of radius (see the main text for further details).
}
\label{fig:firstcoreprop}
\end{figure*}

Three-dimensional calculations are more difficult to analyse because the surfaces involved are complex and we have to deal both with accretion and outflows.  However, one way to examine the amount of energy that is being radiated away compared to that being incorporated into the core is to calculate the kinetic energy flux and compare this to the radiative energy flux in the regions where the cores are accreting.  We calculate the kinetic energy and radiative fluxes through spherical shells as functions of radius.  The radial flux of the kinetic energy of the matter passing through a surface can be written as
\begin{equation}
\Psi_{\rm m} = \frac{1}{2} \rho v^2 v_{\rm r}
\end{equation}
which gives a power per unit area, where $v_{\rm r}$ is the radial velocity.  The radiative flux is given by equation \ref{eddington}.  In the bottom right panel of Fig.~\ref{fig:stellarcoreprop} we plot the ratio of the radiative flux to the kinetic flux as a function of radius.  The kinetic flux is approximately 8--9 orders of magnitude larger than the radiative flux in the vicinity of the surface of the stellar core, indicating that essentially all of the kinetic energy of the accretion flow is advected into the stellar core.  This is a similar conclusion to that drawn by \cite{Vaytetetal2013} from one-dimensional calculations, except that in the three-dimensional calculations presented here the stellar core accretes gas with much lower radial infall speeds  via a boundary layer from a disc or pseudo-disc rather than through a strong radial shock.

By contrast, in Fig.~\ref{fig:firstcoreprop} we plot the same quantities for the first hydrostatic cores when their central densities have reached $10^{-8}$~g~cm$^{-3}$ (before the onset of the second collapse).  In the vicinities of the surfaces of the first cores, the radiative and kinetic fluxes are approximately equal, consistent with the conclusion from the one-dimensional models of \cite{Commerconetal2011b} and \cite{Vaytetetal2012} that essentially all the kinetic energy from the accretion onto the first core is radiated away.

Overall, we find that the properties of the stellar cores are surprisingly uniform across the calculations.  We attribute this to the fact that the stellar core is formed from gas that undergoes the second collapse due to molecular hydrogen dissociation and that the onset of the second collapse occurs at the same temperatures and densities in all the calculations (it is determined by the physics of the dissociation). Thus, the gas that forms the stellar core begins collapsing with the same properties in each of the calculations and, thus, the properties of the stellar cores that result are almost `universal'.  If this is correct, it removes concerns about whether there should be variations in the initial conditions for pre-main-sequence evolution models --- a universal initial condition can be adopted (at least for solar metallicities).  However, the later evolution will still depend on the details of the accretion.

\subsection{The driving of the outflows}

We have seen that outflows are launched from the vicinities of both the first hydrostatic cores and the stellar cores in the three most strongly magnetised calculations.  It is of interest to investigate the driving mechanisms of these outflows.  

\cite{Bate2010, Bate2011} and \cite{SchTsc2011} found in radiation hydrodynamical models of stellar core formation that bipolar outflows could be launched from the vicinity of the stellar core even in the absence of magnetic fields due to the large amount of thermal energy liberated within the first core due to the second collapse, stellar core formation, and accretion onto the stellar core.  Therefore, it is interesting to examine the extent to which the stellar outflows are driven by both magnetic forces and thermal pressure.  We note that the most weakly magnetised calculation ($\mu=100$) does not produce magnetically-driven outflows.  This is likely to be at least partially due to the artificial resistivity within the numerical method which may stop the magnetic field from growing large enough to launch outflows.  As found by \cite{Bate2010, Bate2011}, the unmagnetised calculation ($\mu=\infty$) eventually launches a thermally-driven bipolar outflow from the vicinity of the stellar core which breaks out of the remnant of the first core with a speed $\approx 10$~km/s.  However, this outflow doesn't begin until $\approx 20$~yr after the stellar core forms.  Thus, not only do strong magnetic fields help launch stellar outflows (see below), but they can also begin much sooner than thermally-driven outflows.  We were only able to follow the $\mu=100$ calculation until $\approx 2.5$ yr after stellar core formation due to the shorter timestep required for the magnetic divergence cleaning, but we assume that if we were able to follow it as long as the hydrodynamical calculation similar thermally-driven bipolar outflows would have been generated.

\begin{figure}
\centering \vspace{-0.6cm}
    \includegraphics[width=0.9\textwidth]{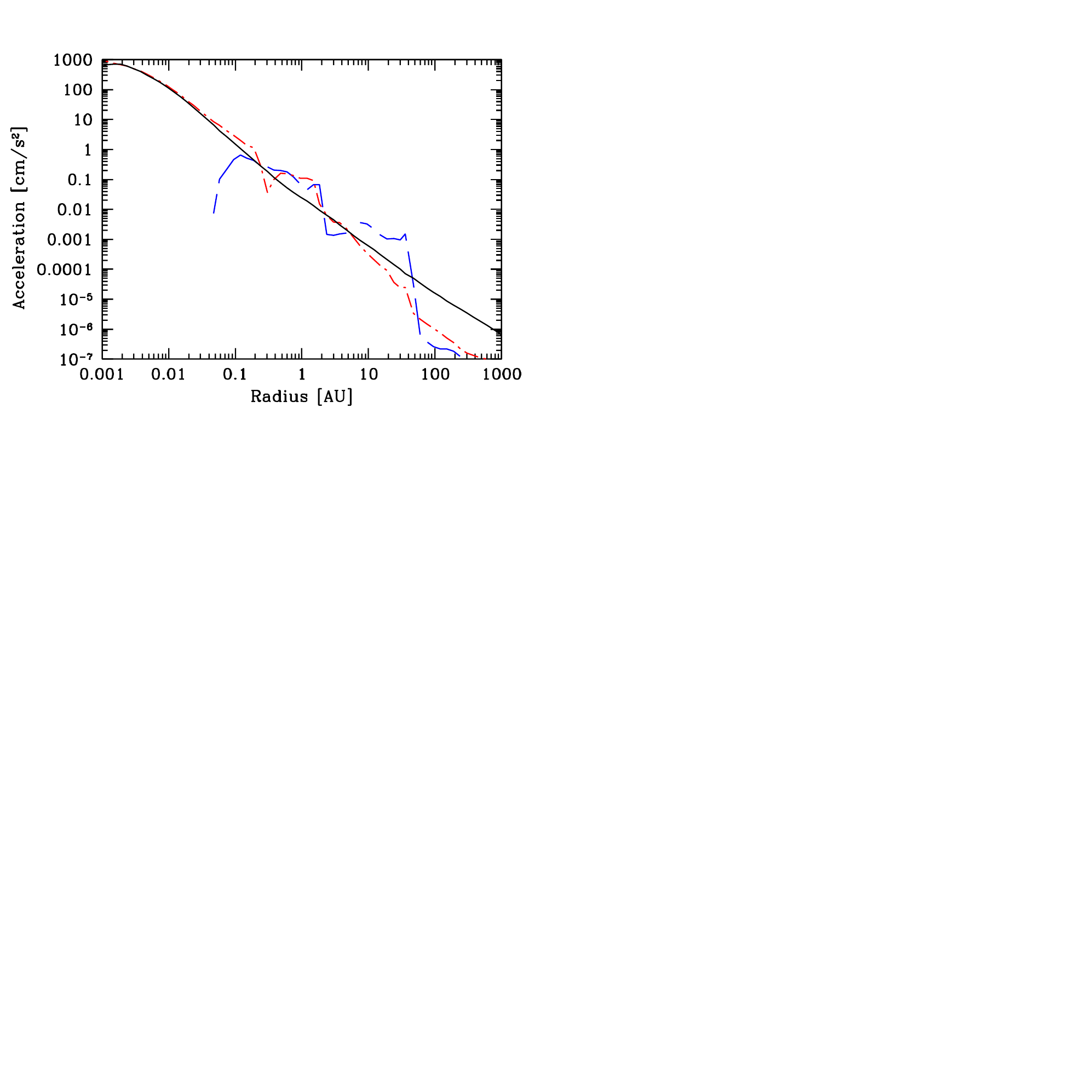} \vspace{-10.5cm}
\caption{One year after stellar core formation in the $\mu=10$ calculation, we plot the magnitudes of the vertical components of the gravitational acceleration (black solid line), the acceleration due to gas pressure (red dot-dashed line), and the acceleration due to the Lorentz force where it is directed outward (blue dashed line). The accelerations are average values computed within 20 degrees of the vertical axis.  The Lorentz force clearly exceeds gravity at radii $\approx 6-40$~AU (the outflow from the first core) and $\approx 0.2-2$~AU (the outflow from the stellar core).  However, the outward acceleration due to gas pressure also exceeds gravity from $\approx 0.4-2$~AU by a similar factor to the Lorentz acceleration and, thus, also plays an important role in launching the stellar outflow.}
\label{fig:lorentz}
\end{figure}

\begin{figure*}
\centering
    \includegraphics[width=\columnwidth]{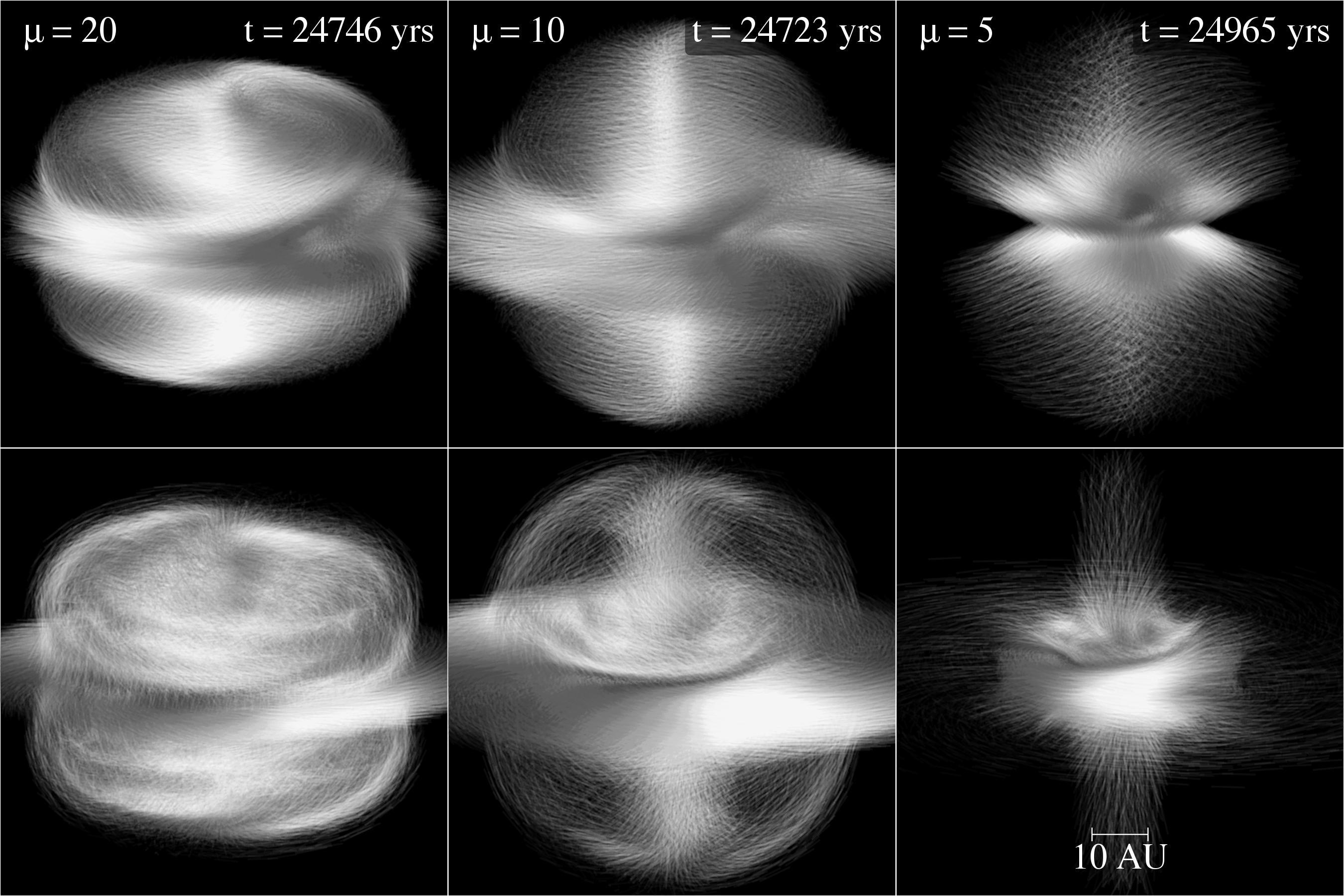}
    \hspace{0.5cm}
    \includegraphics[width=\columnwidth]{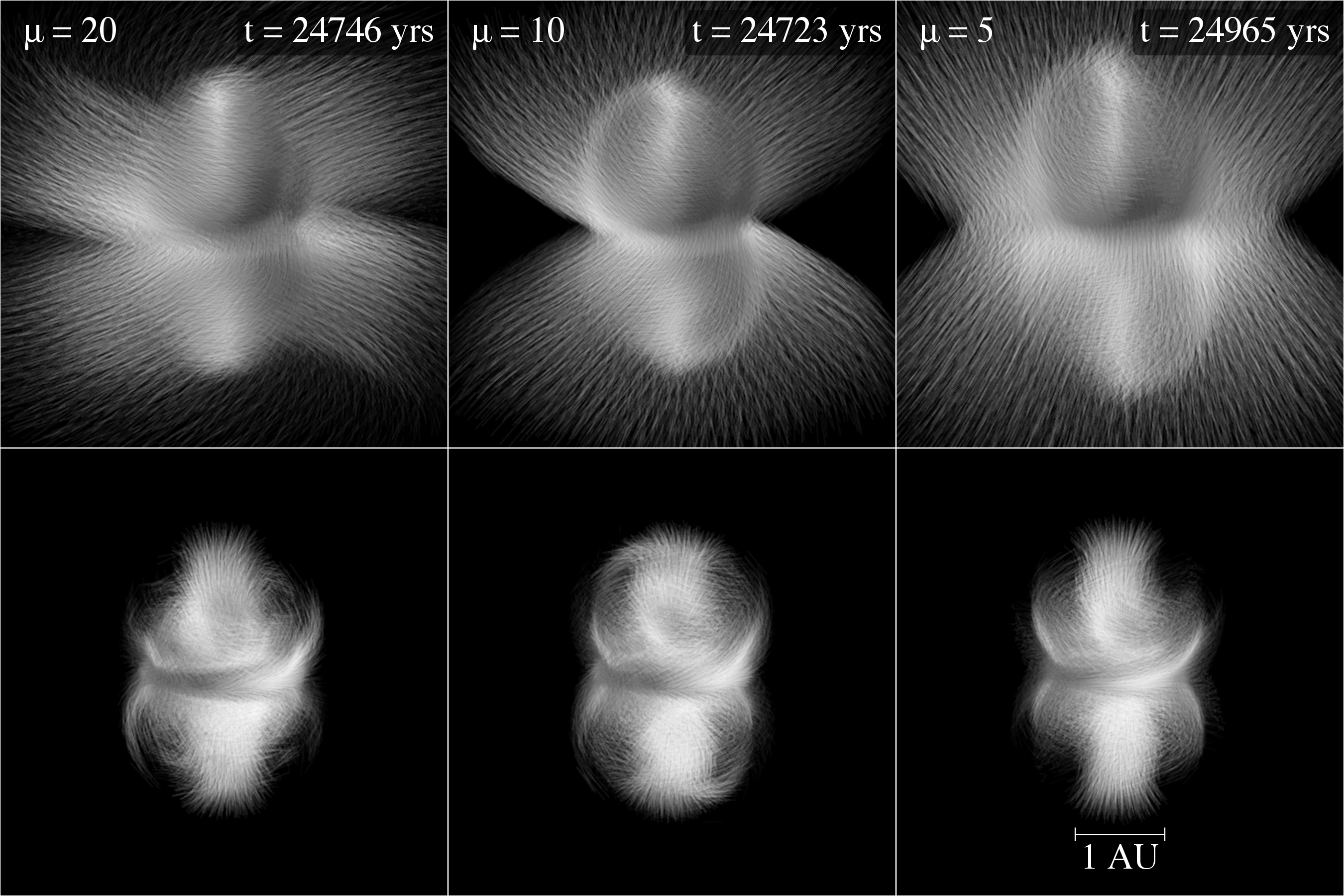}
\caption{Visualisation of the magnetic field lines (top row) and electric current (bottom row) in the outflows from the first core (left figure) and second, stellar core (right figure) in the calculations with initial $\mu = 20$, $\mu=10$ and $\mu=5$ (as indicated), in each case shown 1 year after the formation of the stellar core. The magnetic field is strongly toroidal near the base of both outflows, particularly in the weaker field calculations (top left and top-centre panels in each Figure). A poloidal component is more prominent at higher field strengths in both the first and second core outflows ($\mu=10$ and 5, centre and right columns, see also Fig.~\ref{fig:torpol}). Typical field strengths are $\sim 0.1$--1~G in the first core outflow and $\sim$10-100~G in the outflows from the stellar core (see Fig.~\ref{fig:torpol}).}
\label{fig:fieldlines}
\end{figure*}

\begin{figure*}
\centering
    \includegraphics[width=\columnwidth]{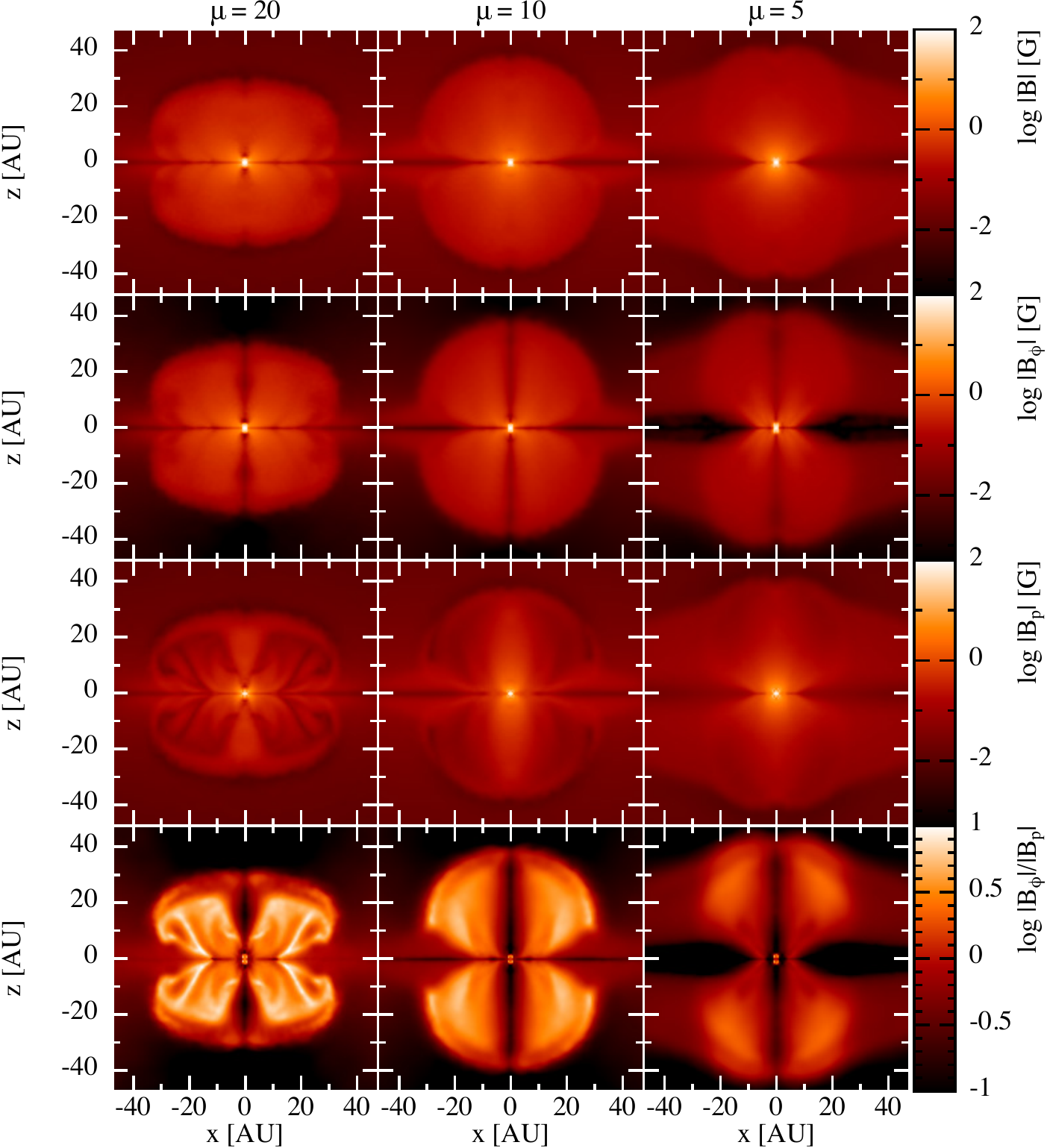}
    \hspace{0.5cm}
    \includegraphics[width=\columnwidth]{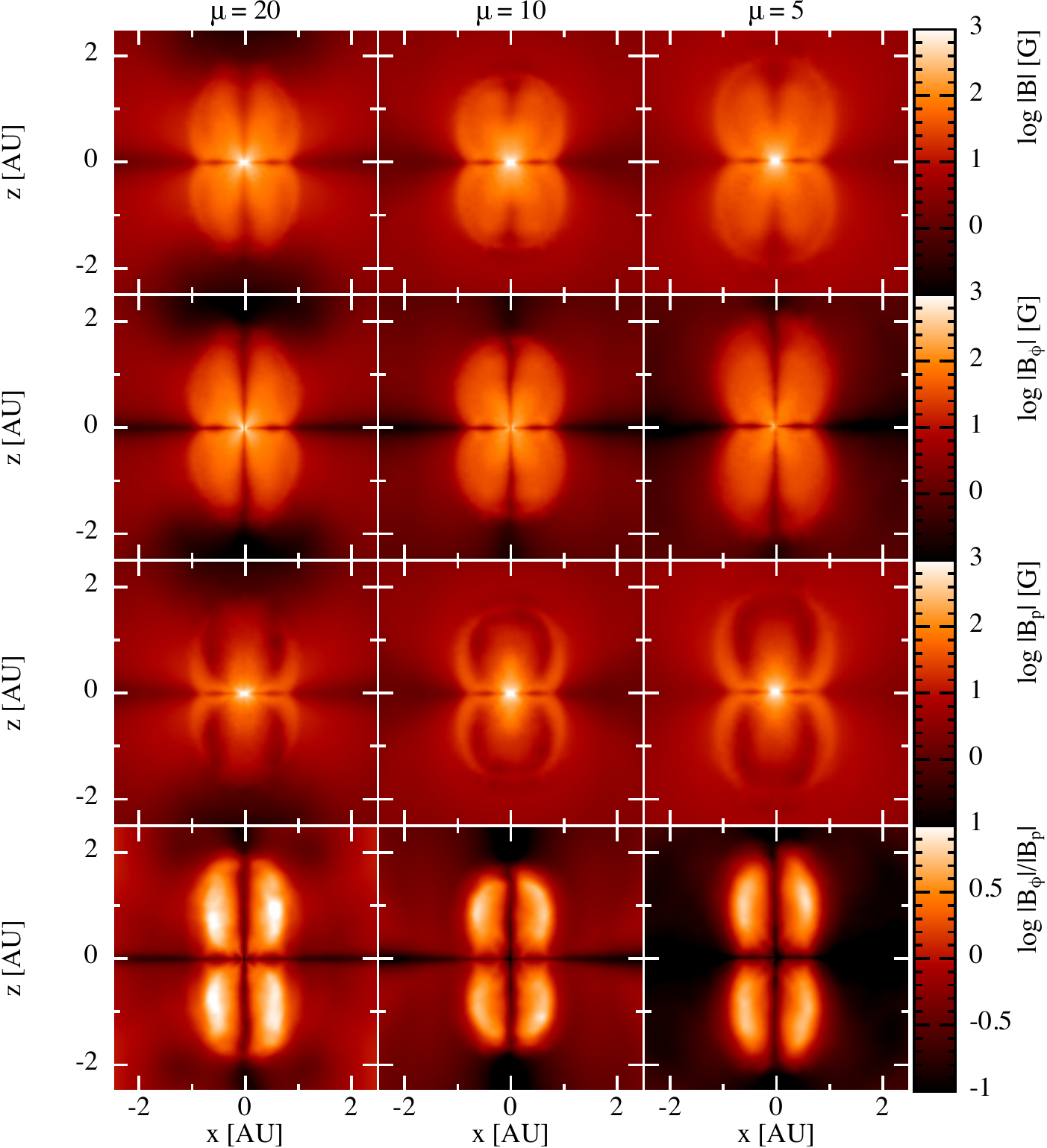}
\caption{Cross sections of the total magnetic field strength (top row), magnitude of the toroidal/azimuthal field $\vert B_{\phi}\vert$ (second row), magnitude of the poloidal field $\vert B_{p}\vert = \sqrt{B_{r}^{2} + B_{z}^{2}}$ (third row) and the ratio $B_{\phi}/B_{p}$ (bottom row) in the outflows from the first core (left figure) and the second core (right figure). The plots are shown 1 year after stellar core formation in the calculations with initial mass-to-flux ratios of $\mu=20$, 10 and 5 times critical (left to right). The poloidal component of the field becomes progressively more important as the interstellar magnetic field strength is increased. The corresponding field structure is shown in Fig.~\ref{fig:fieldlines}.}
\label{fig:torpol}
\end{figure*}

\begin{figure*}
\centering
    \hspace{-1cm}
    \includegraphics[width=0.48\textwidth]{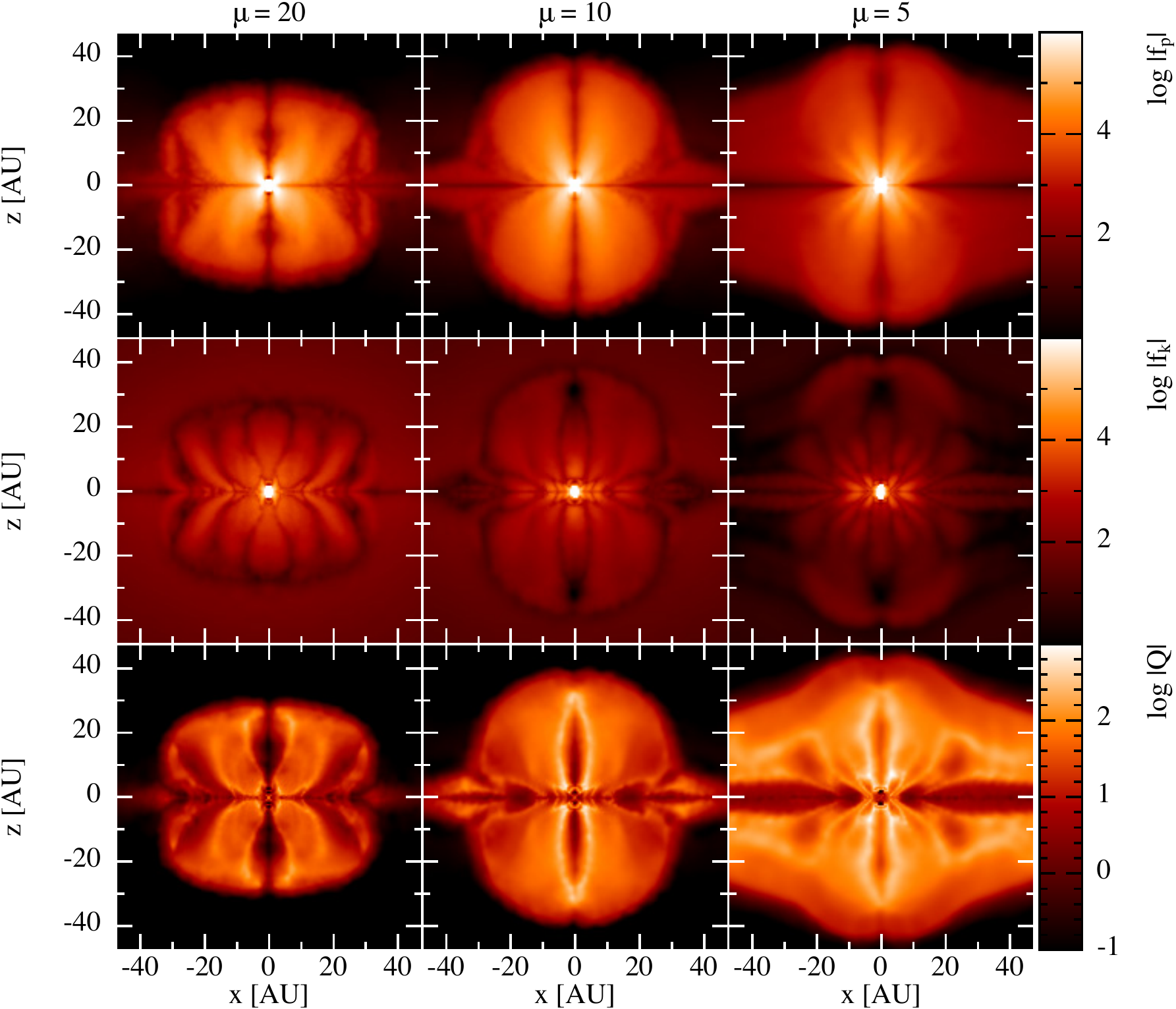}
    \hspace{0.5cm}
    \includegraphics[width=0.48\textwidth]{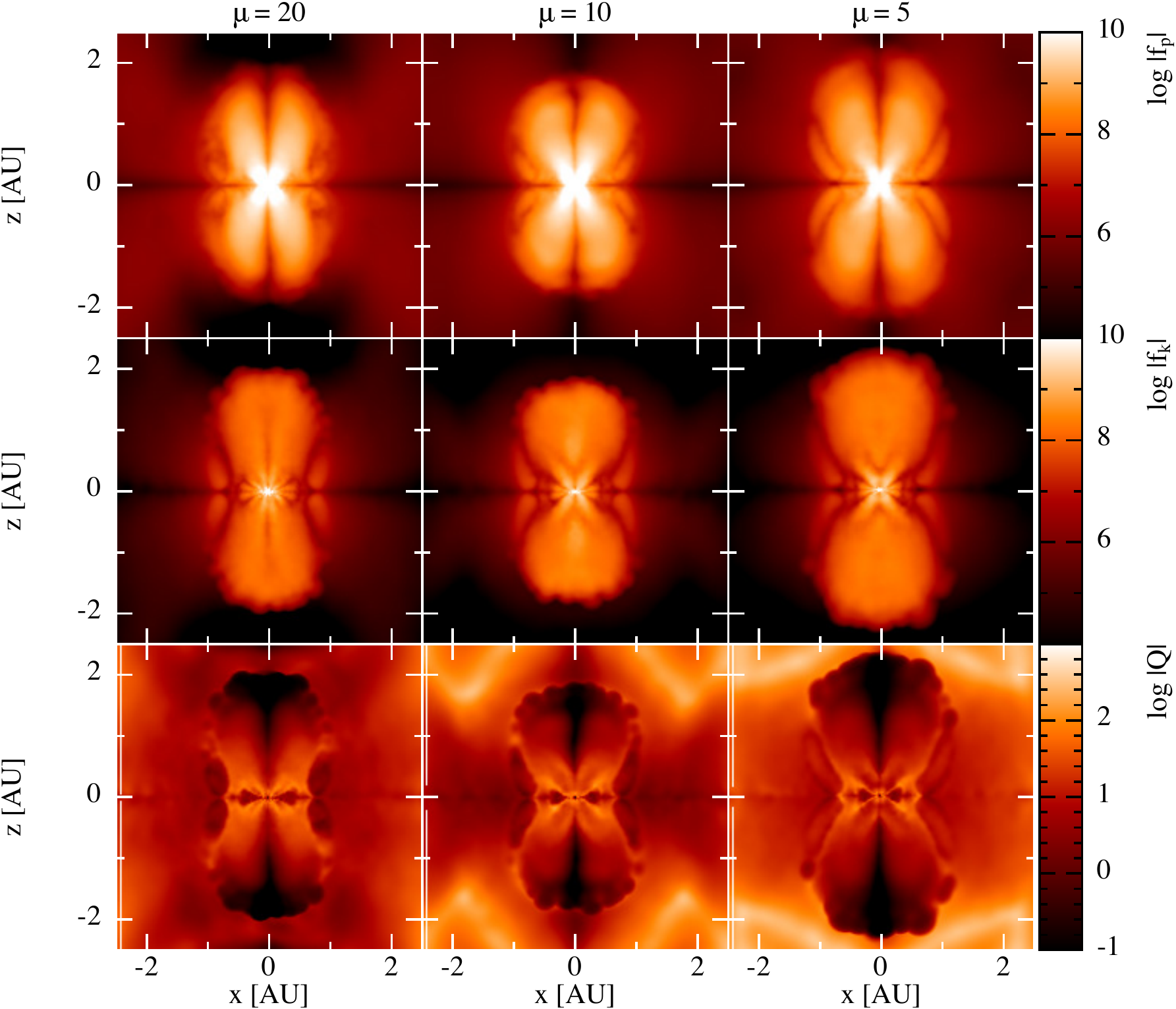}
    \hspace{-1cm}
\caption{Cross sections of the magnitude of the vertical component of the Poynting flux $f_{\rm P}$  (top row), magnitude of the vertical component of the kinetic flux $f_{\rm k}$ (second row), and their ratio $Q$ (bottom row) in the outflows from the first core (left panels) and the second core (right panels). The plots are shown 1 year after stellar core formation in the calculations with initial mass-to-flux ratios of $\mu=20$, 10 and 5 times critical (left to right).  All the outflows are Poynting flux dominated away from the outflow axis.  In the outflows from the first cores, the gas is actually infalling along much of the outflow axis.  In the outflows from the stellar cores, the flow is outwards along the axis but the kinetic flux dominates over the Poynting flux.}
\label{poynting}
\end{figure*}

In Fig.~\ref{fig:lorentz}, for the $\mu =10$ calculation one year after stellar core formation, we plot as functions of distance along the rotation axis the magnitude of the gravitational acceleration (i.e. $GM_{\rm enc} /r^2$) where $M_{\rm enc}$ is the mass enclosed within radius $r$, the magnitude of the radial acceleration due to thermal pressure (i.e. $|{\rm d} P/{\rm d}r|/\rho$), and the magnitude of the vertical component of the Lorentz acceleration (i.e. $| \mbox{\boldmath$J$} \times \mbox{\boldmath$B$}|_z/\rho$).  It can be seen that the Lorentz acceleration exceeds the gravitational acceleration from distances of $\approx 0.2-2$~AU and $\approx 6-40$~AU.  Thus, both the stellar outflow and first core outflow are magnetically driven, as expected.  We note, however, that the outward acceleration due to the thermal pressure gradient also exceeds the gravitational acceleration between $\approx 0.4-2$~AU and also at radii $<0.2$~AU.  Thus, the outflow from the vicinity of the stellar core is both magnetically and thermally driven.

Outflows driven primarily by magnetic forces are expected to be Poynting flux dominated close to where they are launched but how long they remain Poynting flux dominated before becoming essentially hydrodynamic is unknown.  Two magnetically launched classes of outflow are usually considered: magneto-centrifugal jets \citep*{BlaPay1982, OuyPud1997, BlaFraWel2001} or Poynting-flux dominated magnetic tower jets \citep{ShiUch1986, LyndenBell1996, Ustyugovaetal2000, Lovelaceetal2002, NakMei2004}.  In the former, magnetic fields only dominate out to the Alfv\'en radius, while in the latter magnetic fields may dominate to much larger distances.  In this respect, it is also of interest to investigate the poloidal and toroidal components of magnetic field since, conventionally, magneto-centrifugal jets are associated with launching along poloidal field lines, while magnetic towers jets carry large electric currents which generate strong tightly wound helical magnetic fields around the jet axis.  

Fig.~\ref{fig:fieldlines} shows the geometry of the magnetic field and corresponding electric currents in the outflows from both the first and second core for the $\mu = 20$, 10 and 5 calculations. A strong toroidal field is visible at the base of the outflows from both the first and second cores, which is strongest in the weakest field case ($\mu =20$). The poloidal component of the field starts to become significant for the $\mu=10$ calculation and is most strongly visible in the $\mu=5$ calculation. The same general trend occurs in both the first and second core outflows (comparing the left and right parts of Fig.~\ref{fig:fieldlines}). The relative proportions of toroidal to poloidal field in each case are quantified in Fig.~\ref{fig:torpol}, which shows the magnetic field strengths in a $y=0$ cross section through the domain on the scale of the first core (left) and second core (right) outflows. The magnitude of the toroidal field is computed as $\vert B_{\phi} |$, while the magnitude of the poloidal field is computed as $\vert B_{p} \vert = \sqrt{B_{r}^{2} + B_{z}^{2}}$, where $B_{r}, B_{\phi}$ and $B_{z}$ are the components of the field in cylindrical coordinates \citep[e.g.][]{Parker1955}. The lower panel in each figure shows the ratio $\vert B_{\phi} \vert/\vert B_{p} \vert $. Again, the trend seen in Fig.~\ref{fig:fieldlines} is visible, where the relative strength of the toroidal component decreases with increasing initial mass-to-flux ratio. This trend is strongest in the first core outflow (fourth row of left half of Fig.~\ref{fig:torpol}) but is also apparent in the second core outflow. In particular the `halo' of poloidal field around the second core outflow (third row of right half of Fig.~\ref{fig:torpol}) becomes more prominent as the initial interstellar magnetic field strength is increased. This is reflected in the less tightly wrapped field line geometry of the stellar core outflow for the $\mu=10$ and $\mu=5$ calculations seen in Fig.~\ref{fig:fieldlines}. Typical magnetic field strengths in the outflows can also be seen in Fig.~\ref{fig:torpol}, where we find $\vert B \vert \sim 1 $~G in the first core outflows and $\vert B \vert \sim 100$~G in the outflows from the second core.
  
We also compare the vertical component of the Poynting flux 
\begin{equation}
f_{\rm P} = [\mbox{\boldmath$B$} \times ( \mbox{\boldmath$v$} \times \mbox{\boldmath$B$})]_z
\end{equation}
with the vertical component of the kinetic flux
\begin{equation}
f_{\rm k} = \frac{1}{2} \rho v^2 v_z.
\end{equation}
\cite{HuarteEspinosaetal2012} plot the ratio $Q = f_{\rm P}/f_{\rm k}$ of these fluxes and find that the cores of their jets are dominated by kinetic energy flux, while the bulk of the jets is Poynting flux dominated.  In Fig.~\ref{poynting}, we plot the magnitudes of $f_{\rm P}$, $f_{\rm k}$, and $Q$ for both the first core and stellar outflows in each of the $(M/\Phi)_{\rm crit} =5$ and 10 calculations .  
 We also find that the bulk of the outflows are Poynting flux dominated, showing that the outflows are still magnetically-dominated and have not yet reached the hydrodynamic regime.  Near the outflow axis, however, the situation depends on the outflow.  As shown in Fig.~\ref{fig:1stOutflow}, gas is actually infalling along the axis of the outflows from the vicinity of each first core.  For the stellar outflows, though the gas along the axis is outflowing, the kinetic flux dominates over the Poynting flux as in the calculations of \citeauthor{HuarteEspinosaetal2012}

\subsection{Magnetic fields in stars}
  The strength and geometry of magnetic fields implanted in protostars during the star formation process is unknown. Since young low mass stars are fully convective it is generally assumed that the birth fields are quickly diffused and replaced by dynamo-generated fields \citep{ChaKuk2006}. On the other hand, observational studies have so far failed to find any correlation between the measured magnetic field properties in T-Tauri stars and stellar properties thought to be important for dynamo action \citep{JohnsKrull2007,YanJoh2011}, leading to speculation about a primordial or `fossil' origin for stellar magnetic fields even in low mass stars \citep{Tayler1987,Moss2003,YanJoh2011}. Observational measurements of magnetic fields in young T-Tauri stars via Zeeman broadening of spectral lines indicate mean surface magnetic field strengths of $\sim 1-4$~kG, with no strong dependence of the field strength with age, though the magnetic flux appears to decay over time. 
  
  The magnetic field strengths we obtain in the stellar core are consistent with these measurements, reaching maximum values of $10-100$~kG in the stellar cores for the calculations with initial mass-to-flux ratios $\mu=20, 10$ and $5$ (Fig.~\ref{fig:maxdensitytemp}). The rapid decay of the field post-formation of the stellar core seen in Fig.~\ref{fig:maxdensitytemp} can be attributed to the numerical resistivity in our simulations, leading to enhanced diffusion of the magnetic field. In reality the stellar core should be fully ionised, leading to low resistivity and a much slower decay of the field after implantation. On the other hand the effective diffusivity may be enhanced by convective or turbulent motions \citep{Rudigeretal2011}. Although there are many uncertainties involved, \citet{Moss2003} found that at least some of the implanted fossil magnetic field should be able to survive the pre-main sequence, perhaps combining with a dynamo-generated field while the star is fully convective.

   The geometry of the field implanted in the stellar core in the $\mu=5$ calculation is shown in Fig.~\ref{fig:mfstar}, showing the magnetic field (left) and current density (right) 0.2 years after the formation of the stellar core. The field is mainly poloidal, reflective of the entrained interstellar magnetic field. While this is suggestive of the simple dipole-like magnetic field geometries observed in M-dwarfs close to the fully convective limit \citep{Morinetal2008}, some caution is required since details of the stellar magnetic field in our calculations may depend strongly on the numerical resistivity.

\section{Conclusions}
\label{conclusions}

We have presented the results from a series of radiation magnetohydrodynamical calculations of the collapse of rotating, magnetised molecular cloud cores to form protostars.  In each calculation the initial magnetic field was uniform and parallel to the rotation axis and five calculations were performed with different initial field strengths, corresponding to mass-to-flux ratios relative to the critical value required for collapse that varied from $\mu=\infty$ (i.e. no magnetic field) to $\mu=5$.  
Each calculation was followed through the initial isothermal collapse, the formation of the first hydrostatic core, the second collapse, and the formation of the stellar core.  Following \cite{Tomidaetal2013}, this is only the second time that such calculations have been published and we have been able to follow the evolution of the stellar core and its outflow further.  

Our detailed conclusions are as follows.

\begin{enumerate}
\item Stronger initial magnetic fields result in stronger angular momentum transport, producing first hydrostatic cores with smaller radii.  In particular, for our chosen initial conditions the first core in the unmagnetised case rotates rapidly enough to undergo a bar-mode instability, producing a `pre-stellar disc' \citep{Bate2011} with spiral arms and a radius of $\approx 50$~AU before the stellar core forms.  Weak magnetic fields ($\mu=100$) do not brake the gas strongly enough to avoid this dynamical instability, but stronger fields ($\mu \lsim 20$) provide sufficient magnetic braking for the first cores to be rotationally stable and remain axisymmetric.

\item In agreement with previous calculations performed both with and without radiative transfer, we find that calculations with substantial initial magnetic fields ($\mu=5-20$) generate two outflows, a slow outflow ($\approx 2$~km/s) from the vicinity of the first core, and a fast outflow ($\approx 10$~km/s) from the vicinity of the stellar core. In each case, we are able to follow the fast outflow until it has broken out of the remnant of the first core, the first time this has been examined in radiation magnetohydrodynamical calculations.  Without a magnetic field, or with $\mu=100$, no outflow is launched from the first core, but a thermally-driven bipolar outflow is eventually launched from the vicinity of the stellar core in the unmagnetised case \citep[as found by][]{Bate2010, Bate2011, SchTsc2011} and presumably would also be launched in the $\mu=100$ calculation if we were able to follow it further.

\item Both the slow and fast outflows are driven by magnetic forces, but in the case of the stellar outflow both thermal forces and magnetic forces drive the outflow as they both overwhelm the gravitational force and have similar magnitudes.  The toroidal component of the field dominates over the poloidal component in all of the outflows, but the ratio of the toroidal to poloidal components decreases with increasing initial field strength.  As expected, all of the outflows are Poynting flux dominated in the bulk of the volume occupied by the outflows.  However, for those generated by the first cores the outflowing material surrounds the rotation/outflow axis with infall continuing along the axis itself.  For the outflows generated in the vicinities of the stellar cores the outflow includes the rotation axis (perhaps due to the role of the thermal forces), but while the bulk of the outflows are Poynting flux dominated, the kinetic flux dominates along the outflow/rotation axis.

\begin{figure}
\centering
    \includegraphics[width=0.9\columnwidth]{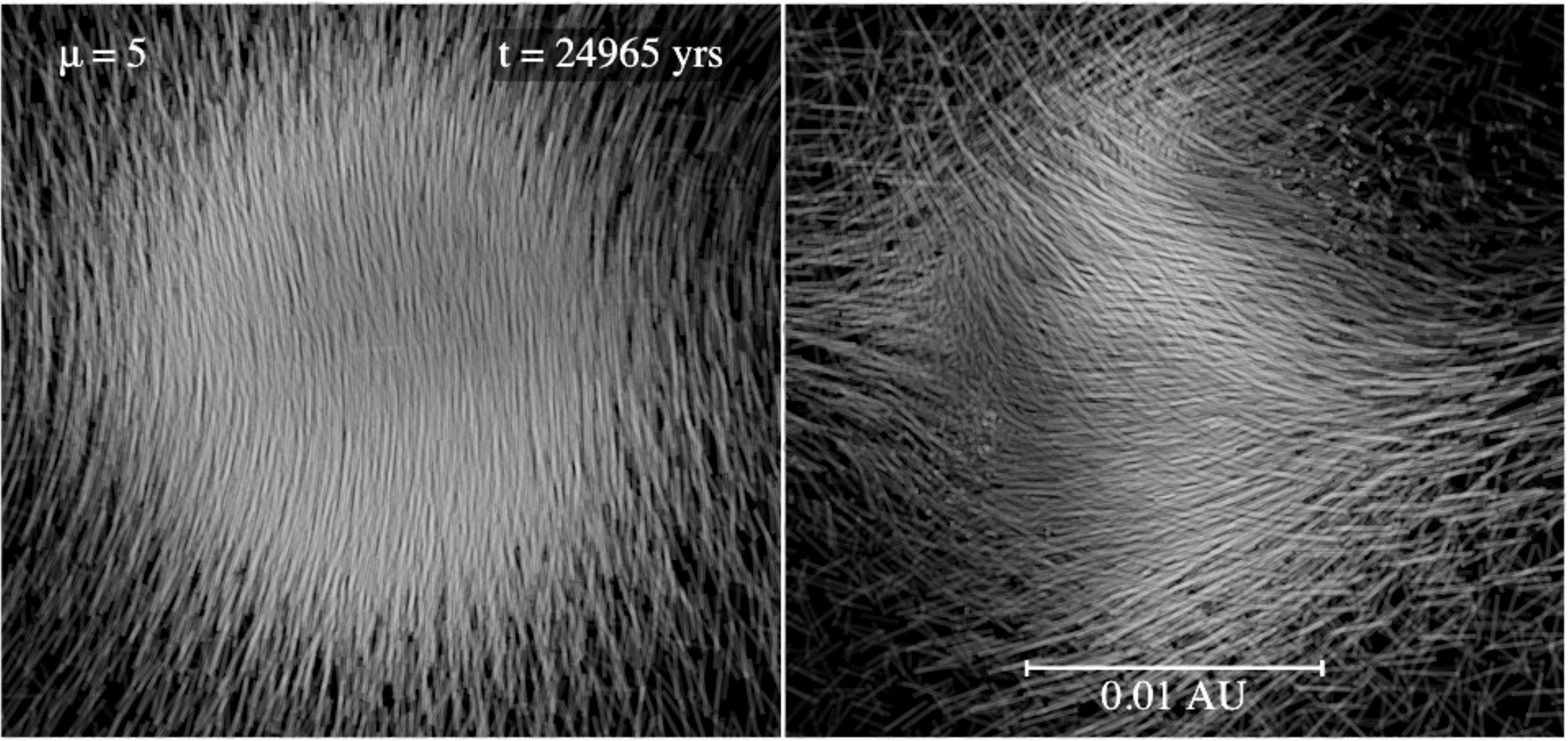}
\caption{Magnetic field (left) and current density (right) in the stellar core from the $\mu = 5$ calculation, shown 0.2 years after the formation of the stellar core. The magnetic field geometry in the stellar core is mainly poloidal (left panel), consistent with the field being entrained in the stellar core from larger scales. The maximum field strength in the stellar core at this stage is $\sim 30$~kG.}
\label{fig:mfstar}
\end{figure}

\item We have examined the rotation of the outflows.  We find that the rotation speeds of the outflows launched from the first core are similar to the outflow speeds, while the ratio of the rotation speed to the outflow speed is much lower in the stellar outflows.  For both outflows we find that the rotation speeds are somewhat greater with lower initial magnetic field strengths.

\item We follow the stellar cores until they have grown to $5-20$ Jupiter-masses (stronger magnetic fields produce faster growth rates).  In the strongly magnetised cases ($\mu=5-20$) the stellar cores accrete from inspiralling pseudo-discs in which the infall speeds are similar to their rotation speeds, while in the unmagnetised and $\mu=100$ calculations the stellar cores accrete from near-Keplerian discs.  However, in all cases, the stellar cores have radii of $\approx 3$~R$_\odot$ and similar structures, regardless of the initial magnetic field strength.  Their radial entropy profiles increase with radius and, thus, they are convectively stable (their central temperatures are insufficient for fusion to occur at this early stage).  We measure their central entropy per baryon to be $s/k_{\rm B} \approx 14$.  It is possible that stellar cores have almost `universal' initial properties that are set by the physics of molecular hydrogen dissociation and the ensuing second collapse.  Further calculations are required to test this hypothesis.

\item We find that the radiative luminosity of the first cores has the same magnitude as the power released by accretion onto the first cores, consistent with one-dimensional calculations that show that essentially all of the kinetic energy from the accretion flow onto the first core is radiated away \citep{Commerconetal2011b, Vaytetetal2012}.  By contrast, however, we show that the radiative losses of the stellar cores are negligible at this early stage of their evolution as the kinetic energy flux exceeds their radiative flux by up to 9 orders of magnitude.   Thus, essentially all of the energy of the accretion flow is advected into the stellar core.  Again, this is consistent with the results of one-dimensional calculations \citep{Vaytetetal2013}, except that here the stellar core accretes via a boundary layer from a disc or pseudo-disc as opposed to through a purely radial shock.

\item Finally, we obtain magnetic fields in the stellar cores that initially have maximum values of $\sim 10- 100$~kG and have a predominantly poloidal structure.  While there are many limitations to our calculations (e.g. they do not treat non-ideal MHD effects), the calculations at least allow for the possibility that fossil magnetic fields implanted during star formation may survive the pre-main-sequence, perhaps combining with a dynamo-generated field once convection begins.

\end{enumerate}

The strongest magnetic field we consider provides a molecular cloud core with an initial mass-to-flux ratio of $\mu=5$ times critical.  In reality, even higher mass-to-flux ratios may be common \citep[e.g.][]{Crutcheretal2010}.  The main effect of even higher initial mass-to-flux ratios would be to constrain the initial collapse to be even more pancake-like than our strongest field case, to produce an even stronger pseudo-disc, and to remove even more angular momentum during the collapse.  However, since the $\mu=5$ case already produces a slowly-rotating almost-spherical first core and we find the properties of the outflows and stellar core do not depend sensitively on the initial field strength, we do not expect the later evolution with an even stronger field to differ greatly from the $\mu=5$ case.

\section*{Acknowledgments}

MRB thanks Monash University for their funding and hospitality during his study leave in 2011 when this work was initiated, and DJP thanks the University of Exeter for a Visiting International Academic Fellowship and hospitality during which this work was completed.  TST acknowledges support by Endeavour IPRS and APA postgraduate research scholarships.  We also  acknowledge funding via the Australian Research Council Discovery Project grant DP1094585.  

The calculations for this paper were performed on the DiRAC Complexity machine, jointly funded by STFC and the Large Facilities Capital Fund of BIS, and the University of Exeter Supercomputer, a DiRAC Facility jointly funded by STFC, the Large Facilities Capital Fund of BIS, and the University of Exeter. \textsc{splash} \citep{Price2007} was used to create the rendered figures and field line visualisations.

\begin{figure*}
\centering \vspace{0.0cm}
    \includegraphics[width=15.0cm]{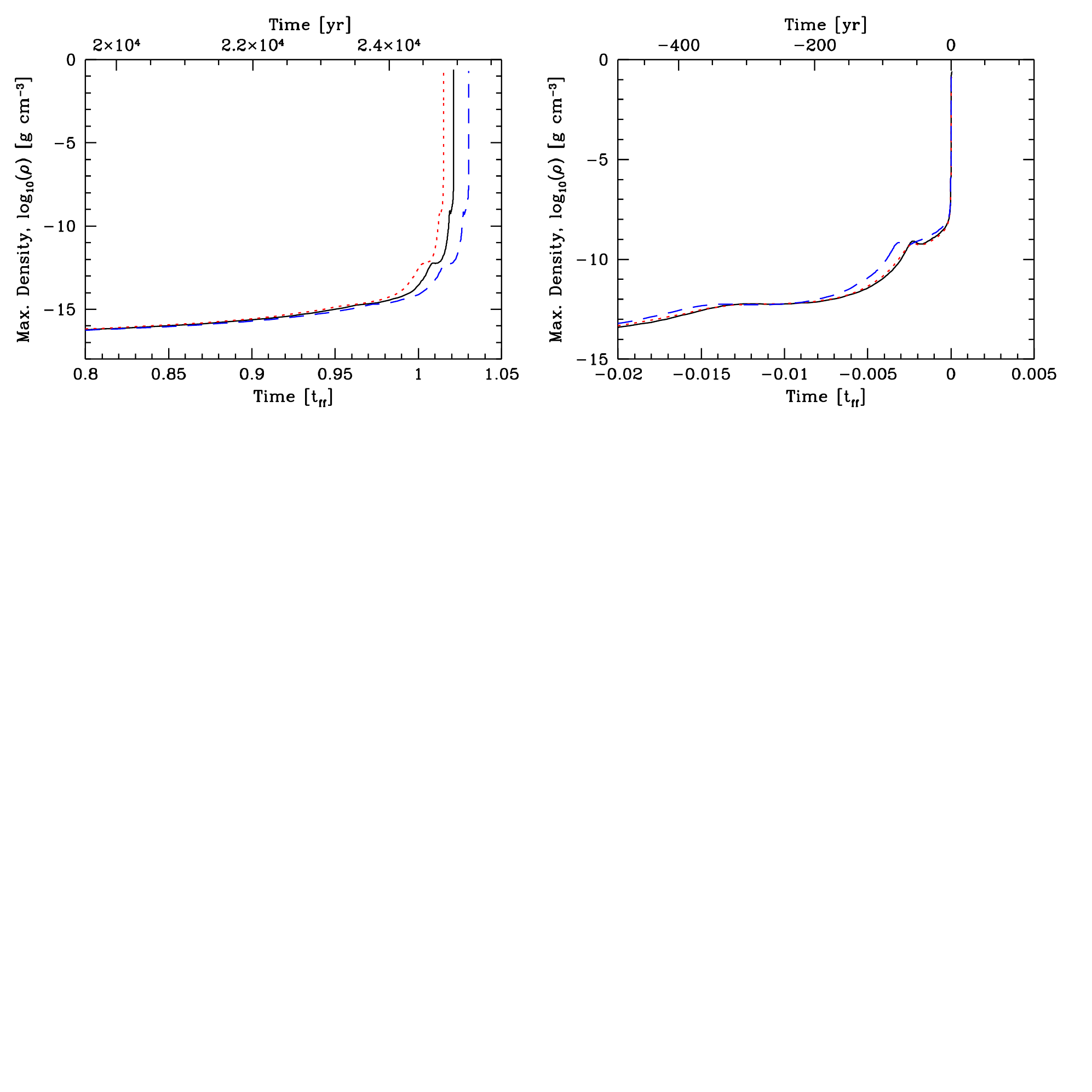}\vspace{-9.5cm}
\caption{The time evolution of the maximum density during radiation magnetohydrodynamical calculations of the collapse of a rotating molecular cloud core with an initial mass-to-flux ratio of $\mu=5$ times critical performed using resolutions of 1 (red dotted line), 3 (black solid line), and 10 (blue dashed line) million particles.  The free-fall time of the initial cloud core, $t_{\rm ff}=7.71\times 10^{11}$~s (24,430 yrs).  In the right panel the time has been set to zero when the stellar core begins to form (i.e. when the maximum density reaches $10^{-4}$~g~cm$^{-3}$). 
}
\label{fig:convergence_denstime}
\end{figure*}

\begin{figure*}
\centering \vspace{-0.5cm}
    \includegraphics[width=15.0cm]{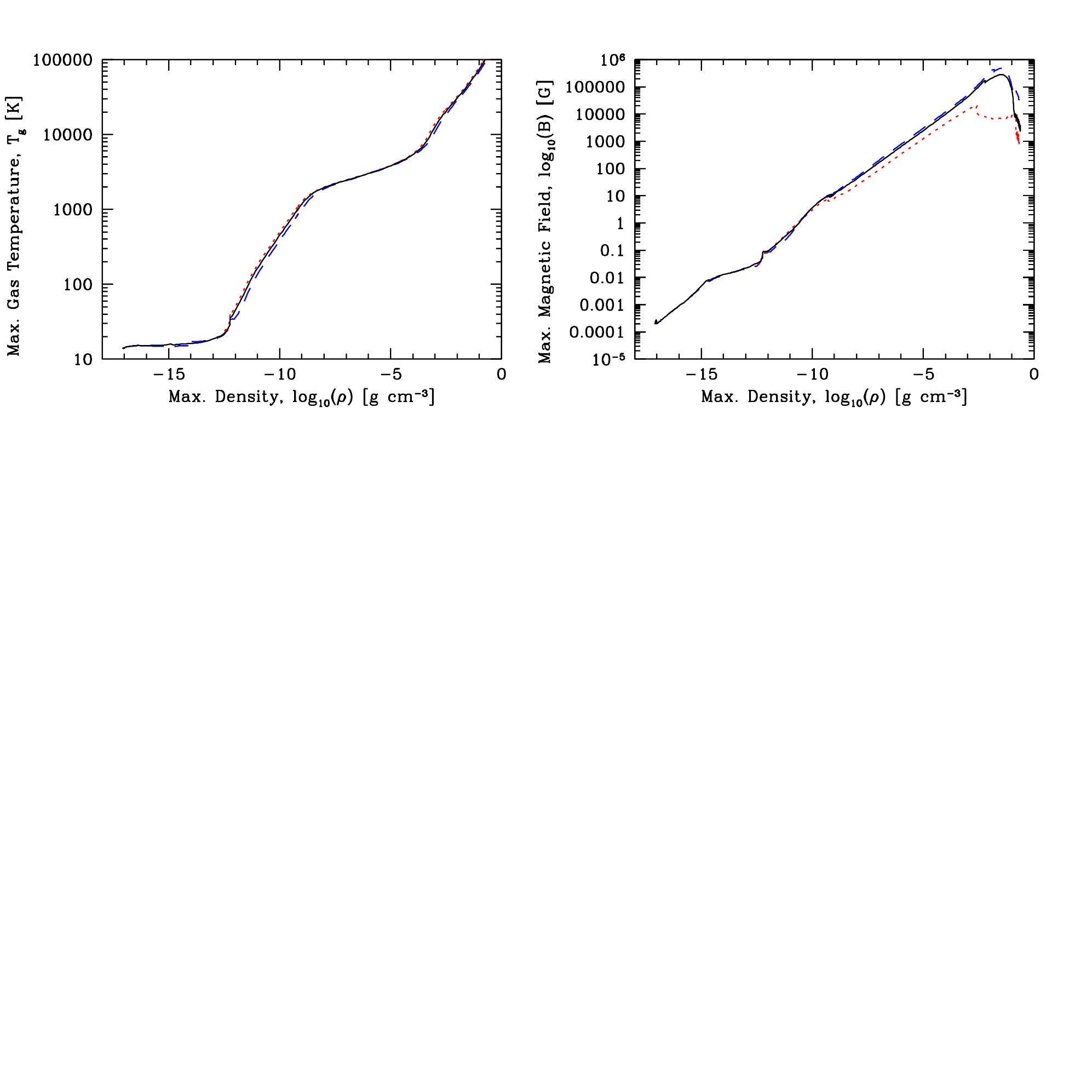}\vspace{-9.5cm}
\caption{The evolution of the maximum gas temperature (left) and maximum magnetic field strength (right) versus maximum density for RMHD calculations of the collapse of a rotating molecular cloud core with an initial mass-to-flux ratio of $\mu=5$ times critical performed using resolutions of 1 (red dotted line), 3 (black solid line), and 10 (blue dashed line) million particles.  With lower resolution the field strength late in the collapse is reduced due to the increased numerical resistivity.
}
\label{fig:convergence_tempfield}
\end{figure*}

\begin{figure*}
\centering \vspace{-0.0cm}
    \includegraphics[height=3.5cm]{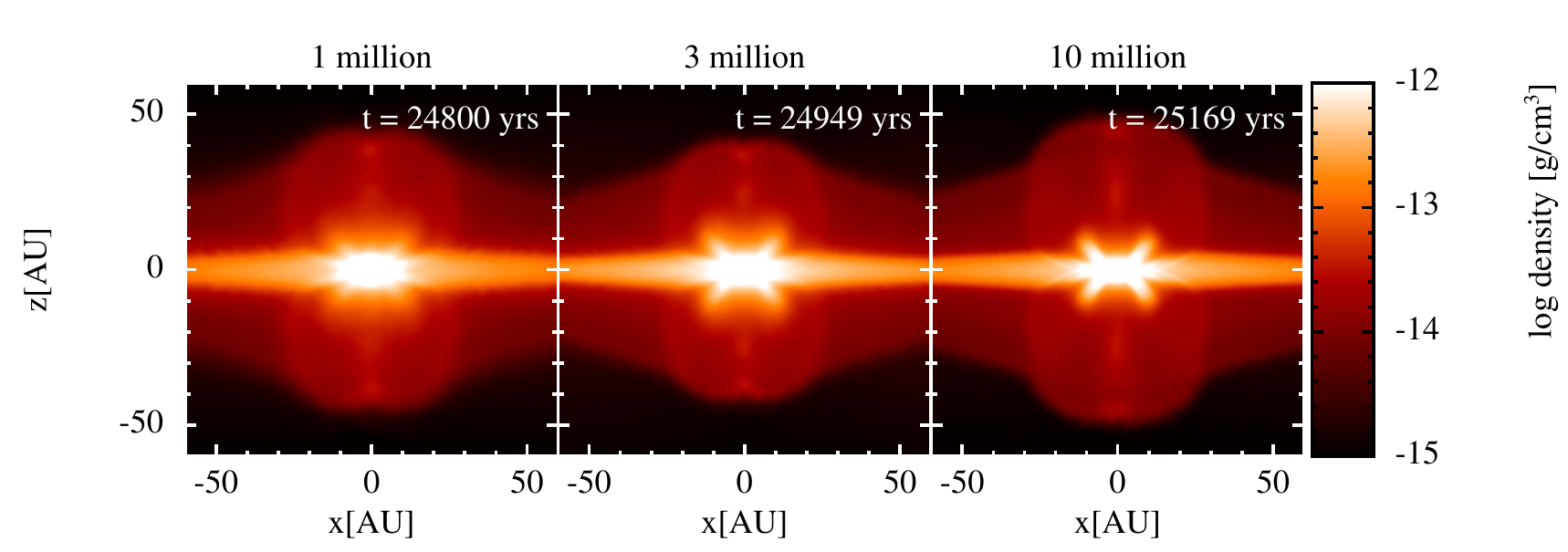}\hspace{0.0cm}
    \includegraphics[height=3.5cm]{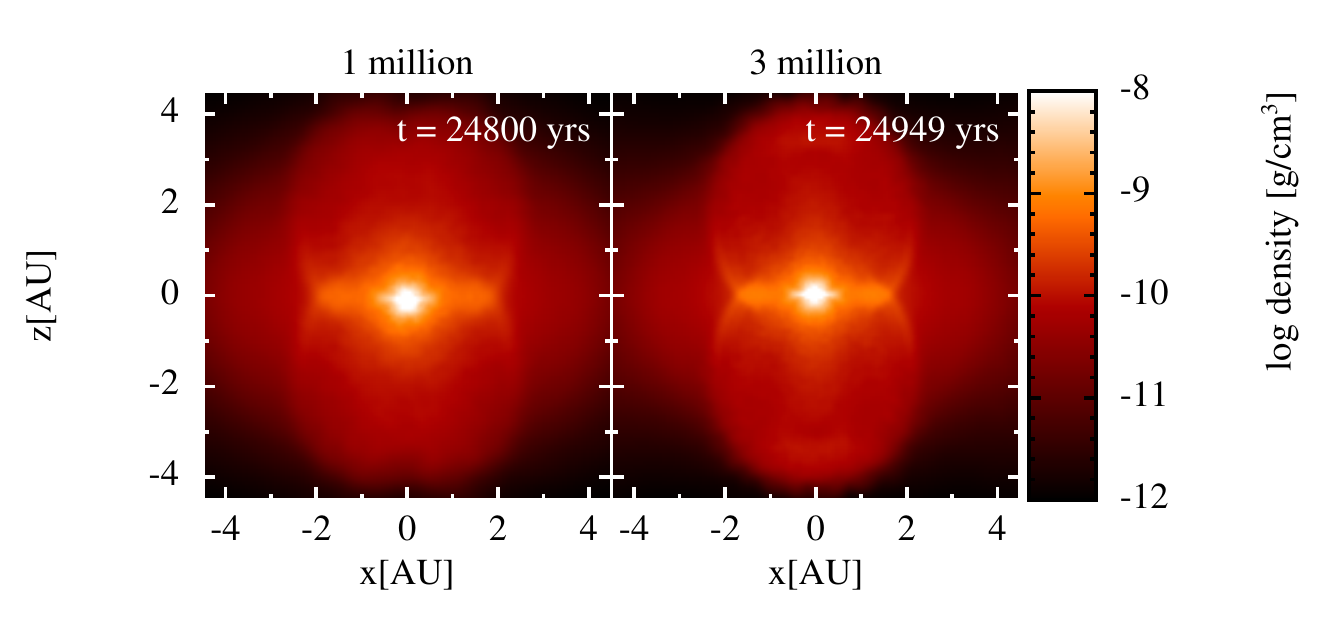}\vspace{-0.0cm}
\caption{Snapshots of the density on slices parallel to the rotation axis showing the development of the outflows that are launched from the first hydrostatic cores (left panels) and stellar cores (right panels) in calculations with initial mass-to-flux ratios of $\mu=5$ times critical performed with different resolutions as labelled above each panel.  }
\label{fig:convergence_images}
\end{figure*}

\begin{figure*}
\centering \vspace{0.0cm}
    \includegraphics[width=15.0cm]{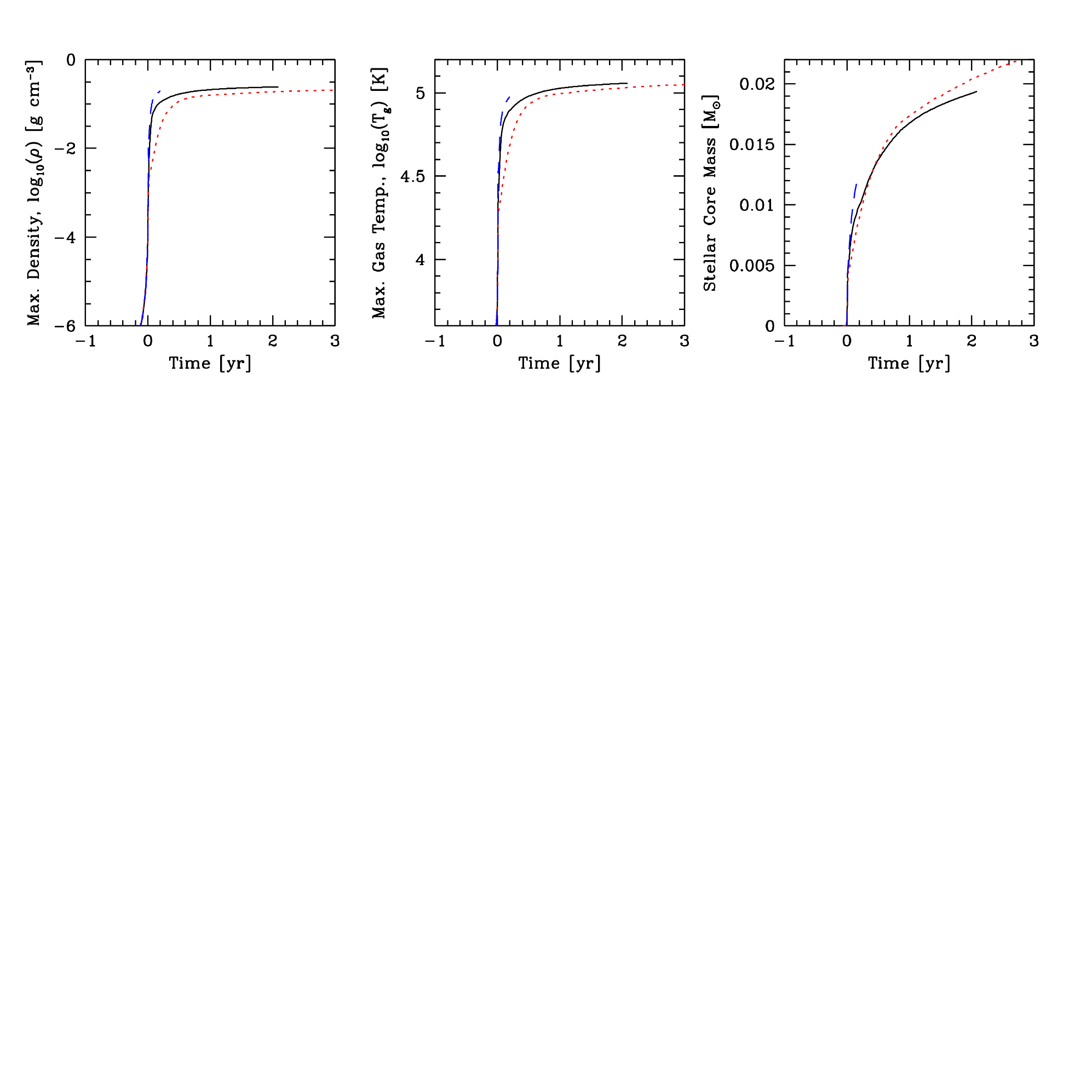}\vspace{-9.5cm}
\caption{The time evolution of the maximum density (left panel), maximum gas temperature (centre panel), and stellar core mass (right panel) during the radiation magnetohydrodynamical calculations of the collapse of a molecular cloud core with initial mass-to-flux ratios $\mu=5$ times critical performed using resolutions of 1 (red dotted line), 3 (black solid line), and 10 (green dashed line) million particles. The time is measured in years from the formation of the stellar core, which is defined as being the moment when the maximum density reaches $10^{-4}$~g~cm$^{-3}$. 
}
\label{fig:stellarmass_convergence}
\end{figure*}

\bibliography{mbate}

\section*{Appendix}

As mentioned in Section \ref{resolution}, calculations of the collapse of the rotating molecular cloud core with initial mass-to-flux ratio $\mu=5$ times critical were performed with three different numerical resolutions: 1, 3, and 10 million particles.  The results are very similar at all three resolutions though, as expected, there are some minor quantitative differences which we briefly illustrate in this Appendix. 

In Fig.~\ref{fig:convergence_denstime}, we plot the evolution of the maximum density versus time for the three resolutions.  The overall collapse of the cloud takes slightly longer as the resolution is increased ($\approx 1$\% for each increase in resolution).  In our experience this is typical for such calculations, even without magnetic fields, presumably due to differences in the accuracy to which pressure gradients are resolved.  However, normalising time to the moment at which the stellar cores begin to form (when the maximum density reaches $10^{-4}$~g~cm$^{-3}$; right panel) it can be seen that the lifetime of the first core is similar in all three calculations (and very short, at approximately 100 yrs).  The evolution of the maximum temperature and maximum magnetic field strength with maximum density is very similar for the three calculations (Fig.~\ref{fig:convergence_tempfield}).  However, the maximum magnetic field strength peaks lower with decreasing resolution towards the end of the collapse (i.e. in the second collapse phase and in the stellar core).  This is unsurprising as it is expected that the numerical resistivity should be higher with reduced resolution.

In Fig.~\ref{fig:convergence_images} we provide images of the outflows that are driven from the first and stellar cores, respectively.  The structures of both type of outflows are very similar between the calculations with different resolutions.  Similarly, the outflow speeds are very similar for each of the resolutions ($\approx 2$ km/s for the first outflow and $\approx 10$ km/s for the outflow from the stellar core).   We do not provide an image of the stellar outflow from the 10 million particle calculation since we are only able to follow it to a fraction of an AU due to the increased computational expense.

\begin{figure}
\centering \vspace{-0.5cm}
    \includegraphics[width=13cm]{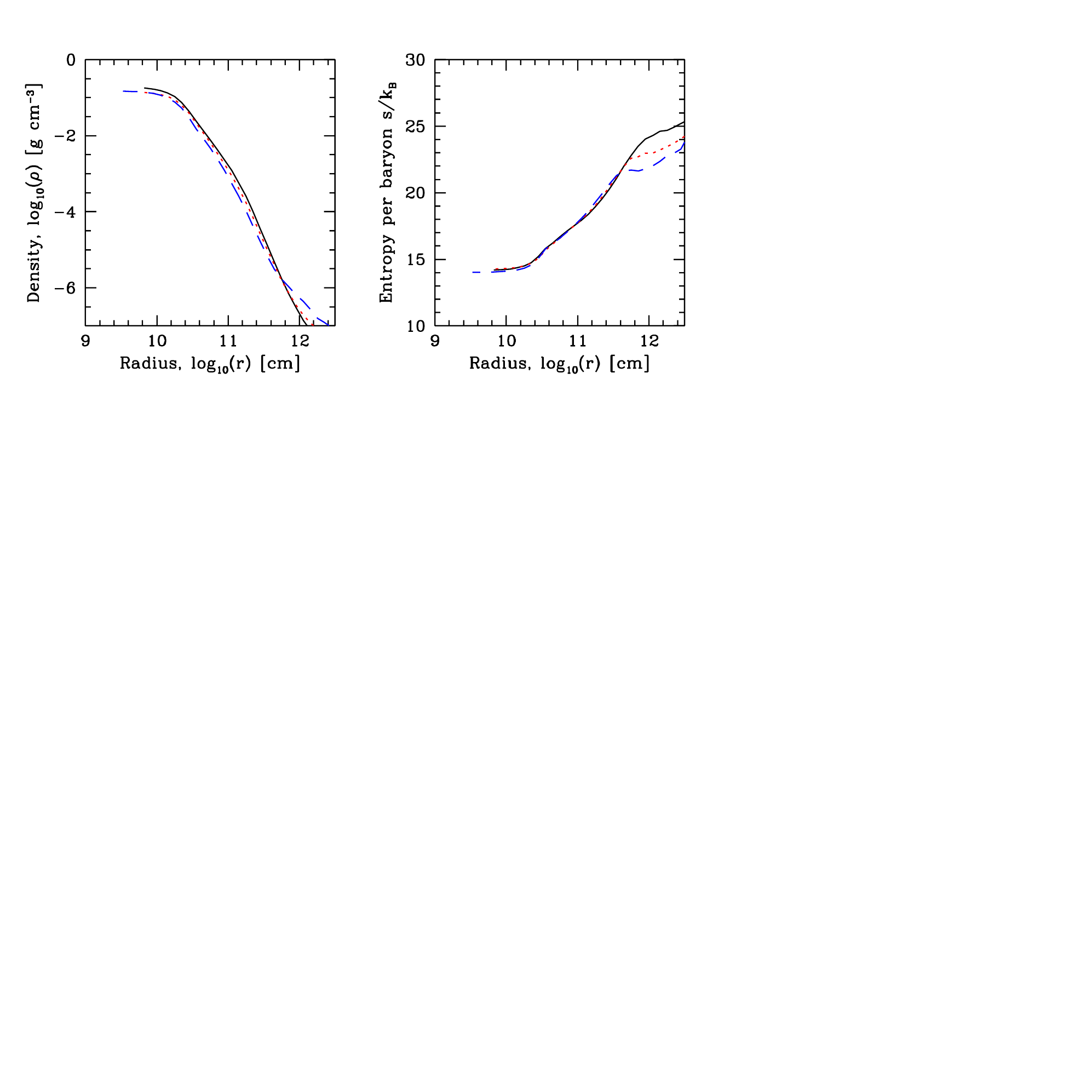}\vspace{-8.2cm}
\caption{The density and entropy structure of the stellar cores formed in radiation magnetohydrodynamical calculations with initial mass-to-flux ratio $\mu=5$ times critical performed using resolutions of 1 (red dotted line), 3 (black solid line), and 10 (blue dashed line) million particles.  The structure is shown approximately 1 year after stellar core formation for the calculations with 1 and 3 million particles, and 0.2 years after stellar core formation for the 10 million particle calculation.  We provide plots of the azimuthally-averaged midplane density and entropy per baryon as functions of radius.  The properties are insensitive to variations of the numerical resolution.
}
\label{fig:stellarcoreprop_convergence}
\end{figure}

Finally, in Figs.~\ref{fig:stellarmass_convergence} and \ref{fig:stellarcoreprop_convergence} we examine the evolution and properties of the stellar cores produced by each of the calculations.  The growth rates of the stellar cores are also similar.  They tend to be slightly higher with higher resolution early in the formation of the stellar core ($\lsim 0.4$~yr), possibly due to the greater field strengths, while are slightly lower with higher resolution later on (possibly due to the increase of numerical viscosity with lower resolution).  Similarly the density and entropy structure of the stellar cores are very similar (Fig.~\ref{fig:stellarcoreprop_convergence}).

In summary, the results described in the main text do not depend significantly on resolution, at least to the highest resolution we are able to test.

\end{document}